\documentclass[12pt,a4paper]{article}
\usepackage[utf8]{inputenc}
\usepackage[T1]{fontenc}
\usepackage{amsmath}
\usepackage{amsfonts}
\usepackage{amssymb}
\usepackage[left=2.00cm, right=2.00cm, top=2.00cm, bottom=2.00cm]{geometry}
\usepackage{mathrsfs}
\usepackage{dsfont}
\usepackage{sectsty}
\usepackage{titling}
\usepackage[colorlinks,citecolor=red,linkcolor=blue]{hyperref}
\numberwithin{equation}{section}
\title{\textbf{The Gelfand-Tsetlin basis for infinite-dimensional representations of} $gl_n(\mathbb{C})$}
\author{Pavel V. Antonenko \\ \normalsize{\textit{St. Petersburg Department of Steklov Mathematical Institute,}} \\ \normalsize{\textit{St. Petersburg, Russia, e-mail:} \texttt{antonenko\_pavel@pdmi.ras.ru}.}}
\date{}

\begin{document}

	\begin{titlingpage}
		\maketitle
		\begin{abstract}
			We consider the problem of determination of the Gelfand-Tsetlin basis for unitary principal series representations of the Lie algebra $gl_n(\mathbb{C})$. The Gelfand-Tsetlin basis for an infinite-dimensional representation can be defined as the basis of common eigenfunctions of corner quantum minors of the corresponding L-operator. The construction is based on the induction with respect to the rank of the algebra: an element of the basis for $gl_n(\mathbb{C})$ is expressed in terms of a Mellin-Barnes type integral of an element of the basis for $gl_{n-1}(\mathbb{C})$. The integration variables are the parameters (in other words, the quantum numbers) setting the eigenfunction. Explicit results are obtained for ranks $3$ and $4$, and the orthogonality of constructed sets of basis elements is demonstrated. For $gl_3(\mathbb{C})$ the kernel of the integral is expressed in terms of gamma-functions of the parameters of eigenfunctions, and in the case of $gl_4(\mathbb{C})$ -- in terms of a hypergeometric function of the complex field at unity. The formulas presented for an arbitrary rank make it possible to obtain the system of finite-difference equations for the kernel. They include expressions for the quantum minors of $gl_n(\mathbb{C})$ L-operator via the minors of $gl_{n-1}(\mathbb{C})$ L-operator for the principal series representations, as well as formulas for action of some non-corner minors on the eigenfunctions of corner ones. The latter hold for any representation of $gl_n(\mathbb{C})$ (not only principal series) in which the corner minors of the L-operator can be diagonalized.
		\end{abstract}
	\end{titlingpage}

	{\hypersetup{linkcolor=black}
	\tableofcontents
	}

	\newpage

	\section{Introduction}

	The representation theory of semi-simple Lie algebras, including the algebra $gl_n(\mathbb{C})$, is of great importance for many areas of mathematics and modern mathematical physics \cite{M_A_Semenov_Tian_Shansky,Kulish_Sklyanin_QSTM,Ryan_Volin,Gromov_Kazakov_et_al,Frassek_Meneghelli,Cavaglia_Gromov_Levkovich-Maslyuk,Mishra_Srivastava}. In the seminal work \cite{Gelfand_Tsetlin}, I.M.~Gelfand and M.L.~Tsetlin introduced a construction which allowed them to derive explicit formulas for bases of finite-dimensional irreducible representations of $gl_n(\mathbb{C})$. Subsequently, an important result was provided by M.L.~Nazarov and V.O.~Tarasov in \cite{Nazarov_Tarasov_Yangians_and_Gelf_Tset_Bases,Nazarov_Tarasov_Reps_of_Yang_with_Gelf_Tset_Bases}, who proved that the Gelfand-Tsetlin basis for a finite-dimensional irreducible representation of $gl_n(\mathbb{C})$ is the basis of common eigenvectors of corner quantum minors of the L-operator.
	
	The L-operator is an $n$ by $n$ matrix $L(u)$ which is constructed from generators of a $gl_n(\mathbb{C})$ representation. It depends on the so-called spectral parameter $u$. The entries of the L-operator satisfy the commutation relations of the Yangian $\mathrm{Y}(gl_n(\mathbb{C}))$ \cite[\S~1.1]{Molev_Yangians_and_classical_Lie_algebras}. Quantum minors of $L(u)$ are defined in the same way as for usual matrices with commutative entries but involve shifts with respect to the spectral parameter \cite[\S~1.6]{Molev_Yangians_and_classical_Lie_algebras}. The definition of the Gelfand-Tsetlin basis from the viewpoint of the theory of Yangians can be used in the case of infinite-dimensional representations, which include the principal series representations considered in the present research.
	
	Principal series representations of the Lie algebra $gl_n(\mathbb{C})$ are derived from the corresponding representations of the Lie Group $SL(n,\mathbb{C})$ \cite{Gelfand_Naimark,Zhelobenko_Shtern_representations_of_Lie_groups,Der_Man_gen_sol_YBE_SL_n_C}. The latter are defined on the space of square-integrable functions on the group of $n$ by $n$ complex lower triangular matrices with units on the diagonal. There are two sets of generators for a principal series representation of $gl_n(\mathbb{C})$: holomorphic and antiholomorphic. These two sets commute with each other and each of them obeys $gl_n$ commutation relations. Therefore, there are holomorphic and antiholomorphic L-operators $L(u)$, $\bar{L}(\bar{u})$ (parameters $u$ and $\bar{u}$ are independent, not complex conjugate), and, consequently, holomorphic and antiholomorphic quantum minors. The eigenvalues of corner minors are polynomials in the spectral parameter $u$. The eigenfunctions are parametrized by the roots these polynomials.
	
	The purpose of the present paper is to derive explicit induction formulas expressing the Gelfand-Tsetlin basis for unitary principal series representations of $gl_n(\mathbb{C})$ in terms of a Mellin-Barnes type integral \cite{Kharchev_Lebedev_Toda_chain_Mellin_Barnes} of similar basis elements for $gl_{n-1}(\mathbb{C})$ (induction on the rank of the algebra). The Mellin-Barnes representations of corner minors' eigenfunctions are obtained in the cases of ranks $3$ and $4$. These results are useful since the principal series representations of $gl_4(\mathbb{C})$ are connected to some conformal field theories \cite{Gromov_Kazakov_et_al,Chich_Der_Isa_conformal_algebra}.
	
	In \cite{Shatashvili_92} it was shown that the expressions for correlation functions in some quantum field theories take the form of integrals over Gelfand-Tsetlin schemes. At the same time, computation of three-point correlation functions is a major problem in conformal field theory. It was recently determined \cite{CGL_18} that this problem is closely related to the quantum method of separation of variables discovered in the works of E.K.~Sklyanin \cite{Sklyanin_85}. And very recently it was observed that the Gelfand-Tsetlin basis for representations of $gl_n(\mathbb{C})$ plays the key role in this construction \cite{Ryan_Volin}. This idea was developed to the full extent in \cite{RV_21}. The separation of variables was first applied for unitary principal series representations of $gl_3(\mathbb{C})$ in \cite{DV_18}, and it is expected that the case of $gl_4(\mathbb{C})$ can be used to compute certain Feynman diagrams \cite{DO_20}, as was demonstrated in $gl_2(\mathbb{C})$ case \cite{DKO_19}.
	
	One way of induction on the rank for principal series representations has already been realized \cite{Valinevich_2019,Val_Der_Kul_Eigenfunc_quantum_minors_SL_n_C}. In those works the Gauss-Givental integral representation \cite{Der_Man_SL_2_C_Gauss_Givental} was used. In the present research an alternative representation of the eigenfunctions is derived.
	
	In the case of finite-dimensional representations of $gl_n$ the functional realization of the Gelfand-Tsetlin basis is known as well. In \cite{Artamonov_unified_approach,Artamonov_CG_coefficients} the realization of representations in the spaces of functions on the corresponding Lie group \cite{Zhelobenko_compact_Lie_gr_rep} is considered.
	
	In our case the induction on the rank of the algebra can be realized thanks to the fact that the generators of principal series representations of $gl_n(\mathbb{C})$ can be expressed in terms of similar generators for $gl_{n-1}(\mathbb{C})$ \cite[eq-s~(2.24--2.27)]{Val_Der_Kul_Eigenfunc_quantum_minors_SL_n_C}. Therefore, the L-operator for $gl_n(\mathbb{C})$ is expressed in terms of the L-operator for $gl_{n-1}(\mathbb{C})$. Consequently, one can derive the formulas expressing the quantum minors of $gl_n(\mathbb{C})$ L-operator in terms of the minors of $gl_{n-1}(\mathbb{C})$ L-operator. If it is known how a special set of non-corner $gl_{n-1}(\mathbb{C})$ minors act on $gl_{n-1}(\mathbb{C})$ eigenfunctions, it is possible to construct $gl_n(\mathbb{C})$ eigenfunctions using $gl_{n-1}(\mathbb{C})$ ones with the help of the Mellin-Barnes representation. A function of the Gelfand-Tsetlin basis for $gl_n(\mathbb{C})$ is expressed as an integral of an eigenfunction for $gl_{n-1}(\mathbb{C})$ with some kernel. The integration variables are the parameters of $gl_{n-1}(\mathbb{C})$ corner minors' eigenvalues corresponding to the $gl_{n-1}(\mathbb{C})$ eigenfunction. The condition that the obtained ansatz is the eigenfunction for $gl_n(\mathbb{C})$ corner quantum minors is equivalent to a system of equations for the kernel. These equations can be obtained from the formulas for action of $gl_{n-1}(\mathbb{C})$ minors on the $gl_{n-1}(\mathbb{C})$ eigenfunction under the integral. Similar idea has already been realized for a resembling model -- the quantum Toda chain -- in the works of S.M.~Kharchev and D.R.~Lebedev \cite{Kharchev_Lebedev_Toda_chain_Mellin_Barnes,Kharchev_Lebedev_Int_rep_Toda_chain_from_QISM}. In these articles the induction on the number of sites of the chain was considered.
	
	This paper is organized as follows. In Section~\ref{section_Gelfand_Tsetlin_basis_definition} we give the basic information about the Yangian of $gl_n(\mathbb{C})$, the quantum minors, principal series representations of $gl_n(\mathbb{C})$ and recurrence formulas for them, the L-operators and the Gelfand-Tsetlin basis. Section~\ref{section_formulae_for_induction} contains the expressions for the minors of $gl_n(\mathbb{C})$ L-operator in terms of the minors of $gl_{n-1}(\mathbb{C})$ L-operator as well as the formulas for action of needed non-corner minors on the eigenfunctions of corner ones. The idea how to perform the calculations for arbitrary rank, including the ansatz for elements of the Gelfand-Tsetlin basis, is given in Section~\ref{section_formulae_for_induction} and in the beginning of Section~\ref{section_induction_steps}. The remaining part of Section~\ref{section_induction_steps} is devoted to the particular cases of ranks 3 and 4 containing the formulae for basis elements and the orthogonality relations for them. Technical details are contained in Appendices.

	\section{Definition of the Gelfand-Tsetlin basis}   \label{section_Gelfand_Tsetlin_basis_definition}

	\subsection{Yangian of $gl_n(\mathbb{C})$}   \label{subsection_Yangian}

	Yangian $\mathrm{Y}(gl_n(\mathbb{C}))$ of the Lie algebra $gl_n(\mathbb{C})$ is an associative $\mathbb{C}$-algebra with unity generated by $t_{ij}^{(r)}$ (where $r\in\mathbb{Z}_+$ and $i,j\in\{1,\ldots,n\}$) satisfying the defining relations \cite[\S~1.1]{Molev_Yangians_and_classical_Lie_algebras}
	\begin{equation} \label{Yangian_defining_relations}
		[t_{ij}^{(r+1)},t_{kl}^{(s)}]-[t_{ij}^{(r)},t_{kl}^{(s+1)}]=t_{kj}^{(s)}t_{il}^{(r)}-t_{kj}^{(r)}t_{il}^{(s)}.
	\end{equation}
	Here $t_{ij}^{(0)}=\delta_{ij}$ and $\delta_{ij}$ is the Kronecker delta. Introduce the formal power series $T(u)^i_j=\sum_{r=0}^\infty t_{ij}^{(r)}u^{-r}$ in $u$ and consider it as an element $(i,j)$ of $n\times n$ matrix $T(u)$. Space  of dimension $n$ on which $T(u)$ acts is called the auxiliary space. The variable $u$ is called the spectral parameter. Relations \eqref{Yangian_defining_relations} are equivalent to the matrix relation \cite{Val_Der_Kul_Eigenfunc_quantum_minors_SL_n_C,Molev_Yangians_and_classical_Lie_algebras}
	\begin{equation} \label{Yangian_defining_relation_RTT}
		R(u-v)T_1(u)T_2(v)=T_2(v)T_1(u)R(u-v).
	\end{equation}
	It is called the RTT-relation. The matrix $R(u-v)$ is called the Yang R-matrix. It acts in the tensor product of copies $1$ and $2$ of the auxiliary space and has the form
	\begin{equation} \label{Yang_R_matrix}
		R(u-v) = \mathds{1} + \frac{P}{u-v},
	\end{equation}
	where $\mathds{1}$ is the identity matrix and $P$ is the permutation matrix: $P\,v\otimes w=w\otimes v$. The matrix $T_j(u)$ acts as $T(u)$ in the $j$-th copy and trivially in the other copy.
	
	Define the quantum minor $T(u)^{i_1\ldots i_m}_{j_1\ldots j_m}$ of order $m$ as \cite{Val_Der_Kul_Eigenfunc_quantum_minors_SL_n_C,Molev_Yangians_and_classical_Lie_algebras}
	\begin{equation} \label{quantum_minor_definition}
		T(u)^{i_1\ldots i_m}_{j_1\ldots j_m} \equiv \sum\limits_{\tau\in S_m}\mathrm{sgn}(\tau)T(u-m+1)^{i_{\tau(1)}}_{j_1}T(u-m+2)^{i_{\tau(2)}}_{j_2}\ldots T(u)^{i_{\tau(m)}}_{j_m},
	\end{equation}
	where $S_m$ is the symmetric group, $\mathrm{sgn}(\tau)$ is the sign of permutation $\tau$. Quantum minors are antisymmetric under the permutations of upper and lower indices: for any $\kappa$ in $S_m$
	\begin{equation} \label{antisymmetry_of_quantum_minors}
		T(u)^{i_{\kappa(1)}\ldots i_{\kappa(m)}}_{j_1\ldots j_m} = \mathrm{sgn}(\kappa)\,T(u)^{i_1\ldots i_m}_{j_1\ldots j_m}, \qquad T(u)^{i_1\ldots i_m}_{j_{\kappa(1)}\ldots j_{\kappa(m)}} = \mathrm{sgn}(\kappa)\,T(u)^{i_1\ldots i_m}_{j_1\ldots j_m}.
	\end{equation}
	Coefficients of powers of $u$ in corner minors $A_1(u),\ldots,A_n(u)$, where
	\begin{equation*}
		A_m(u) \equiv T(u)^{1\ldots m}_{1\ldots m},
	\end{equation*}
	generate the maximal commutative subalgebra in $\mathrm{Y}(gl_n(\mathbb{C}))$ called the Gelfand-Tsetlin subalgebra \cite[\S~1.13]{Molev_Yangians_and_classical_Lie_algebras}. The quantum determinant $A_n(u)$ generates the center of $\mathrm{Y}(gl_n(\mathbb{C}))$.
	
	If one considers a representation of $\mathrm{Y}(gl_n(\mathbb{C}))$, the elements $t_{ij}^{(r)}$ become operators in linear space which is called the quantum space. Let $\{E_{ab}\}_{1\leq a,b\leq n}$ be the set of standard generators of the universal enveloping algebra $U(gl_n(\mathbb{C}))$. The map
	\begin{equation*}
		T(u)^i_j\mapsto \delta_{ij}+E_{ji}u^{-1}
	\end{equation*}
	is a surjective homomorphism from $\mathrm{Y}(gl_n(\mathbb{C}))$ to $U(gl_n(\mathbb{C}))$ \cite[\S~2]{Nazarov_Tarasov_Yangians_and_Gelf_Tset_Bases}. Using it one can build a representation of $\mathrm{Y}(gl_n(\mathbb{C}))$ from a representation of $gl_n(\mathbb{C})$. The matrix $T(u)$ in this representation is called the L-operator. More complicated representations of the Yangian can be obtained by multiplication of $k$ L-operators acting on the same auxiliary space but with matrix entries acting on different copies of the quantum space. It is common to refer to this construction as the $k$-site $GL(n,\mathbb{C})$-invariant spin chain \cite{Val_Der_Kul_Eigenfunc_quantum_minors_SL_n_C}. The obtained representation of $T(u)$ is called the monodromy matrix.

	\subsection{Principal series representations}

	Denote by $Z_n$ the set of $n\times n$ complex lower triangular matrices with units on the diagonal and by $H_n$ -- the set of $n\times n$ complex upper triangular matrices with determinant equal to $1$. The principal series representation $T^\sigma$ of $SL(n,\mathbb{C})$ with parameters $\sigma=\{(\sigma_1,\ldots,\sigma_n),(\bar{\sigma}_1,\ldots,\bar{\sigma}_n)\}$ is defined on the space $L^2(Z_n)$ of square integrable functions on $Z_n$ \cite{Gelfand_Naimark,Zhelobenko_Shtern_representations_of_Lie_groups,Der_Man_gen_sol_YBE_SL_n_C}
	\begin{equation} \label{principal_series_rep_definition}
		[T^\sigma(g)\Phi](z) \equiv \alpha_\sigma\left(h_g(z)^{-1}\right)\Phi(z\bar{g}).
	\end{equation}
	where $z\bar{g}$ and $h_g(z)$ are the matrices appearing from the LU-decomposition of $g^{-1}z$:
	\begin{equation*}
		g^{-1}z = z\bar{g}\cdot h_g(z), \qquad z\bar{g}\in Z_n, \; h_g(z)\in H_n,
	\end{equation*}
	$\alpha_\sigma$ is the characher on $H_n$ of the form
	\begin{equation*}
		\alpha_\sigma(h) \equiv \prod\limits_{k=1}^n h_{kk}^{-\sigma_k-k}\overline{h_{kk}}^{-\bar{\sigma}_k-k}.
	\end{equation*}
	Call the parameters $\sigma_1,\ldots,\sigma_n$ holomorphic, $\bar{\sigma}_1,\ldots,\bar{\sigma}_n$ -- antiholomorphic. The numbers $\sigma_j$ and $\bar{\sigma}_j$ are not complex conjugate in the general case. The principal series representation is defined correctly if \cite[Section~5.1]{Der_Man_gen_sol_YBE_SL_n_C}
	\begin{equation} \label{condition_on_sigma_k,k+1_alpha_correctness}
		\bar{\sigma}_{k,k+1}-\sigma_{k,k+1}=n_k \in \mathbb{Z}, \qquad k=1,\ldots,n-1,
	\end{equation}
	where $\sigma_{k,k+1}\equiv \sigma_k-\sigma_{k+1}$.
	
	The scalar product on $L^2(Z_n)$ has the form
	\begin{equation*}
		\langle\Phi|\Psi\rangle = \int\limits_{\mathbb{C}} \prod\limits_{1\leq j<i\leq n} d^2z_{ij} \, \Phi^\ast(z)\Psi(z),
	\end{equation*}
	where the symbol $\ast$ donotes the complex conjugation. In the present research unitary principal series representations are considered. The condition of unitarity of the representation $T^\sigma$ gives the following system of equations on $\sigma$ \cite[Section~5.2]{Der_Man_gen_sol_YBE_SL_n_C}:
	\begin{equation} \label{condition_on_sigma_k,k+1_unitarity}
		\sigma_{k,k+1}^\ast+\bar{\sigma}_{k,k+1}=0, \qquad k=1,2\ldots,n-1.
	\end{equation}
	Combination of conditions \eqref{condition_on_sigma_k,k+1_alpha_correctness} and \eqref{condition_on_sigma_k,k+1_unitarity} gives the following representation for the parameters $\sigma$:
	\begin{equation} \label{principal_series_correctness_unitarity_reseriction_on_parameters}
		\sigma_{k,k+1}=-\frac{n_k}{2}+i\nu_k, \quad \bar{\sigma}_{k,k+1}=\frac{n_k}{2}+i\nu_k, \qquad k=1,\ldots,n-1,
	\end{equation}
	where $n_k\in\mathbb{Z}$, $\nu_k\in\mathbb{R}$. Unitary principal series representations are irreducible \cite[\S~5,~thm~2]{Gelfand_Naimark}.
	
	Consider $\sigma=\{(\sigma_1,\ldots,\sigma_n),(\bar{\sigma}_1,\ldots,\bar{\sigma}_n)\}$, $\sigma'=\{(\sigma'_1,\ldots,\sigma'_n),(\bar{\sigma}'_1,\ldots,\bar{\sigma}'_n)\}$. Representations $T^\sigma$ and $T^{\sigma'}$ are unitarily equivalent iff $(\sigma_1,\ldots,\sigma_n)$ and $(\bar{\sigma}_1,\ldots,\bar{\sigma}_n)$ differ from $(\sigma'_1,\ldots,\sigma'_n)$ and $(\bar{\sigma}'_1,\ldots,\bar{\sigma}'_n)$, accordingly, by one and the same permutation \cite{Gelfand_Naimark}.
	
	To obtain principal series representations of the Lie algebra $gl_n(\mathbb{C})$, one needs to consider the infinitesimal limit of the formula \eqref{principal_series_rep_definition}. The generators of $sl_n(\mathbb{C})$ are matrices $\mathbf{e}_{ij}$ defined as
	\begin{equation*}
		(\mathbf{e}_{ij})^a_b = \delta_{ia}\delta_{jb}-\frac{\delta_{ij}\delta_{ab}}{n}
	\end{equation*}
	(the index $a$ denotes the row and the index $b$ denotes the column). Take $g=\mathds{1}+\varepsilon\mathbf{e}_{ki}$ in \eqref{principal_series_rep_definition}, where $\mathds{1}$ is the unity matrix, and decompose
	\begin{equation*}
		[T^\sigma(\mathds{1}+\epsilon\mathbf{e}_{ki})\Phi](z) = \Phi(z) + \epsilon\,[E_{ki}\Phi](z) + \bar{\epsilon}\,[\bar{E}_{ki}\Phi](z) + O(\epsilon^2),
	\end{equation*}
	The generators $E_{ki}$ are called holomorphic, the generators $\bar{E}_{ki}$ -- antiholomorphic. One can find the following expression for the holomorphic generators of the representation \cite[\S~5, prop.~7]{Der_Man_gen_sol_YBE_SL_n_C}:
	\begin{equation} \label{holomorphic_generators_matrix_form}
		E = -z(D+\sigma)z^{-1},
	\end{equation}
	where $z\equiv\|z_{ij}\|_{1\leq j<i\leq n}$ is a lower triangular matrix with units on the diagonal, $E$, $\sigma$ and $D$ are $n$ by $n$ matrices of the form: $E^i_k = E_{ki}$ (the upper index $i$ denotes the row of the matrix, the lower index $k$ -- the column), $\sigma^c_d = \delta_{dc}\sigma_c$, $D^c_d = D_{dc}$, and the operators $D_{dc}$ are defined as
	\begin{equation*}
		D_{dc} \equiv \begin{cases}
			\sum\limits_{k=i}^nz_{kd}\partial_{kc}, & d>c \\
			0, & d\leq c
		\end{cases}, \qquad \partial_{kc}\equiv\frac{\partial}{\partial z_{kc}}.
	\end{equation*}
	An expression similar to \eqref{holomorphic_generators_matrix_form} holds for the antiholomorphic generators, one just needs to replace the holomorphic variables and parameters by the antiholomorphic ones, and the notation $E$ -- by $\bar{E}$. One can verify that for any values of parameters $\sigma_1,\ldots,\sigma_n$, $\bar{\sigma}_1,\ldots,\bar{\sigma}_n$ the obtained generators obey the commutation relations of the form $[E_{ij},E_{kl}]=\delta_{jk}E_{il}-\delta_{il}E_{kj}$, $[\bar{E}_{ij},\bar{E}_{kl}]=\delta_{jk}\bar{E}_{il}-\delta_{il}\bar{E}_{kj}$, that is, they define a representation of the Lie algebra $gl_n(\mathbb{C})$.
	
	Holomorphic and antiholomorphic L-operators $L(u)$ and $\bar{L}(\bar{u})$ are expressed in terms of matrices $E$ and $\bar{E}$ as\footnotemark
	\begin{equation} \label{L_operators_holomorphic_and_antiholomorphic}
		L(u) = u\mathds{1} + E, \qquad \bar{L}(\bar{u}) = \bar{u}\mathds{1} + \bar{E},
	\end{equation}
	where $\mathds{1}$ is the unity matrix. Spectral parameters $u$ and $\bar{u}$ are independent, they are not complex conjugate.
	\footnotetext{Since in the representations of Yangian under consideration only two first terms of the series $T(u)$ survive, one can multiply it by $u$ and obtain the expression with nonnegative powers of $u$. This does not affect the relation \eqref{Yangian_defining_relation_RTT}.}
	
	The crucial part in construction of Gelfand-Tsetlin basis by induction on the rank of the algebra play the fact that the generators of $gl_n(\mathbb{C})$ representation $T^\sigma$ can be expressed through the generators of $gl_{n-1}(\mathbb{C})$ representation $T^{\sigma'}$, where $\sigma=\{(\sigma_1,\sigma_2\ldots,\sigma_n),(\bar{\sigma}_1,\bar{\sigma}_2,\ldots,\bar{\sigma}_n)\}$ and $\sigma'=\{(\sigma_2,\ldots,\sigma_n),(\bar{\sigma}_2,\ldots,\bar{\sigma}_n)\}$. The explicit recurrence formulas read \cite[eq-s (3.5--3.8)]{Val_Der_Kul_Eigenfunc_quantum_minors_SL_n_C}
	\begin{equation} \label{L_operator_recurrence_relation_explicit}
		\begin{split}
			& L(u)^1_1=u-\sigma_1+n-1+\sum\limits_{k=2}^n z_{k1}\partial_{k1}, \\
			& L^1_j=-\partial_{j1}, \\
			& L(u)^i_1=z_{i1}(u-\sigma_1+n-1)-\sum\limits_{k=2}^n z_{k1}\left(\mathcal{L}(u)^{i-1}_{k-1}-z_{i1}\partial_{k1}\right), \\
			& L(u)^i_j=\mathcal{L}(u)^{i-1}_{j-1}-z_{i1}\partial_{j1},
		\end{split}
	\end{equation}
	where $2\leq i,j\leq n$, the matrix $L(u)$ is the $gl_n(\mathbb{C})$ L-operator from \eqref{L_operators_holomorphic_and_antiholomorphic}, the matrix $\mathcal{L}(u)$ is the $gl_{n-1}(\mathbb{C})$ L-operator: $\mathcal{L}(u) = u\mathds{1}+\mathcal{E}$. Here $\mathcal{E}$ is the matrix with $gl_{n-1}(\mathbb{C})$ generators having the same form as in \eqref{holomorphic_generators_matrix_form}: $\mathcal{E} = -z'(D'+\sigma')z'^{-1}$, where the matrices $z'=\|z_{i+1,j+1}\|_{1\leq j<i\leq n-1}$, $(D')^d_c=D^{d+1}_{c+1}, \; (\sigma')^d_c=\sigma^{d+1}_{c+1}$ are $(n-1)\times(n-1)$ submatrices of $z$, $D$, $\sigma$ from \eqref{holomorphic_generators_matrix_form}. Of course, the formulas similar to \eqref{L_operator_recurrence_relation_explicit} hold for the antiholomorphic L-operator as well.

	\subsection{The Gelfand-Tsetlin basis}

	Define the Gelfand-Tsetlin basis for a unitary principal series representation of $gl_n(\mathbb{C})$ as the basis of common eigenfunctions of corner quantum minors $\{A_m(u), \bar{A}_m(\bar{u})\}_{m=1}^n$ of the L-operator. From definitions of quantum minors \eqref{quantum_minor_definition} and the L-operator \eqref{L_operators_holomorphic_and_antiholomorphic} it follows that $A_m(u)$ is a polynomial in $u$ of degree $m$. As it was mentioned in Section~\ref{subsection_Yangian}, coefficients of polynomials $\{A_m(u), \bar{A}_m(\bar{u})\}_{m=1}^n$ form a set of commuting operators \cite[\S~1.13]{Molev_Yangians_and_classical_Lie_algebras}, and the eigenfunctions of corner minors are in fact the common eigenfunctions of these coefficients, therefore they do not depend on spectral parameters $u$, $\bar{u}$. Consequently, the eigenvalues of $A_m(u)$ are also polynomials in $u$ of degree $m$. The highest degree term in $A_m(u)$ is $u^mI$, where $I$ is the identity operator. Similarly for the antiholomorphic minors $\{\bar{A}_m(\bar{u})\}_{m=1}^n$.
	
	The eigenvalues of $\{A_m(u),\bar{A}_m(\bar{u})\}_{m=1}^n$, as polynomials in $u$ and $\bar{u}$, are parametrized by their roots:
	\begin{equation} \label{Gelfand_Tsetlin_basis_definition}
		A_m(u)\Psi_{\boldsymbol{\lambda}} = \prod\limits_{k=1}^m (u-\lambda_{mk}) \, \Psi_{\boldsymbol{\lambda}}, \qquad \bar{A}_m(\bar{u})\Psi_{\boldsymbol{\lambda}} = \prod\limits_{k=1}^m (\bar{u}-\bar{\lambda}_{mk}) \, \Psi_{\boldsymbol{\lambda}},
	\end{equation}
	where $\boldsymbol{\lambda}=(\lambda,\bar{\lambda})$, $\lambda=\{\lambda_{mk}\}_{1\leq k\leq m\leq n}$, $\bar{\lambda}=\{\bar{\lambda}_{mk}\}_{1\leq k\leq m\leq n}$, and the quantities $\lambda_{mk}$ and $\bar{\lambda}_{mk}$ are not complex conjugate in the general case. Since, as we have already mentioned, uniraty principal series representations are irreducible, the quantum determinant $A_n(u)$ generating the center of $\mathrm{Y}(gl_n(\mathbb{C}))$ is proportional to the unity operator in these representations. With the help of \eqref{L_m_th_corner_minor_in_terms_of_mathcal_L} it can be shown by induction on the rank that
	\begin{equation} \label{A_n_in_principal_series_representation}
		A_n(u) = \prod\limits_{k=1}^n (u-\sigma_k)\,I, \qquad \bar{A}_n(\bar{u}) = \prod\limits_{k=1}^n (\bar{u}-\bar{\sigma}_k)\,I,
	\end{equation}
	where $I$ is the identity operator. From \eqref{A_n_in_principal_series_representation} it follows that $\lambda_{nk}=\sigma_k$, $\bar{\lambda}_{nk}=\bar{\sigma}_k$.
	
	Depict the arrays $\lambda$ and $\bar{\lambda}$ in the form
	\begin{equation} \label{Gelfand_Tsetlin_schemes}
		\begin{split}
			& \lambda = \begin{pmatrix}
				\sigma_1 & & \sigma_2 & \ldots & \sigma_{n-1} & & \sigma_n \\
				& \lambda_{n-1,1} & & \ldots & & \lambda_{n-1,n-1} & \\
				& & \ldots & \ldots & \ldots & & \\
				& & \lambda_{21} & & \lambda_{22} & & \\
				& & & \lambda_{11} & & &
			\end{pmatrix}, \\
			& \bar{\lambda} = \begin{pmatrix}
				\bar{\sigma}_1 & & \bar{\sigma}_2 & \ldots & \bar{\sigma}_{n-1} & & \bar{\sigma}_n \\
				& \bar{\lambda}_{n-1,1} & & \ldots & & \bar{\lambda}_{n-1,n-1} & \\
				& & \ldots & \ldots & \ldots & & \\
				& & \bar{\lambda}_{21} & & \bar{\lambda}_{22} & & \\
				& & & \bar{\lambda}_{11} & & &
			\end{pmatrix}
		\end{split}
	\end{equation}
	and call them the Gelfand-Tsetlin schemes. The roots of eigenvalues of operators $A_m(u)$, $\bar{A}_m(\bar{u})$ are located in the $(n-m+1)$-th rows of the corresponding arrays from \eqref{Gelfand_Tsetlin_schemes}. The $k$-th element $\rho_{mk}$ in the $(n-m+1)$-th row of the conventionally defined Gelfand-Tsetlin scheme is expressed in terms of the entry of the array from \eqref{Gelfand_Tsetlin_schemes} as \cite[\S~2]{Nazarov_Tarasov_Yangians_and_Gelf_Tset_Bases}
	\begin{equation*}
		\rho_{mk} = -\lambda_{mk}+k-1.
	\end{equation*}
	In the finite-dimensional case $\rho_{mk}$ are integer, and there is a condition
	\begin{equation*}
		\rho_{mk}\geq\rho_{m-1,k}\geq\rho_{m,k+1}.
	\end{equation*}
	All these conditions vanish in the infinite-dimensional case. This fact was observed by M.I.~Graev in his papers \cite{Graev_Gelf_Tset_basis_inf_dim_rep_gl_n_C,Graev_a_continuous_analogue_of_Gelfand_Tsetlin_schemes} on infinite-dimensional representations of $gl_n(\mathbb{C})$.

	\section{Formulas for induction on the rank of the algebra}   \label{section_formulae_for_induction}

	For the induction step from $gl_{n-1}(\mathbb{C})$ to $gl_n(\mathbb{C})$ one needs the expressions of corner quantum minors of $gl_n(\mathbb{C})$ L-operator $L(u)$ in terms of the minors of $gl_{n-1}(\mathbb{C})$ L-operator $\mathcal{L}(u)$. For the next step from $gl_n(\mathbb{C})$ to $gl_{n+1}(\mathbb{C})$ one also needs to know the action of some non-corner $gl_n(\mathbb{C})$ minors on the Gelfand-Tsetlin basis. To obtain this information the expressions of these $gl_n(\mathbb{C})$ minors in terms of $gl_{n-1}(\mathbb{C})$ ones are needed.
	
	Using the formulas \eqref{L_operator_recurrence_relation_explicit} one can find all the expressions indicated above. They have the form
	\begin{multline} \label{L_m_th_corner_minor_in_terms_of_mathcal_L}
		L(u)^{1\ldots m}_{1\ldots m} = \mathcal{L}(u)^{1,\ldots,m-1}_{1,\ldots,m-1}\left(u-\sigma_1+n-m+\sum\limits_{k=m+1}^nz_{k1}\partial_{k1}\right) \\
		+ \sum\limits_{b=m}^{n-1}\sum\limits_{a=1}^{m-1} (-1)^{m+a}\mathcal{L}(u)^{1,\ldots,m-1}_{1,\ldots,\widehat{a},\ldots,m-1,b}\,z_{b+1,1}\partial_{a+1,1},
	\end{multline}
	where $1\leq m\leq n$, the definition of quantum minors is given in \eqref{quantum_minor_definition}, the notation $\widehat{a}$ means the skipped index $a$;
	\begin{multline} \label{L_m_th_non_corner_minor_1_in_terms_of_mathcal_L}
		L(u)^{1\ldots m}_{1,i_1,\ldots, i_{m-1}} = \mathcal{L}(u)^{1,\ldots,m-1}_{i_1-1,\ldots,i_{m-1}-1}\left(u-\sigma_1+n-m+\sum\limits_{\begin{smallmatrix}
				k=2 \\ k\neq i_1,\ldots,i_{m-1}
		\end{smallmatrix}}^n z_{k1}\partial_{k1}\right) \\
		+ \sum\limits_{\begin{smallmatrix}
				b=1 \\ b\neq i_1-1,\ldots,i_{m-1}-1
		\end{smallmatrix}}^{n-1}\sum\limits_{a=1}^{m-1} (-1)^{m+a}\mathcal{L}(u)^{1,\ldots,m-1}_{i_1-1,\ldots\widehat{i_a-1}\ldots i_{m-1}-1,b}\,z_{b+1,1}\partial_{i_a1},
	\end{multline}
	where $2\leq m\leq n$ and $2\leq i_1<\ldots<i_{m-1}\leq n$;
	\begin{equation} \label{L_m_th_non_corner_minor_2_in_terms_of_mathcal_L}
		L(u)^{1\ldots m}_{i_1\ldots i_m} = \sum\limits_{a=1}^m (-1)^a\mathcal{L}(u)^{1,\ldots,m-1}_{i_1-1,\ldots,\widehat{i_a-1},\ldots,i_m-1}\,\partial_{i_a1},
	\end{equation}
	where $1\leq m\leq n$ and $2\leq i_1<\ldots<i_m\leq n$. These expressions are derived in Appendix~\ref{appendix_minors}. Similar formulas hold for the minors of the antiholomorphic L-operator, one just needs to replace the variables and parameters with the antiholomorphic ones.
	
	Consider the algebra $gl_{n'}(\mathbb{C})$ of rank $n'$. For any natural numbers $r$, $a$, $b$ such that\\
	$1\leq a\leq r<b\leq n'$ denote by $B_{rab}(u)$ the following minor of the $gl_{n'}(\mathbb{C})$ L-operator with one skipped lower index among $1,\ldots,r$:
	\begin{equation} \label{B_rab_minor_definition}
		B_{rab}(u) \equiv L(u)^{1\ldots r}_{1,\ldots,\widehat{a},\ldots,r,b}.
	\end{equation}
	Minors of this type for $gl_{n-1}(\mathbb{C})$ are located in the double sum in \eqref{L_m_th_corner_minor_in_terms_of_mathcal_L}. In accordance with \cite[\S~1]{Nazarov_Tarasov_Yangians_and_Gelf_Tset_Bases} denote
	\begin{equation} \label{B_r_definition}
		B_r(u) \equiv B_{r,r,r+1}(u) = L(u)^{1\ldots r}_{1,\ldots,r-1,r+1}, \qquad 1\leq r\leq n'-1.
	\end{equation}
	The action of any minor of type \eqref{B_rab_minor_definition} on an eigenfunction \eqref{Gelfand_Tsetlin_basis_definition} of $\{A_k(u),\bar{A}_k(\bar{u})\}_{k=1}^{n'}$ can be expressed in terms of the action of a product of operators $\{B_r(u)\}_{r=1}^{n'-1}$:
	\begin{equation} \label{B_rab_acts_on_Psi_lambda_bar_lambda}
		B_{rab}(u)\Psi_{\boldsymbol{\lambda}} = (-1)^{b-r-1}\sum\limits_{\begin{smallmatrix}
				s_a,s_{a+1},\ldots,s_{b-1} \\ s_k=1,\ldots,k
		\end{smallmatrix}} \frac{\prod\limits_{\begin{smallmatrix}
			l=1 \\ l\neq s_r
	\end{smallmatrix}}^r(u-\lambda_{rl}) \prod\limits_{i=a}^{\overset{r}{\longleftarrow}}B_i(\lambda_{is_i})\prod\limits_{j=r+1}^{\overset{b-1}{\longrightarrow}}B_j(\lambda_{js_j})\;\Psi_{\boldsymbol{\lambda}}}{\prod\limits_{k=a}^{b-2}(\lambda_{k+1,s_{k+1}}-\lambda_{ks_k}-1_{k<r}) \prod\limits_{k=a}^{b-1}\prod\limits_{\begin{smallmatrix}
	l=1 \\ l\neq s_k
	\end{smallmatrix}}^k (\lambda_{ks_k}-\lambda_{kl})},
	\end{equation}
	where
	\begin{equation} \label{ordered_operator_products_and_1_k<r_definitions}
		\begin{split}
			&  \prod\limits_{i=a}^{\overset{r}{\longleftarrow}}B_i(\lambda_{is_i}) \equiv B_r(\lambda_{rs_r})B_{r-1}(\lambda_{r-1,s_{r-1}})\ldots B_a(\lambda_{as_a}), \\
			& \prod\limits_{j=r+1}^{\overset{b-1}{\longrightarrow}}B_j(\lambda_{js_j}) \equiv B_{r+1}(\lambda_{r+1,s_{r+1}})B_{r+2}(\lambda_{r+2,s_{r+2}})\ldots B_{b-1}(\lambda_{b-1,s_{b-1}}), \\
			& 1_{k<r}\equiv\begin{cases}
				1, & k<r \\
				0, & k\geq r
			\end{cases}.
		\end{split}
	\end{equation}	
	The result of the action of $B_r(\lambda_{ri})$ on $\Psi_{\boldsymbol{\lambda}}$ is an eigenfunction of $\{A_j\}_{j=1}^{n'}$ with the root $\lambda_{ri}$ of $A_r(u)$ eigenvalue shifted by $1$
	\begin{equation} \label{B_r_acts_on_Psi_lambda_bar_lambda}
		A_r(u)[B_r(\lambda_{ri})\Psi_{\boldsymbol{\lambda}}] = \prod\limits_{k=1}^r (u-\lambda_{rk}-\delta_{ki})\,\Psi_{\boldsymbol{\lambda}}, \qquad A_j(u)[B_r(\lambda_{ri})\Psi_{\boldsymbol{\lambda}}] = \prod\limits_{k=1}^j (u-\lambda_{jk})\,\Psi_{\boldsymbol{\lambda}}, \; j \neq r.
	\end{equation}
	Formulas similar to \eqref{B_rab_acts_on_Psi_lambda_bar_lambda} and \eqref{B_r_acts_on_Psi_lambda_bar_lambda} hold for the antiholomorphic minors as well.
	
	Expressions \eqref{B_rab_acts_on_Psi_lambda_bar_lambda} and \eqref{B_r_acts_on_Psi_lambda_bar_lambda} are derived in Appendix~\ref{appendix_B_action}. It is worth mentioning that these formulas hold for any representation of $gl_{n'}(\mathbb{C})$ (not only principal series) in which the corner minors of the L-operator can be diagonalized. It follows from the fact that for their derivation one needs only \eqref{Gelfand_Tsetlin_basis_definition}, the commutation relations \eqref{quantum_minors_commutation_relations_final} between the minors and Lagrange interpolation.
	
	Therefore, in the case of the induction step $gl_{n-1}(\mathbb{C})\to gl_n(\mathbb{C})$ for the next step $gl_n(\mathbb{C})\to gl_{n+1}(\mathbb{C})$ one needs to deduce only the action of $gl_n(\mathbb{C})$ minors $\{B_m(u),\bar{B}_m(\bar{u})\}_{m=1}^{n-1}$ on the Gelfand-Tsetlin basis for $gl_n(\mathbb{C})$. The minor $B_1(u)=L(u)^1_2$ equals to $-\partial_{21}$. For $m\geq 2$ taking $i_1=2,\ldots,i_{m-2}=m-1,i_{m-1}=m+1$ in \eqref{L_m_th_non_corner_minor_1_in_terms_of_mathcal_L} one obtains
	\begin{multline} \label{B_m_for_L_in_terms_of_mathcal_L}
		B_m(u) = L(u)^{1\ldots m}_{1,\ldots,m-1,m+1} = \mathcal{L}(u)^{1,\ldots,m-1}_{1,\ldots,m-2,m}\left(u-\sigma_1+n-m+z_{m1}\partial_{m1}+\sum\limits_{k=m+2}^n z_{k1}\partial_{k1}\right) \\
		- \sum\limits_{k=m+1}^{n-1} \mathcal{L}(u)^{1,\ldots,m-1}_{1,\ldots,m-2,k}\,z_{k+1,1}\partial_{m+1,1} + \sum\limits_{a=1}^{m-2} (-1)^{m+a+1}\mathcal{L}(u)^{1,\ldots,m-1}_{1,\ldots,\widehat{a},\ldots,m-1,m}\,z_{m1}\partial_{a+1,1} \\
		- \mathcal{L}(u)^{1,\ldots,m-1}_{1,\ldots,m-1}\,z_{m1}\partial_{m+1,1} + \sum\limits_{k=m+1}^{n-1}\sum\limits_{a=1}^{m-2} (-1)^{m+a}\mathcal{L}(u)^{1,\ldots,m-1}_{1,\ldots,\widehat{a},\ldots,m-2,m,k}\,z_{k+1,1}\partial_{a+1,1}.
	\end{multline}
	In the double sum in \eqref{B_m_for_L_in_terms_of_mathcal_L} there are minors $\mathcal{L}(u)^{1,\ldots,m-1}_{1,\ldots,\widehat{a},\ldots,m-2,m,k}$ of order $m-1$ with two lower indices skipped among $1,\ldots,m-1$ (these skipped indices are $a$ and $m-1$). For any rank $n'$ of algebra $gl_{n'}(\mathbb{C})$ the action of a minor of this kind on an eigenfunction $\Psi_{\boldsymbol{\lambda}}$ can be expressed in terms of the action of operators \eqref{B_rab_minor_definition}. For natural numbers $r,a_1,a_2,b_1,b_2$ such that $1\leq a_1<a_2\leq r<b_1<b_2\leq n'$ holds
	\begin{multline} \label{B_r,a1,a2,b1,b2(lambda_ir)_action_on_Psi_lambda}
		L(\lambda_{ri})^{1\ldots r}_{1,\ldots,\widehat{a_1},\ldots,\widehat{a_2},\ldots,r,b_1,b_2}\Psi_{\boldsymbol{\lambda}} \\
		= \frac{1}{2}\sum\limits_{j=1}^r \frac{\sum\limits_{k=1}^2[B_{ra_kb_k}(\lambda_{ri})B_{ra_{\widetilde{k}}b_{\widetilde{k}}}(\lambda_{rj})-B_{ra_kb_{\widetilde{k}}}(\lambda_{ri})B_{ra_{\widetilde{k}}b_k}(\lambda_{rj})]\Psi_{\boldsymbol{\lambda}}}{(\lambda_{rj}-\lambda_{ri}+1)\prod\limits_{\begin{smallmatrix}
					l=1 \\ l\neq j
			\end{smallmatrix}}^r (\lambda_{rj}-\lambda_{rl})},
	\end{multline}
	where $1\leq i\leq r$, $\widetilde{1}=2$, $\widetilde{2}=1$. This formula is derived in Appendix~\ref{appendix_B_action}. Of course, a similar relation holds for the antiholomorphic minors. For fixed $r,a_1,a_2,b_1,b_2$ one has $r$ expressions of type \eqref{B_r,a1,a2,b1,b2(lambda_ir)_action_on_Psi_lambda}. From definitions of the L-operator \eqref{L_operators_holomorphic_and_antiholomorphic} and quantum minors \eqref{quantum_minor_definition} it follows that $L(u)^{1\ldots r}_{1,\ldots,\widehat{a_1},\ldots,\widehat{a_2},\ldots,r,b_1,b_2}$ is a polynomial in $u$ of degree $r-2$. Therefore, $L(u)^{1\ldots r}_{1,\ldots,\widehat{a_1},\ldots,\widehat{a_2},\ldots,r,b_1,b_2}\Psi_{\boldsymbol{\lambda}}$ can be calculated with the help of relations \eqref{B_r,a1,a2,b1,b2(lambda_ir)_action_on_Psi_lambda} and Lagrange interpolation. As it was already mentioned for \eqref{B_rab_acts_on_Psi_lambda_bar_lambda} and \eqref{B_r_acts_on_Psi_lambda_bar_lambda}, the equality \eqref{B_r,a1,a2,b1,b2(lambda_ir)_action_on_Psi_lambda} is valid not only for principal series representations, but for any representation of $gl_{n'}(\mathbb{C})$ in which the corner minors of the L-operator can be diagonalized.

	\section{Induction steps from $gl_2(\mathbb{C})$ to $gl_3(\mathbb{C})$ and from $gl_3(\mathbb{C})$ to $gl_4(\mathbb{C})$}   \label{section_induction_steps}

	From the expression for $A_1(u)=L(u)^1_1$ in \eqref{L_operator_recurrence_relation_explicit} and the definition of the Gelfand-Tsetlin basis \eqref{Gelfand_Tsetlin_basis_definition} it follows that the elements of the basis for $gl_n(\mathbb{C})$ are functions homogeneous in variables $\{z_{k1}\}_{k=2}^n$. Holomorphic and antiholomorphic degrees of homogeneity are $\sigma_1-n+1-\lambda_{11}$ and $\bar{\sigma}_1-n+1-\bar{\lambda}_{11}$, correspondingly. Denote
	\begin{equation} \label{Lambda_C}
		\Lambda_\mathbb{C} \equiv \{(a,\bar{a})\in\mathbb{C}\times\mathbb{C} | a-\bar{a}\in\mathbb{Z}\},
	\end{equation}
	where $a$ and $\bar{a}$ are not complex conjugate in the general case. At the step from $gl_{n-1}(\mathbb{C})$ to $gl_n(\mathbb{C})$ the following ansatz is considered:
	\begin{equation} \label{gl_n_ansatz}
		\Psi_{\boldsymbol{\lambda}}(z) = \int D\boldsymbol{\gamma} \, K_{\boldsymbol{\gamma}}(\boldsymbol{\lambda})\,\Phi_{\boldsymbol{\gamma}}(\mathbf{z}) \, z_{21}^{\boldsymbol{\sigma}_1-\boldsymbol{n}+\boldsymbol{1}-\boldsymbol{\lambda}_{11}-\sum\limits_{k=3}^n \boldsymbol{b}_k(\boldsymbol{\gamma},\boldsymbol{\lambda})}z_{31}^{\boldsymbol{b}_3(\boldsymbol{\gamma},\boldsymbol{\lambda})}\ldots z_{n1}^{\boldsymbol{b}_n(\boldsymbol{\gamma},\boldsymbol{\lambda})}.
	\end{equation}
	Here $z\equiv\{z_{ij}\}_{1\leq j<i\leq n}$, $\mathbf{z}\equiv\{z_{ij}\}_{2\leq j<i\leq n}$, the function $\Psi_{\boldsymbol{\lambda}}(z)$ is an element of the Gelfand-Tsetlin basis for $gl_n(\mathbb{C})$, the function $\Phi_{\boldsymbol{\gamma}}(\mathbf{z})$ is an element of the Gelfand-Tsetlin basis for $gl_{n-1}(\mathbb{C})$, the quantity $K_{\boldsymbol{\gamma}}(\boldsymbol{\lambda})$ is an unknown function of $\gamma$, $\bar{\gamma}$ with parameters $\lambda$, $\bar{\lambda}$, the notation
	\begin{equation} \label{double_power_notation}
		z^{\boldsymbol{a}} \equiv z^a\bar{z}^{\bar{a}}, \qquad \boldsymbol{a}=(a,\bar{a}) \in \Lambda_\mathbb{C}
	\end{equation}
	is called the double power \cite[Section~1.2.3]{Neretin_func_complex_field}, the notation $\boldsymbol{\lambda}_{11}$ means $(\lambda_{11},\bar{\lambda}_{11})$ (similarly for other parameters), the notation $\boldsymbol{r}$ means $(r,r)$ for any $r\in\mathbb{R}$, and $\boldsymbol{b}_j(\boldsymbol{\gamma},\boldsymbol{\lambda})\equiv(b_j(\gamma,\lambda),\bar{b}_j(\bar{\gamma},\bar{\lambda}))\in\Lambda_\mathbb{C}$. The symbol $D\boldsymbol{\gamma}$ in \eqref{gl_n_ansatz} stands for
	\begin{equation} \label{gl_n_integration_measure}
		\int D\boldsymbol{\gamma} \equiv \prod\limits_{1\leq j<i\leq n}\int D\boldsymbol{\gamma}_{ij}, \quad \int D\boldsymbol{\gamma}_{ij} \equiv \sum\limits_{k_{ij}\in\mathbb{Z}}\,\int\limits_{\mathscr{C}_{ij}}d\nu_{ij}, \quad \gamma_{ij}=\frac{k_{ij}}{2}+i\nu_{ij}, \; \bar{\gamma}_{ij}=-\frac{k_{ij}}{2}+i\nu_{ij}.
	\end{equation}
	Here the contours $\mathscr{C}_{ij}$ must be chosen in such a way that all changes of variables of integration in subsequent calculations can be done. Define the linear combination of two elements from $\Lambda_\mathbb{C}$ as
	\begin{equation*}
		\alpha\boldsymbol{a} + \beta\boldsymbol{b} = (\alpha a+\beta b,\alpha\bar{a}+\beta\bar{b}), \qquad \boldsymbol{a}=(a,\bar{a})\in\Lambda_\mathbb{C}, \; \boldsymbol{b}=(b,\bar{b})\in\Lambda_\mathbb{C}, \; \alpha,\beta\in\mathbb{Z}.
	\end{equation*}
	The function \eqref{gl_n_ansatz} satisfies the condition of homogeneity. An ansatz similar to \eqref{gl_n_ansatz} was considered by S.M.~Kharchev and D.R.~Lebedev in the works on inductive construction of eigenfunctions for a resembling model -- the Toda chain (the induction on the number of sites of the chain) \cite{Kharchev_Lebedev_Toda_chain_Mellin_Barnes,Kharchev_Lebedev_Int_rep_Toda_chain_from_QISM}. 
	
	Find the quantities $\{b_i(\gamma,\lambda)\}_{i=3}^n$ and $\{\bar{b}_i(\bar{\gamma},\bar{\lambda})\}_{i=3}^n$. From the definition of quantum minors \eqref{quantum_minor_definition} and the L-operator \eqref{L_operators_holomorphic_and_antiholomorphic} it follows that $\mathcal{L}(u)^{1,\ldots,m-1}_{1,\ldots,\widehat{a},\ldots,m-1,b}$ in the double sum in \eqref{L_m_th_corner_minor_in_terms_of_mathcal_L} is a polynomial of degree $m-2$ in $u$. Therefore, in \eqref{L_m_th_corner_minor_in_terms_of_mathcal_L} only the LHS and the first term in the RHS have nonzero coefficients of $u^{m-1}$. Substituting \eqref{gl_n_ansatz} into the eigenvalue equation \eqref{Gelfand_Tsetlin_basis_definition} for $L(u)^{1\ldots m}_{1\ldots m}$, $m\in\{2,\ldots,n\}$, using the expression \eqref{L_m_th_corner_minor_in_terms_of_mathcal_L} for this minor and equating the coefficients of $u^{m-1}$ one finds
	\begin{equation*}
		\sum\limits_{k=m+1}^n b_k(\gamma,\lambda) = \sum\limits_{l=1}^{m-1}\gamma_{m-1,l} - \sum\limits_{s=1}^m \lambda_{ms} + \sigma_1-n+m, \qquad m\in\{2,\ldots,n-1\}.
	\end{equation*}
	From this system of equations it follows that
	\begin{equation} \label{b_i_formulas}
		\begin{split}
			& b_i(\gamma,\lambda) = \sum\limits_{l=1}^{i-2}\gamma_{i-2,l} - \sum\limits_{l=1}^{i-1}\gamma_{i-1,l} - \sum\limits_{s=1}^{i-1}\lambda_{i-1,s} + \sum\limits_{s=1}^i \lambda_{is} - 1, \qquad i=3,\ldots,n-1, \\
			& b_n(\gamma,\lambda) = \sum\limits_{l=1}^{n-2}\gamma_{n-2,l}-\sum\limits_{s=1}^{n-1}\lambda_{n-1,s}+\sigma_1-1.
		\end{split}
	\end{equation}
	The derivation of $\{\bar{b}_i(\bar{\gamma},\bar{\lambda})\}_{i=3}^n$ is absolutely the same, these functions have similar form, one just needs to replace the holomorphic parameters with antiholomorphic.
	
	To find $K_{\boldsymbol{\gamma}}(\boldsymbol{\lambda})$ one needs to substitute \eqref{gl_n_ansatz} into eigenvalue equations \eqref{Gelfand_Tsetlin_basis_definition}, use recurrence formulas \eqref{L_m_th_corner_minor_in_terms_of_mathcal_L} for $gl_n(\mathbb{C})$ corner minors and act by $gl_{n-1}(\mathbb{C})$ minors from these formulas on $gl_{n-1}(\mathbb{C})$ eigenfunctions under the integral \eqref{gl_n_ansatz}. Since the highest-degree term in $A_m(u)$ is the identity operator and the equality corresponding to $u^{m-1}$ has already been solved, the eigenvalue problem for $A_m(u)$ yields $m-1$ equations for $K_{\boldsymbol{\gamma}}(\boldsymbol{\lambda})$. Therefore, in the $gl_n(\mathbb{C})$ case the eigenvalue problems for all holomorphic corner minors yield $\frac{(n-1)(n-2)}{2}$ equations for $K_{\boldsymbol{\gamma}}(\boldsymbol{\lambda})$ as a function of $\frac{(n-1)(n-2)}{2}$ ``holomorphic'' variables $\{\gamma_{ij}\}_{1\leq j<i\leq n-1}$. Similarly, the eigenvalue problems for all antiholomorphic corner minors yield $\frac{(n-1)(n-2)}{2}$ equations for $K_{\boldsymbol{\gamma}}(\boldsymbol{\lambda})$ as a function of $\frac{(n-1)(n-2)}{2}$ ``antiholomorphic'' variables $\{\bar{\gamma}_{ij}\}_{1\leq j<i\leq n-1}$.
	
	This idea of induction on the rank of the algebra is illustrated by the example of ranks $3$ and $4$.
	
	Touching on the issues of contours of integration in the Mellin-Barnes representation and orthogonality of obtained sets of eigenfunctions we considered the parameters of principal series representations of $gl_n(\mathbb{C})$ to be of the form
	\begin{equation} \label{principal_series_rep_parameters_form}
		\sigma_j = \frac{s_j+\varkappa+i\eta_j}{2}, \quad \bar{\sigma}_j = \frac{-s_j+\varkappa+i\eta_j}{2}, \qquad s_j\in\mathbb{Z}, \; \varkappa,\eta_j\in\mathbb{R}, \quad j=1,\ldots,n.
	\end{equation}
	The parameters \eqref{principal_series_rep_parameters_form} obey the condition \eqref{principal_series_correctness_unitarity_reseriction_on_parameters}.

	\subsection{The Gelfand-Tsetlin basis for $gl_2(\mathbb{C})$}

	The holomorphic L-operator for $gl_2(\mathbb{C})$ has the form (similar expression holds for the antiholomorphic one)
	\begin{equation*}
		L(u) = \begin{pmatrix}
			u-\sigma_1+1+z\partial & -\partial \\
			z(z\partial+\sigma_2+1-\sigma_1) & u-\sigma_2-z\partial
		\end{pmatrix},
	\end{equation*}
	where $z\equiv z_{21}$, $\partial\equiv\frac{\partial}{\partial z}$, $\bar{\partial}\equiv\frac{\partial}{\partial \bar{z}}$. The minor $A_1$ reads $A_1(u) = u-\sigma_1+1+z\partial$. The eigenvalue equation for $A_1$ has the form $A_1(u)\Psi_{\boldsymbol{\lambda}} = (u-\lambda_{11})\Psi_{\boldsymbol{\lambda}}$, same equation holds for the antiholomorphic minor $\bar{A}_1(\bar{u})$. The function $\Psi_{\boldsymbol{\lambda}}$ corresponds to the following pair of Gelfand-Tsetlin schemes (see \eqref{Gelfand_Tsetlin_basis_definition}, \eqref{Gelfand_Tsetlin_schemes})
	\begin{equation*}
		\lambda=\begin{pmatrix}
			\sigma_1 & & \sigma_2 \\
			& \lambda_{11} &
		\end{pmatrix}, \qquad
		\bar{\lambda}=\begin{pmatrix}
			\bar{\sigma}_1 & & \bar{\sigma}_2 \\
			& \bar{\lambda}_{11} &
		\end{pmatrix}
	\end{equation*}
	and has the form
	\begin{equation} \label{gl_2_eigenfunction}
		\Psi_{\boldsymbol{\lambda}}(z) = \Gamma^\mathbb{C}(\boldsymbol{1}-\boldsymbol{\sigma}_1+\boldsymbol{\lambda}_{11})\,z^{\boldsymbol{\sigma}_1-\boldsymbol{1}-\boldsymbol{\lambda}_{11}},
	\end{equation}
	where the notation for the power function was introduced in \eqref{double_power_notation}, and for $\boldsymbol{\mu}=(\mu,\bar{\mu})\in\Lambda_\mathbb{C}$ (see \eqref{Lambda_C}) the expression
	\begin{equation} \label{Gamma^C}
		\Gamma^\mathbb{C}(\boldsymbol{\mu}) \equiv \frac{\Gamma(\mu)}{\Gamma(1-\bar{\mu})}
	\end{equation}
	is called the gamma-function of the complex field \cite{Gelfand_Graev_hypergeom_func_arbitrary_field,Neretin_func_complex_field}. The normalization factor $\Gamma^\mathbb{C}(\boldsymbol{1}-\boldsymbol{\sigma}_1+\boldsymbol{\lambda}_{11})$ in \eqref{gl_2_eigenfunction} is chosen in such a way that
	\begin{equation} \label{gl_2_minors_action_on_Psi_lambda}
		B_1(u)\Psi_{\boldsymbol{\lambda}} = \Psi_{\boldsymbol{\lambda}+e_{11}}, \qquad \bar{B}_1(\bar{u})\Psi_{\boldsymbol{\lambda}} = -\Psi_{\boldsymbol{\lambda}+\bar{e}_{11}},
	\end{equation}
	where $B_1(u)=-\partial$ (see \eqref{B_r_definition}), $\bar{B}_1(\bar{u})=-\bar{\partial}$, and the following short notations are used:
	\begin{equation} \label{gl_2_short_notations}
		\Psi_{\boldsymbol{\lambda}\pm e_{11}} \equiv \Psi_{\left(\begin{smallmatrix}
				\sigma_1 & & \sigma_2 \\
				& \lambda_{11}\pm 1 &
			\end{smallmatrix}\right),\left(\begin{smallmatrix}
				\bar{\sigma}_1 & & \bar{\sigma}_2 \\
				& \bar{\lambda}_{11} &
			\end{smallmatrix}\right)}, \qquad
		\Psi_{\boldsymbol{\lambda}\pm \bar{e}_{11}} \equiv \Psi_{\left(\begin{smallmatrix}
				\sigma_1 & & \sigma_2 \\
				& \lambda_{11} &
			\end{smallmatrix}\right),\left(\begin{smallmatrix}
				\bar{\sigma}_1 & & \bar{\sigma}_2 \\
				& \bar{\lambda}_{11}\pm 1 &
			\end{smallmatrix}\right)}.
	\end{equation}
	
	Complete orthogonal set of functions \eqref{gl_2_eigenfunction} is indexed by parameters of the form
	\begin{equation} \label{gl_2_parameters_orthogonal_set}
		\lambda_{11} = \frac{k+\varkappa-1+iv}{2}, \quad \bar{\lambda}_{11} = \frac{-k+\varkappa-1+iv}{2}, \qquad k\in\mathbb{Z}, \; v\in\mathbb{R},
	\end{equation}
	where $\varkappa\in\mathbb{R}$ is a constant from the parameters of principal series representation \eqref{principal_series_rep_parameters_form}. Then from the property of gamma-function $\Gamma(x^\ast)=\Gamma(x)^\ast$ (the symbol $\ast$ denotes the complex conjugation) it follows that $|\Gamma^\mathbb{C}(\boldsymbol{1}-\boldsymbol{\sigma}_1+\boldsymbol{\lambda}_{11})|=1$. Consider the formula \cite[eq. (67)]{Der_Man_SL_2_C_Gauss_Givental}
	\begin{equation} \label{delta_as_integral_of_power_functions}
		\int\limits_{\mathbb{C}} d^2z \, z^{\boldsymbol{\mu}-\boldsymbol{\mu}'-\boldsymbol{1}} = 2\pi^2\delta^{(2)}(\boldsymbol{\mu}-\boldsymbol{\mu}'),
	\end{equation}
	where $\boldsymbol{\mu}=(\mu,\bar{\mu})$, $\boldsymbol{\mu}'=(\mu',\bar{\mu}')$, and
	\begin{eqnarray*}
		 \mu=\frac{a+ib}{2}, \; \bar{\mu}=\frac{-a+ib}{2}, \quad \mu'=\frac{a'+ib'}{2}, \; \bar{\mu}'=\frac{-a'+ib'}{2}, \qquad a, a' \in \mathbb{Z}, \; b, b' \in \mathbb{R}.
	\end{eqnarray*}
	The notation in the RHS of \eqref{delta_as_integral_of_power_functions} reads
	\begin{equation} \label{parameters'_delta_function_definition}
		\delta^{(2)}(\boldsymbol{\mu}-\boldsymbol{\mu}') \equiv \delta_{a,a'}\,\delta\left(\frac{b}{2}-\frac{b'}{2}\right),
	\end{equation}
	where $\delta_{a,a'}$ is the Kronecker delta and $\delta\left(\frac{b}{2}-\frac{b'}{2}\right)$ is the delta-function. From \eqref{delta_as_integral_of_power_functions} it follows that the orthogonality relation reads
	\begin{equation} \label{gl_2_orthogonality}
		\int\limits_{\mathbb{C}} d^2z \, \Psi_{\boldsymbol{\lambda}}^\ast(z) \Psi_{\boldsymbol{\lambda}}(z) = 2\pi^2\delta^{(2)}(\boldsymbol{\lambda}_{11}-\boldsymbol{\lambda}_{11}'),
	\end{equation}
	where $\lambda_{11},\bar{\lambda}_{11}$ are defined in \eqref{gl_2_parameters_orthogonal_set}, for $\lambda_{11}'$ and $\bar{\lambda}_{11}'$ the symbols $k$ and $v$ from \eqref{gl_2_parameters_orthogonal_set} must be replaced by $k'$ and $v'$. The completeness relation has the form \cite[\S~4]{Bel_Der_completeness_3j_symb_SL_2_C}
	\begin{equation*}
		\sum\limits_{k\in\mathbb{Z}}\int\limits_{-\infty}^\infty \frac{dv}{2} \, \Psi_{\boldsymbol{\lambda}}^\ast(z)\Psi_{\boldsymbol{\lambda}}(z') = 2\pi^2\delta^{(2)}(z-z'),
	\end{equation*}
	where $\delta^{(2)}(z-z') \equiv \delta(\mathrm{Re}(z-z'))\,\delta(\mathrm{Im}(z-z'))$, the notations $\mathrm{Re}$ and $\mathrm{Im}$ mean the real and the imaginary part of a complex number, correspondingly.

	\subsection{Induction step from $gl_2(\mathbb{C})$ to $gl_3(\mathbb{C})$}

	Denote
	\begin{equation*}
		(x,y,z) \equiv (z_{21},z_{31},z_{32}).
	\end{equation*}
	Denote by $L(u)$ the L-operator for $gl_3(\mathbb{C})$. Expressions of holomorphic $gl_3(\mathbb{C})$ minors in terms of $gl_2(\mathbb{C})$ follow from (\ref{L_m_th_corner_minor_in_terms_of_mathcal_L}--\ref{L_m_th_non_corner_minor_2_in_terms_of_mathcal_L}):
	\begin{equation} \label{gl_3_minors_in_terms_of_gl_2_minors}
		\begin{split}
			& A_1(u) = u-\sigma_1+2+x\partial_x+y\partial_y, \\
			& A_2(u) = \mathcal{L}(u)^1_1\,(u-\sigma_1+1+y\partial_y)-\mathcal{L}(u)^1_2\,y\partial_x \\
			& L(u)^1_2 = -\partial_x, \\
			& L(u)^1_3 = -\partial_y, \\
			& L(u)^{12}_{13} = \mathcal{L}(u)^1_2\,(u-\sigma_1+1+x\partial_x) - \mathcal{L}(u)^1_1\,x\partial_y, \\
			& L(u)^{12}_{23} = -\mathcal{L}(u)^1_2\,\partial_x + \mathcal{L}(u)^1_1\,\partial_y,
		\end{split}
	\end{equation}
	where $\mathcal{L}(u)$ is the L-operator for $gl_2(\mathbb{C})$ with parameters $\sigma_2,\sigma_3$. The entries of $\mathcal{L}(u)$ act on the variable $z$. The variables $x,y$ are ``new'', they appear in $gl_3(\mathbb{C})$. Same expressions hold for the antiholomorphic minors.
	
	Substituting $A_1$, $A_2$ from \eqref{gl_3_minors_in_terms_of_gl_2_minors} into \eqref{Gelfand_Tsetlin_basis_definition} one obtains the eigenvalue equations for them. In accordance with \eqref{gl_n_ansatz}, \eqref{gl_n_integration_measure} and \eqref{b_i_formulas}, consider the following ansatz:
	\begin{equation} \label{gl_3_eigenfunction}
		\Psi_{\boldsymbol{\lambda}}(x,y,z)=\int D\boldsymbol{\gamma}_{11}\,K_{\boldsymbol{\gamma}}({\boldsymbol{\lambda}})\,\Phi_{\boldsymbol{\gamma}}(z)x^{\boldsymbol{\sigma}_1-\boldsymbol{2}-\boldsymbol{\lambda}_{11}-\boldsymbol{b}_3(\boldsymbol{\gamma},\boldsymbol{\lambda})}y^{\boldsymbol{b}_3(\boldsymbol{\gamma},\boldsymbol{\lambda})},
	\end{equation}
	where $\Phi_{\boldsymbol{\gamma}}$ is an element of the Gelfand-Tsetlin basis for $gl_2(\mathbb{C})$, the quantity $K_{\boldsymbol{\gamma}}(\boldsymbol{\lambda})$ is an unknown function of $\gamma$, $\bar{\gamma}$ with parameters $\lambda$, $\bar{\lambda}$, and
	\begin{equation} \label{gamma_11_step_from_gl_2_to_gl_3}
		\gamma_{11}=\frac{k+\varkappa-1+iv}{2}, \; \bar{\gamma}_{11}=\frac{-k+\varkappa-1+iv}{2}, \qquad \int D\boldsymbol{\gamma}_{11} \equiv \sum\limits_{k\in\mathbb{Z}}\int\limits_{-\infty}^\infty\frac{dv}{2}.
	\end{equation}
	The real number $\varkappa$ in \eqref{gamma_11_step_from_gl_2_to_gl_3} is a constant from parameters \eqref{principal_series_rep_parameters_form} of the principal series representation. The reason for choosing the domain of integration \eqref{gamma_11_step_from_gl_2_to_gl_3} is that it is the domain of parameters \eqref{gl_2_parameters_orthogonal_set} indexing the complete orthogonal set of $gl_2(\mathbb{C})$ eigenfunctions. Substituting \eqref{gl_3_eigenfunction} into the left part of the eigenvalue equation for $A_2(u)$ and using the formula \eqref{gl_2_minors_action_on_Psi_lambda} for $gl_2(\mathbb{C})$ minors action on $\Phi_{\boldsymbol{\gamma}}$ one obtains
	\begin{multline} \label{gl_3_A_2_action_on_Psi_lambda}
		A_2(u)\Psi_{\boldsymbol{\lambda}} = \int D\boldsymbol{\gamma}_{11}\,K_{\boldsymbol{\gamma}}\,(u-\gamma_{11})(u-\sigma_1+1+b_3)\Phi_{\boldsymbol{\gamma}}(z)x^{\boldsymbol{\sigma}_1-\boldsymbol{2}-\boldsymbol{\lambda}_{11}-\boldsymbol{b}_3}y^{\boldsymbol{b}_3} \\
		- \int D\boldsymbol{\gamma}_{11}\,K_{\boldsymbol{\gamma}}\,(\sigma_1-2-\lambda_{11}-b_3)\Phi_{\boldsymbol{\gamma}+e_{11}}(z)x^{\boldsymbol{\sigma}_1-\boldsymbol{2}-\boldsymbol{\lambda}_{11}-\boldsymbol{b}_3-e}y^{\boldsymbol{b}_3+e},
	\end{multline}
	where the notation $\Phi_{\boldsymbol{\gamma}+e_{11}}(z)$ is similar to \eqref{gl_2_short_notations}, and
	\begin{equation} \label{e_and_bar_e_definition}
		e\equiv(1,0), \qquad \bar{e}\equiv(0,1).
	\end{equation}
	Consider in the second integral in \eqref{gl_3_A_2_action_on_Psi_lambda} the change of variables of summation and integration which is equivalent to the substitution $k\to k-1, \; v\to v+i$. Then \eqref{gl_3_A_2_action_on_Psi_lambda} is transformed to
	\begin{multline} \label{gl_3_A_2_action_on_Psi_lambda_1}
		A_2(u)\Psi_{\boldsymbol{\lambda}} = \int D\boldsymbol{\gamma}_{11}\,\{K_{\boldsymbol{\gamma}}\,(u-\gamma_{11})(u-\sigma_1+1+b_3)-K_{\boldsymbol{\gamma}-e_{11}}\,(\sigma_1-1-\lambda_{11}-b_3)\} \\
		\times \Phi_{\boldsymbol{\gamma}}(z)x^{\boldsymbol{\sigma}_1-\boldsymbol{2}-\boldsymbol{\lambda}_{11}-\boldsymbol{b}_3}y^{\boldsymbol{b}_3},
	\end{multline}
	where the notation $K_{\boldsymbol{\gamma}-e_{11}}$ is similar to \eqref{gl_2_short_notations}. From \eqref{gl_3_A_2_action_on_Psi_lambda_1} follows the equation for the ``holomorphic'' part of $K_{\boldsymbol{\gamma}}$
	\begin{equation} \label{gl_3_equation_for_K_gamma_with_spectral_parameter}
		(u-\gamma_{11})(u-\sigma_1+1+b_3)K_{\boldsymbol{\gamma}} + (-\sigma_1+1+\lambda_{11}+b_3)K_{\boldsymbol{\gamma}-e_{11}} = (u-\lambda_{21})(u-\lambda_{22})K_{\boldsymbol{\gamma}}.
	\end{equation}
	The equality with coefficients of $u^2$ in \eqref{gl_3_equation_for_K_gamma_with_spectral_parameter} is trivial. The equation with coefficients of $u^1$ is solved already. Therefore, it is sufficient to consider \eqref{gl_3_equation_for_K_gamma_with_spectral_parameter} with $u$ taking one value. Considering $u=\gamma_{11}$ one obtains a finite-difference equation for $K_\gamma$
	\begin{equation} \label{gl_3_K_gamma_holomorphic_equation}
		K_{\boldsymbol{\gamma}-e_{11}}(\boldsymbol{\lambda}) = \frac{(\gamma_{11}-\lambda_{21})(\gamma_{11}-\lambda_{22})}{\gamma_{11}+\lambda_{11}-\lambda_{21}-\lambda_{22}} K_{\boldsymbol{\gamma}}(\boldsymbol{\lambda}).
	\end{equation}
	Doing the same calculations with antiholomorphic minors $\bar{A}_1(\bar{u})$, $\bar{A}_2(\bar{u})$ one obtains the antiholomorphic analogue of \eqref{gl_3_K_gamma_holomorphic_equation}, which differs only by the sign of the RHS. The function
	\begin{equation*}
		K_{\boldsymbol{\gamma}}(\boldsymbol{\lambda}) = (-1)^{\sum(\lambda_{2j}-\bar{\lambda}_{2j})}\Gamma^\mathbb{C}(\boldsymbol{\gamma}_{11}+\boldsymbol{\lambda}_{11}-\sum\boldsymbol{\lambda}_{2j}+\boldsymbol{1})\prod\limits_{l=1}^2\Gamma^\mathbb{C}(\boldsymbol{\lambda}_{2l}-\boldsymbol{\gamma}_{11})
	\end{equation*}
	gives a solution of \eqref{gl_3_K_gamma_holomorphic_equation} and of its antiholomorphic analogue. The gamma-function of the complex field $\Gamma^\mathbb{C}$ is defined in \eqref{Gamma^C}.
	
	In some terms of the sum by $k$ in \eqref{gl_3_eigenfunction} the contour of integration with respect to $v$ contains poles, therefore the $i\varepsilon$-prescription must be added:
	\begin{multline} \label{gl_3_eigenfunction_orthogonal_set}
		\Psi_{\boldsymbol{\lambda}}(x,y,z) = (-1)^{\sum(\lambda_{2j}-\bar{\lambda}_{2j})}\prod\limits_{k=1}^2 \Gamma^\mathbb{C}(\boldsymbol{\lambda}_{2k}+\boldsymbol{1}-\boldsymbol{\sigma}_1) \sum\limits_{k\in\mathbb{Z}}\lim\limits_{\varepsilon\to0_+}\int\limits_{-\infty}^\infty \frac{dv}{2} \, \Phi_{\boldsymbol{\gamma}}(z) \\
		\times \Gamma^\mathbb{C}(\boldsymbol{\gamma}_{11}+\boldsymbol{\lambda}_{11}-\sum\boldsymbol{\lambda}_{2j}+\boldsymbol{1})\prod\limits_{l=1}^2\Gamma^\mathbb{C}(\boldsymbol{\lambda}_{2l}-\boldsymbol{\gamma}_{11}+\frac{\boldsymbol{\varepsilon}}{\boldsymbol{2}})\,x^{-\boldsymbol{\gamma}_{11}-\boldsymbol{1}-\boldsymbol{\lambda}_{11}+\sum\boldsymbol{\lambda}_{2j}}y^{\boldsymbol{\gamma}_{11}+\boldsymbol{\sigma}_1-\boldsymbol{1}-\sum\boldsymbol{\lambda}_{2j}}.
	\end{multline}
	The analysis of poles originating from complex gamma-functions under the integral in \eqref{gl_3_eigenfunction_orthogonal_set} shows that the change of integration variable, which was done between \eqref{gl_3_A_2_action_on_Psi_lambda} and \eqref{gl_3_A_2_action_on_Psi_lambda_1}, is legal. With the pre-factor $\prod_{k=1}^2 \Gamma^\mathbb{C}(\boldsymbol{\lambda}_{2k}+\boldsymbol{1}-\boldsymbol{\sigma}_1)$ added in \eqref{gl_3_eigenfunction_orthogonal_set} the action of non-corner minors $B_j$ (see \eqref{B_r_definition}) on the functions \eqref{gl_3_eigenfunction_orthogonal_set} reads
	\begin{equation} \label{gl_3_non_corner_minors_action}
		B_1(u)\Psi_{\boldsymbol{\lambda}} = \Psi_{\boldsymbol{\lambda}+e_{11}}, \qquad B_2(\lambda_{2i})\Psi_{\boldsymbol{\lambda}} = (\lambda_{2i}-\lambda_{11})\Psi_{\boldsymbol{\lambda}+e_{2i}},
	\end{equation}
	where the short notations for the shifts of eigenfunction parameters are similar to \eqref{gl_2_short_notations}. The action of $B_1(u)$, $B_2(u)$ on $\Psi_{\boldsymbol{\lambda}}$ can be easily found using the formulas \eqref{gl_3_minors_in_terms_of_gl_2_minors}. The action of $L(u)^1_3=B_{113}(u)$ (see \eqref{B_rab_minor_definition}) and $L(u)^{12}_{23}=B_{213}(u)$ can be expressed from \eqref{B_rab_acts_on_Psi_lambda_bar_lambda} and verified by the direct calculation using \eqref{gl_3_minors_in_terms_of_gl_2_minors}. The antiholomorphic analogue of \eqref{gl_3_non_corner_minors_action} differs only by the sign of the RHS.
	
	Consider
	\begin{equation} \label{gl_3_parameters_orthogonal_set}
		\lambda_{lj} = \frac{n_{lj}+\varkappa-3+l+i\mu_{lj}}{2}, \; \bar{\lambda}_{lj} = \frac{-n_{lj}+\varkappa-3+l+i\mu_{lj}}{2}, \qquad 1\leq j<l\leq 2,
	\end{equation}
	where $n_{lj}\in\mathbb{Z}$, $\mu_{lj}\in\mathbb{R}$, and $\varkappa$ is a constant from parameters \eqref{principal_series_rep_parameters_form} of the principal series representation. In Appendix~\ref{appendix_orthogonality} it is shown that the functions \eqref{gl_3_eigenfunction_orthogonal_set} indexed by parameters \eqref{gl_3_parameters_orthogonal_set} form an orthogonal set. The orthogonality relation takes the form
	\begin{multline} \label{gl_3_orthogonality}
		\int\limits_{\mathbb{C}} d^2x\,d^2y\,d^2z \, \|\boldsymbol{\lambda}_{21}-\boldsymbol{\lambda}_{22}\|^2\Psi_{\boldsymbol{\lambda}}^\ast(x,y,z) \, \|\boldsymbol{\lambda}_{21}'-\boldsymbol{\lambda}_{22}'\|^2\Psi_{\boldsymbol{\lambda}'}(x,y,z) = 32\pi^8 \|\boldsymbol{\lambda}_{21}-\boldsymbol{\lambda}_{22}\|^2 \\
		\times \delta^{(2)}(\boldsymbol{\lambda}_{11}-\boldsymbol{\lambda}_{11}')\left[\delta^{(2)}(\boldsymbol{\lambda}_{21}-\boldsymbol{\lambda}_{21}') \delta^{(2)}(\boldsymbol{\lambda}_{22}-\boldsymbol{\lambda}_{22}') + \delta^{(2)}(\boldsymbol{\lambda}_{21}-\boldsymbol{\lambda}_{22}') \delta^{(2)}(\boldsymbol{\lambda}_{22}-\boldsymbol{\lambda}_{21}')\right],
	\end{multline}
	where the symbol $\ast$ denotes the complex conjugation, the notation $\delta^{(2)}(\boldsymbol{\lambda}_{lj}-\boldsymbol{\lambda}_{ls}')$ is similar to \eqref{parameters'_delta_function_definition}, and
	\begin{equation} \label{squared_norm_of_lambda_difference}
		 \|\boldsymbol{\lambda}_{21}-\boldsymbol{\lambda}_{22}\|^2 \equiv -(\lambda_{21}-\lambda_{22})(\bar{\lambda}_{21}-\bar{\lambda}_{22})=\frac{(n_{22}-n_{21})^2+(\mu_{22}-\mu_{21})^2}{4}.
	\end{equation}
	Both in cases of $gl_3(\mathbb{C})$ and $gl_4(\mathbb{C})$ the orthogonality relations for the eigenfunctions are derived with the help of generalization of Gustafson's integrals \cite{Gustafson_94,Gustafson_92} developed in \cite{Der_Man_on_complex_gamma_func_integrals,Der_Man_Val_SL_2_C_Gustafson_integrals,DMV_17}.
	
	The expression \eqref{gl_3_eigenfunction_orthogonal_set} for elements of the Gelfand-Tsetlin basis coincide, up to a factor, with the definition of the function ${}_2G_2^\mathbb{C}$ from Section~1.4 in \cite{Neretin_func_complex_field}. With the help of statement~b) from Lemma~2.3 in \cite{Neretin_analog_Dougall_formula} this function can be expressed in terms of the Gauss hypergeometric function of the complex field \cite{Gelfand_Graev_hypergeom_func_arbitrary_field,Neretin_analog_Dougall_formula}. Using the integral representation of Euler type for the latter \cite[eq.~(2.2)]{Neretin_analog_Dougall_formula} one can compare \eqref{gl_3_eigenfunction_orthogonal_set} with the Gauss-Givental representation for the Gelfand-Tsetlin basis which is given by formula (57) in \cite{Valinevich_2019}.

	\subsection{Induction step from $gl_3(\mathbb{C})$ to $gl_4(\mathbb{C})$}

	Denote
	\begin{equation*}
		\xi\equiv z_{21}, \quad \eta\equiv z_{31}, \quad \zeta\equiv z_{41}, \quad \mathbf{z} \equiv (z_{32},z_{42},z_{43}).
	\end{equation*}
	From (\ref{L_m_th_corner_minor_in_terms_of_mathcal_L}--\ref{L_m_th_non_corner_minor_2_in_terms_of_mathcal_L}) it follows that the expressions of $gl_4(\mathbb{C})$ minors in terms of $gl_3(\mathbb{C})$ minors read
	\begin{equation} \label{gl_4_minors_in_terms_of_gl_3_minors}
		\begin{split}
			& A_1(u) = u-\sigma_1+3+\xi\partial_{\xi}+\eta\partial_{\eta}+\zeta\partial_{\zeta}, \\
			& A_2(u) = \mathcal{L}(u)^1_1\,(u-\sigma_1+2+\eta\partial_{\eta}+\zeta\partial_{\zeta})-\mathcal{L}(u)^1_2\,\eta\partial_{\xi}-\mathcal{L}(u)^1_3\,\zeta\partial_{\xi}, \\
			& A_3(u) = \mathcal{L}(u)^{12}_{12}\,(u-\sigma_1+1+\zeta\partial_{\zeta})+\mathcal{L}(u)^{12}_{23}\,\zeta\partial_{\xi}-\mathcal{L}(u)^{12}_{13}\,\zeta\partial_{\eta}, \\
			& L(u)^1_2 = -\partial_{\xi}, \\
			& L(u)^{12}_{13} = \mathcal{L}(u)^1_2\,(u-\sigma_1+2+\xi\partial_{\xi}+\zeta\partial_{\zeta})-\mathcal{L}(u)^1_1\,\xi\partial_{\eta}-\mathcal{L}(u)^1_3\,\zeta\partial_{\eta}, \\
			& L(u)^{123}_{124} = \mathcal{L}(u)^{12}_{13}\,(u-\sigma_1+1+\eta\partial_{\eta})-\mathcal{L}(u)^{12}_{23}\,\eta\partial_{\xi}-\mathcal{L}(u)^{12}_{12}\,\eta\partial_\zeta.
		\end{split}
	\end{equation}
	Substituting $A_1$, $A_2$, $A_3$ from \eqref{gl_4_minors_in_terms_of_gl_3_minors} into \eqref{Gelfand_Tsetlin_basis_definition} one obtains the eigenvalue equations for them.
	
	In accordance with \eqref{gl_n_ansatz}, \eqref{gl_n_integration_measure} and \eqref{b_i_formulas}, consider the following ansatz:
	\begin{equation} \label{gl_4_eigenfunction}
		\Psi_{\boldsymbol{\lambda}}(\xi,\eta,\zeta,\mathbf{z}) = \int D\boldsymbol{\gamma}\,K_{\boldsymbol{\gamma}}(\boldsymbol{\lambda})\,\Phi_{\boldsymbol{\gamma}}(\mathbf{z})\xi^{\boldsymbol{\sigma}_1-\boldsymbol{3}-\boldsymbol{\lambda}_{11}-\boldsymbol{b}_3(\boldsymbol{\gamma},\boldsymbol{\lambda})-\boldsymbol{b}_4(\boldsymbol{\gamma},\boldsymbol{\lambda})}\eta^{\boldsymbol{b}_3(\boldsymbol{\gamma},\boldsymbol{\lambda})}\zeta^{\boldsymbol{b}_4(\boldsymbol{\gamma},\boldsymbol{\lambda})},
	\end{equation}
	where $\Phi_{\boldsymbol{\gamma}}$ is an element of the Gelfand-Tsetlin basis for $gl_3(\mathbb{C})$, the quantity $K_{\boldsymbol{\gamma}}(\boldsymbol{\lambda})$ is an unknown function of $\gamma$, $\bar{\gamma}$ with parameters $\lambda$, $\bar{\lambda}$. Similarly to the case of $gl_3(\mathbb{C})$, substituting \eqref{gl_4_eigenfunction} into the eigenvalue equations for $A_2$ and $A_3$ one obtains the system of ``holomorphic'' finite-difference equations for $K_{\boldsymbol{\gamma}}$:
	\begin{gather}
		\frac{(\gamma_{11}-\lambda_{21})(\gamma_{11}-\lambda_{22})}{\gamma_{11}+\lambda_{11}-\sum\limits_{k=1}^2\lambda_{2k}}K_{\boldsymbol{\gamma}} = K_{\boldsymbol{\gamma}-e_{11}}-\frac{K_{\boldsymbol{\gamma}-e_{11}-e_{21}}}{\gamma_{21}-\gamma_{22}-1}-\frac{K_{\boldsymbol{\gamma}-e_{11}-e_{22}}}{\gamma_{22}-\gamma_{21}-1},  \label{gl_4_system_K_gamma_hol_1} \\
		\begin{split}
			& (\gamma_{21}-\gamma_{11}-1)\left(\gamma_{11}-\sum\limits_{k=1}^2\gamma_{2k}-\sum\limits_{k=1}^2\lambda_{2k}+\sum\limits_{k=1}^3\lambda_{3k}\right)K_{\boldsymbol{\gamma}-e_{21}} \\
			& + \left(\gamma_{11}+\lambda_{11}-\sum\limits_{k=1}^2\lambda_{2k}\right)K_{\boldsymbol{\gamma}-e_{11}-e_{21}} = -\frac{\prod\limits_{k=1}^3(\gamma_{21}-\lambda_{3k}) \; (\gamma_{21}-\gamma_{22}-1)}{\gamma_{21}-\gamma_{22}}K_{\boldsymbol{\gamma}},
		\end{split} \label{gl_4_system_K_gamma_hol_2} \\
		\begin{split}
			& (\gamma_{22}-\gamma_{11}-1)\left(\gamma_{11}-\sum\limits_{k=1}^2\gamma_{2k}-\sum\limits_{k=1}^2\lambda_{2k}+\sum\limits_{k=1}^3\lambda_{3k}\right)K_{\boldsymbol{\gamma}-e_{22}} \\
			& + \left(\gamma_{11}+\lambda_{11}-\sum\limits_{k=1}^2\lambda_{2k}\right)K_{\boldsymbol{\gamma}-e_{11}-e_{22}} = -\frac{\prod\limits_{k=1}^3(\gamma_{22}-\lambda_{3k}) \; (\gamma_{22}-\gamma_{21}-1)}{\gamma_{22}-\gamma_{21}}K_{\boldsymbol{\gamma}},
		\end{split}  \label{gl_4_system_K_gamma_hol_3}
	\end{gather}
	where the notations like $K_{\boldsymbol{\gamma}-e_{11}}$ are similar to \eqref{gl_2_short_notations}. The eigenvalue equations for the antiholomorphic minors $\bar{A}_2$, $\bar{A}_3$ lead to the system of equations for $K_{\boldsymbol{\gamma}}$ with the shifts of antiholomorphic parameters. It is similar to (\ref{gl_4_system_K_gamma_hol_1}--\ref{gl_4_system_K_gamma_hol_3}) up to the signs of some terms in equations.
	
	The system (\ref{gl_4_system_K_gamma_hol_1}--\ref{gl_4_system_K_gamma_hol_3}) and its antiholomorphic counterpart are solved in Appendix~\ref{appendix_gl4_system_solution} with the help of the complex Mellin transform \cite[Section~1.2.4]{Neretin_func_complex_field}. Their solution has the form
	\begin{multline} \label{K_gamma_}
		K_{\boldsymbol{\gamma}}(\boldsymbol{\lambda}) = c(\boldsymbol{\lambda})\,(-1)^{\sum(\gamma_{2j}-\bar{\gamma}_{2j})}(\gamma_{21}-\gamma_{22})(\bar{\gamma}_{21}-\bar{\gamma}_{22}) \, \Gamma^\mathbb{C}\left(\boldsymbol{\gamma}_{11}+\boldsymbol{\lambda}_{11}-\sum\boldsymbol{\lambda}_{2j}+\boldsymbol{1}\right) \\
		\times \Gamma^\mathbb{C}\left(-\boldsymbol{\gamma}_{11}+\sum\boldsymbol{\gamma}_{2j}+\sum\boldsymbol{\lambda}_{2j}-\sum\boldsymbol{\lambda}_{3j}+\boldsymbol{1}\right) \prod\limits_{l=1}^3\prod\limits_{j=1}^2\Gamma^\mathbb{C}(\boldsymbol{\lambda}_{3l}-\boldsymbol{\gamma}_{2j}) \\
		\times {}_4G^{\mathbb{C}}_4\left[\begin{array}{c}
			\boldsymbol{\lambda}_{21}, \boldsymbol{\lambda}_{22}, \boldsymbol{\gamma}_{21}, \boldsymbol{\gamma}_{22} \\
			\boldsymbol{1}-\boldsymbol{\lambda}_{31}, \boldsymbol{1}-\boldsymbol{\lambda}_{32}, \boldsymbol{1}-\boldsymbol{\lambda}_{33}, -\boldsymbol{\gamma}_{11}
		\end{array}; 1\right],
	\end{multline}
	where $c(\boldsymbol{\lambda})$ is the normalization coefficient of the Gelfand-Tsetlin basis depending only on $\lambda$ and $\bar{\lambda}$, the function $\Gamma^\mathbb{C}$ is defined in \eqref{Gamma^C}, the expression ${}_4G^{\mathbb{C}}_4$ is the hypergeometric function of the complex field \cite[Section~1.4]{Neretin_func_complex_field}. With the help of technique of calculations with ${}_pG^{\mathbb{C}}_q$ similar to that used for Meijer G-function in Section 3 of \cite{contiguous_rel_Meijer_G} it is verified in Appendix~\ref{appendix_gl4_system_solution} that \eqref{K_gamma_} indeed solves (\ref{gl_4_system_K_gamma_hol_1}--\ref{gl_4_system_K_gamma_hol_3}) and its antiholomorphic analogue.
	
	In accordance with \eqref{B_r_acts_on_Psi_lambda_bar_lambda}, holomorphic and antiholomorphic minors $B_2(\lambda_{2i})=L(\lambda_{2i})^{12}_{13}$ and $B_3(\lambda_{3i})=L(\lambda_{3i})^{123}_{124}$ shift by $1$ the parameters $\lambda_{2i}$ and $\lambda_{3i}$. From this follow the conditions on $c(\boldsymbol{\lambda})$. For $i=1,2$:
	\begin{equation} \label{gl_4_condition_on_c_shift_2i}
		c(\boldsymbol{\lambda}+e_{2i}) = -\frac{\lambda_{2i}-\lambda_{11}}{\mathcal{M}_i(\lambda)}\,c(\boldsymbol{\lambda}), \qquad c(\boldsymbol{\lambda}+\bar{e}_{i2}) = \frac{\bar{\lambda}_{2i}-\bar{\lambda}_{11}}{\widetilde{\mathcal{M}}_i(\bar{\lambda})}\,c(\boldsymbol{\lambda}),
	\end{equation}
	for $i=1,2,3$:
	\begin{equation} \label{gl_4_condition_on_c_shift_3i}
		\begin{split}
			& c(\boldsymbol{\lambda}+e_{3i})=\frac{(\lambda_{3i}-\sigma_1+1)\prod\limits_{k=1}^2(\lambda_{3i}-\lambda_{2k})}{\mathcal{M}'_i(\lambda)}\,c(\boldsymbol{\lambda}), \\
			& c(\boldsymbol{\lambda}+\bar{e}_{3i})=\frac{(\bar{\lambda}_{3i}-\bar{\sigma}_1+1)\prod\limits_{k=1}^2(\bar{\lambda}_{3i}-\bar{\lambda}_{2k})}{\widetilde{\mathcal{M}}'_i(\bar{\lambda})}\,c(\boldsymbol{\lambda}),
		\end{split}
	\end{equation}
	where
	\begin{equation} \label{mathcal_M_definition}
		\begin{split}
			& B_2(\lambda_{2i})\Psi_{\boldsymbol{\lambda}} = \mathcal{M}_i(\lambda)\Psi_{\boldsymbol{\lambda}+e_{2i}}, \qquad \bar{B}_2(\bar{\lambda}_{2i})\Psi_{\boldsymbol{\lambda}} = \widetilde{\mathcal{M}}_i(\bar{\lambda})\Psi_{\boldsymbol{\lambda}+\bar{e}_{2i}}, \\
			& B_3(\lambda_{3i})\Psi_{\boldsymbol{\lambda}} = \mathcal{M}'_i(\lambda)\Psi_{\boldsymbol{\lambda}+e_{3i}}, \qquad \bar{B}_3(\bar{\lambda}_{3i})\Psi_{\boldsymbol{\lambda}} = \widetilde{\mathcal{M}}'_i(\bar{\lambda})\Psi_{\boldsymbol{\lambda}+\bar{e}_{3i}}.
		\end{split}
	\end{equation}
	The conditions \eqref{gl_4_condition_on_c_shift_2i} and \eqref{gl_4_condition_on_c_shift_3i} are derived in Appendix~\ref{appendix_gl4_coefficient_conditions}.
	
	Consider
	\begin{equation*}
		\begin{split}
			& \mathcal{M}_i(\lambda) = \lambda_{2i}-\lambda_{11}, \qquad \widetilde{\mathcal{M}}_i(\bar{\lambda}) = -(\bar{\lambda}_{2i}-\bar{\lambda}_{11}), \\
			& \mathcal{M}'_i(\lambda) = \prod\limits_{k=1}^2(\lambda_{3i}-\lambda_{2k}), \qquad \widetilde{\mathcal{M}}'_i(\bar{\lambda}) = -\prod\limits_{k=1}^2(\bar{\lambda}_{3i}-\bar{\lambda}_{2k}).
		\end{split}
	\end{equation*}
	Then,
	\begin{equation*}
		c(\boldsymbol{\lambda}) = (-1)^{1+\sum(\lambda_{2j}-\bar{\lambda}_{2j})}\prod\limits_{j=1}^3\Gamma^\mathbb{C}(\boldsymbol{\lambda}_{3j}-\boldsymbol{\sigma}_1+\boldsymbol{1})
	\end{equation*}
	is a solution of \eqref{gl_4_condition_on_c_shift_2i} and \eqref{gl_4_condition_on_c_shift_3i}. Choose the appropriate domain of integration in \eqref{gl_4_eigenfunction}. It comes out that it is the domain of parameters \eqref{gl_3_parameters_orthogonal_set} indexing the orthogonal set of $gl_3(\mathbb{C})$ eigenfunctions:
	\begin{multline} \label{gl_4_eigenfunction_orthogonal_set}
		\Psi_{\boldsymbol{\lambda}}(\xi,\eta,\zeta,\mathbf{z}) = (-1)^{\sum(\lambda_{2j}-\bar{\lambda}_{2j})}\prod\limits_{j=1}^3\Gamma^\mathbb{C}(\boldsymbol{\lambda}_{3j}-\boldsymbol{\sigma}_1+\boldsymbol{1}) \lim\limits_{\varepsilon,\varepsilon_1\to 0_+} \int D\boldsymbol{\gamma}\,(-1)^{\sum(\gamma_{2j}-\bar{\gamma}_{2j})}\|\boldsymbol{\gamma}_{21}-\boldsymbol{\gamma}_{22}\|^2 \\
		\times \Phi_{\boldsymbol{\gamma}}(\mathbf{z}) \Gamma^\mathbb{C}\left(-\boldsymbol{\gamma}_{11}+\sum\boldsymbol{\gamma}_{2j}+\sum\boldsymbol{\lambda}_{2j}-\sum\boldsymbol{\lambda}_{3j}+\boldsymbol{1}\right) \prod\limits_{k,j} \Gamma^\mathbb{C}(\boldsymbol{\lambda}_{3k}-\boldsymbol{\gamma}_{2j}+\frac{\boldsymbol{\varepsilon}}{\boldsymbol{2}}) \\
		\times \Gamma^\mathbb{C}\left(\boldsymbol{\gamma}_{11}+\boldsymbol{\lambda}_{11}-\sum\boldsymbol{\lambda}_{2j}+\boldsymbol{1}\right) {}_4G^{\mathbb{C}}_4\left[\begin{array}{c}
			\boldsymbol{\lambda}_{21}+\frac{\boldsymbol{\varepsilon}_1}{\boldsymbol{2}}, \boldsymbol{\lambda}_{22}+\frac{\boldsymbol{\varepsilon}_1}{\boldsymbol{2}}, \boldsymbol{\gamma}_{21}, \boldsymbol{\gamma}_{22} \\
			\boldsymbol{1}-\boldsymbol{\lambda}_{31}, \boldsymbol{1}-\boldsymbol{\lambda}_{32}, \boldsymbol{1}-\boldsymbol{\lambda}_{33}, -\boldsymbol{\gamma}_{11}
		\end{array}; 1\right] \\
		\times \xi^{-\boldsymbol{\gamma}_{11}-\boldsymbol{\lambda}_{11}+\sum\boldsymbol{\lambda}_{2j}-\boldsymbol{1}} \eta^{\boldsymbol{\gamma}_{11}-\sum\boldsymbol{\gamma}_{2j}-\sum\boldsymbol{\lambda}_{2j}+\sum\boldsymbol{\lambda}_{3j}-\boldsymbol{1}} \zeta^{\boldsymbol{\sigma}_1-\boldsymbol{1}+\sum\boldsymbol{\gamma}_{2j}-\sum\boldsymbol{\lambda}_{3j}},
	\end{multline}
	where
	\begin{equation} \label{gl_4_eigenfunction_orthogonal_set_notations}
		\gamma_{lj} = \frac{n_{lj}+\varkappa-3+l+i\mu_{lj}}{2}, \; \bar{\gamma}_{lj} = \frac{-n_{lj}+\varkappa-3+l+i\mu_{lj}}{2}, \qquad 1\leq j< l\leq 2
	\end{equation}
	with $n_{lj}\in\mathbb{Z}$, $\mu_{lj}\in\mathbb{R}$ and $\varkappa$ the constant from parameters \eqref{principal_series_rep_parameters_form} of the principal series representation. The notation $\|\boldsymbol{\gamma}_{21}-\boldsymbol{\gamma}_{22}\|^2$ is similar to \eqref{squared_norm_of_lambda_difference}. The symbol $D\boldsymbol{\gamma}$ stands for
	\begin{equation*}
		\int D\boldsymbol{\gamma} \equiv \prod\limits_{1\leq j< l\leq 2}\left(\sum\limits_{n_{lj}}\int\limits_{-\infty}^\infty \frac{d\mu_{lj}}{2}\right).
	\end{equation*}
	The expression for the hypergeometric function of the complex field reads
	\begin{multline} \label{4G4}
		{}_4G^{\mathbb{C}}_4\left[\begin{array}{c}
			\boldsymbol{\lambda}_{21}+\frac{\boldsymbol{\varepsilon}_1}{\boldsymbol{2}}, \boldsymbol{\lambda}_{22}+\frac{\boldsymbol{\varepsilon}_1}{\boldsymbol{2}}, \boldsymbol{\gamma}_{21}, \boldsymbol{\gamma}_{22} \\
			\boldsymbol{1}-\boldsymbol{\lambda}_{31}, \boldsymbol{1}-\boldsymbol{\lambda}_{32}, \boldsymbol{1}-\boldsymbol{\lambda}_{33}, -\boldsymbol{\gamma}_{11}
		\end{array}; 1\right] = \frac{1}{2\pi}\sum\limits_{k\in\mathbb{Z}} \lim\limits_{\varepsilon_2\to0_+} \int\limits_{-\infty}^\infty dv \prod\limits_{l=1}^2\Gamma^\mathbb{C}(\boldsymbol{\lambda}_{2l}+\boldsymbol{s}+\frac{\boldsymbol{\varepsilon}_1}{\boldsymbol{2}}) \\
		\times \prod\limits_{l=1}^2\Gamma^\mathbb{C}(\boldsymbol{\gamma}_{2l}+\boldsymbol{s})\prod\limits_{l=1}^3\Gamma^\mathbb{C}(\boldsymbol{1}-\boldsymbol{\lambda}_{3l}-\boldsymbol{s}) \; \Gamma^\mathbb{C}(-\boldsymbol{\gamma}_{11}-\boldsymbol{s}+\frac{\boldsymbol{\varepsilon}_2}{\boldsymbol{2}}),
	\end{multline}
	where
	\begin{equation*}
		s=\frac{k-\varkappa+2+iv}{2}, \qquad \bar{s}=\frac{-k-\varkappa+2+iv}{2}.
	\end{equation*}
	In \eqref{gl_4_eigenfunction_orthogonal_set} the $i\varepsilon$-prescription is used in integration with respect to $\mu_{21}$ and $\mu_{22}$, the $i\varepsilon_1$-prescription is used in integration with respect to $\mu_{11}$. Consider
	\begin{equation} \label{gl_4_parameters_orthogonal_set}
		\lambda_{lj} = \frac{m_{lj}+\varkappa-4+l+i\beta_{lj}}{2}, \; \bar{\lambda}_{lj} = \frac{-m_{lj}+\varkappa-4+l+i\beta_{lj}}{2}, \qquad 1\leq j<l\leq 3,
	\end{equation}
	where $m_{lj}\in\mathbb{Z}$, $\beta_{lj}\in\mathbb{R}$. In Appendix~\ref{appendix_orthogonality} it is shown that the functions \eqref{gl_4_eigenfunction_orthogonal_set} indexed by parameters \eqref{gl_4_parameters_orthogonal_set} form an orthogonal set. The orthogonality relation reads
	\begin{multline} \label{gl_4_orthogonality}
		\int\limits_{\mathbb{C}}\prod\limits_{1\leq j<l\leq 4}d^2z_{lj} \, \rho(\boldsymbol{\lambda})\Psi_{\boldsymbol{\lambda}}^\ast(z_{21},z_{31},z_{41},\mathbf{z}) \, \rho(\boldsymbol{\lambda}')\Psi_{\boldsymbol{\lambda}'}(z_{21},z_{31},z_{41},\mathbf{z}) = 2^{13}\,3^2\,\pi^{18} \rho(\boldsymbol{\lambda}) \\
		\times \delta^{(2)}(\boldsymbol{\lambda}_{11}-\boldsymbol{\lambda}_{11}') \left(\sum\limits_{\tau\in S_2}\prod\limits_{j=1}^2\delta^{(2)}(\boldsymbol{\lambda}_{2j}-\boldsymbol{\lambda}_{2\tau(j)}')\right) \left(\sum\limits_{\tau\in S_3}\prod\limits_{j=1}^3\delta^{(2)}(\boldsymbol{\lambda}_{3j}-\boldsymbol{\lambda}_{3\tau(j)}')\right),
	\end{multline}
	where $S_2$ and $S_3$ are permutation groups, the notation $\delta^{(2)}(\boldsymbol{\lambda}_{lj}-\boldsymbol{\lambda}_{lp}')$ is similar to \eqref{parameters'_delta_function_definition}, the function $\rho(\boldsymbol{\lambda})$ has the form
	\begin{equation} \label{rho(lambda)}
		\rho(\boldsymbol{\lambda}) \equiv \prod\limits_{r=1}^3 \prod\limits_{1\leq l<j\leq r}\|\boldsymbol{\lambda}_{rl}-\boldsymbol{\lambda}_{rj}\|^2.
	\end{equation}

	\section{Conclusion}

	As the result, the Mellin-Barnes integral representation for the Gelfand-Tsetlin bases of unitary principal series representations was derived for ranks 3 and 4. The corresponding formulas are \eqref{gl_3_eigenfunction_orthogonal_set} and \eqref{gl_4_eigenfunction_orthogonal_set}. The orthogonality relations for the obtained sets of basis elements are \eqref{gl_3_orthogonality} and \eqref{gl_4_orthogonality}. In the case of $gl_4(\mathbb{C})$ the kernel of the Mellin-Barnes type integral is expressed via the hypergeometric function of the complex field ${}_4G^{\mathbb{C}}_4$ at unity \cite[Section~1.4]{Neretin_func_complex_field}. This form is more complicated than for the rank $3$ where only gamma functions appear in the expression. Thereon, it can be supposed that in the case of higher ranks the kernel is expressed in terms of hypergeometric functions of the complex field or their multivariable generalizations.
	
	The advantage of the Mellin-Barnes representation is that it can be used to prove completeness of obtained sets of eigenfunctions by analogy with what was done in the case of Toda chain with the help of integral representation of the same type (see, for example, Section~2 in \cite{DKM_21}). The derivation for the Toda lattice relies on Gustafson's integral identities \cite{Gustafson_94,Gustafson_92}. Their analogues with domain of integration \eqref{gl_n_integration_measure} were obtained in \cite{Der_Man_on_complex_gamma_func_integrals,Der_Man_Val_SL_2_C_Gustafson_integrals,DMV_17} and can be used to derive completeness relations for Gelfand-Tsetlin bases considered in the present paper.
	
	As a separate result for arbitrary rank it is worth noting the formulae (\ref{L_m_th_corner_minor_in_terms_of_mathcal_L}--\ref{L_m_th_non_corner_minor_2_in_terms_of_mathcal_L}) expressing the minors of $gl_n(\mathbb{C})$ L-operator via the minors of $gl_{n-1}(\mathbb{C})$ L-operator for principal series representations, as well as formulas \eqref{B_rab_acts_on_Psi_lambda_bar_lambda} and \eqref{B_r,a1,a2,b1,b2(lambda_ir)_action_on_Psi_lambda} of action of non-corner minors on the eigenfunctions of corner ones. The latter hold for any representation of $gl_n(\mathbb{C})$ (not only principal series) in which the corner minors of the L-operator can be diagonalized. It follows from the fact that for their derivation one needs only \eqref{Gelfand_Tsetlin_basis_definition}, commutation relations \eqref{quantum_minors_commutation_relations_final} between the minors and Lagrange interpolation.
	
	On the grounds of relations \eqref{gl_3_non_corner_minors_action} for $gl_3(\mathbb{C})$ and (\ref{gl_4_condition_on_c_shift_2i}--\ref{gl_4_condition_on_c_shift_3i}) for $gl_4(\mathbb{C})$ it is possible to formulate an induction proposition that in any rank $n'$ there exists a normalization of Gelfand-Tsetlin basis such that the minors $B_r$ and $\bar{B}_r$ from \eqref{B_r_definition} act in the following way:
	\begin{equation} \label{B_r_action_induction_assumption}
		\begin{split}
			& B_1(u)\Psi_{\boldsymbol{\lambda}} = \Psi_{\boldsymbol{\lambda}+e_{11}}, \qquad B_r(\lambda_{ri})\Psi_{\boldsymbol{\lambda}} = \prod\limits_{j=1}^{r-1}(\lambda_{ri}-\lambda_{r-1,j})\;\Psi_{\boldsymbol{\lambda}+e_{ri}}, \quad 2\leq r\leq n'-1, \\
			& \bar{B}_1(\bar{u})\Psi_{\boldsymbol{\lambda}} = -\Psi_{\boldsymbol{\lambda}+\bar{e}_{11}}, \qquad \bar{B}_r(\bar{\lambda}_{ri})\Psi_{\boldsymbol{\lambda}} = -\prod\limits_{j=1}^{r-1}(\bar{\lambda}_{ri}-\bar{\lambda}_{r-1,j})\;\Psi_{\boldsymbol{\lambda}+\bar{e}_{ri}}, \quad 2\leq r\leq n'-1.
		\end{split}
	\end{equation}
	The notations like $\Psi_{\boldsymbol{\lambda}+e_{ri}}$ are similar to \eqref{gl_2_short_notations}. Suppose that \eqref{B_r_action_induction_assumption} is correct for $gl_{n-1}(\mathbb{C})$. Then with the help of \eqref{L_m_th_corner_minor_in_terms_of_mathcal_L} and \eqref{B_rab_acts_on_Psi_lambda_bar_lambda} one can obtain a system of finite-difference equations for the kernel $K_{\boldsymbol{\gamma}}(\boldsymbol{\lambda})$ of the Mellin-Barnes representation \eqref{gl_n_ansatz} for $gl_n(\mathbb{C})$. With $K_{\boldsymbol{\gamma}}(\boldsymbol{\lambda})$ obtained it is possible to prove or disprove the proposition \eqref{B_r_action_induction_assumption} for $gl_n(\mathbb{C})$ with the help of \eqref{B_m_for_L_in_terms_of_mathcal_L}, \eqref{B_rab_acts_on_Psi_lambda_bar_lambda} and \eqref{B_r,a1,a2,b1,b2(lambda_ir)_action_on_Psi_lambda}.
	
	The objectives of the further research include the completeness of obtained sets of basis elements, solution of the system of finite-difference equations for $K_{\boldsymbol{\gamma}}(\boldsymbol{\lambda})$ for arbitrary rank and the question of equivalence of the Mellin-Barnes and the Gauss-Givental \cite{Valinevich_2019,Val_Der_Kul_Eigenfunc_quantum_minors_SL_n_C} integral representations.

	\section*{Acknowledgements}

	This paper was supported in part by the Russian Science Foundation, under grant 18-11-00297. I would like to express my deepest gratitude to P.A.~Valinevich for valuable insights, fruitful discussions and critical remarks. I am grateful to A.G.~Pronko for helpful comments on the paper. I would like to thank S.E.~Derkachov and P.G.~Gavrylenko for discussions and references. I would like to express my appreciation to faculty members of Skoltech Center for Advanced Studies and the Faculty of Mathematics of HSE University for teaching me and organizing student schools at one of which I got familiar with the topic of the present article.

	\appendix

	\section{Derivation of expressions for the minors of $gl_n(\mathbb{C})$ L-operator in terms of the minors of $gl_{n-1}(\mathbb{C})$ L-operator}   \label{appendix_minors}

	In order to obtain the expressions the columnwise form of $gl_n(\mathbb{C})$ L-operator $L(u)$ is needed. In accordance with \eqref{L_operator_recurrence_relation_explicit},
	\begin{equation*}
		L(u) = \begin{pmatrix}
			(u-\sigma_1+n-1)\mathrm{z}-\sum\limits_{b=1}^{n-1} z_{b+1,1}(\mathcal{L}'_b(u)-\mathrm{z}\partial_{b+1,1}) & \mathcal{L}'_1(u)-\mathrm{z}\partial_{21} & \ldots & \mathcal{L}'_{n-1}(u)-\mathrm{z}\partial_{n1}
		\end{pmatrix},
	\end{equation*}
	where
	\begin{equation} \label{mathrm_z_and_mathcal_L'_p_definition}
		\mathrm{z}\equiv\begin{pmatrix}
			1 \\ z_{21} \\ \vdots \\ z_{n1}
		\end{pmatrix}, \qquad \mathrm{z}\partial_{k1} =
		\begin{pmatrix}
			\partial_{k1} \\ z_{21}\partial_{k1} \\ \vdots \\ z_{n1}\partial_{k1}
		\end{pmatrix}, \qquad
		\mathcal{L}'_p(u)\equiv\begin{pmatrix}
			0 \\ \mathcal{L}_p(u)
		\end{pmatrix}, \quad p=1,\ldots,n-1,
	\end{equation}
	and $\mathcal{L}_p(u)$ is the $p$-th column of $gl_{n-1}(\mathbb{C})$ L-operator $\mathcal{L}(u)$.
	
	For any natural number $k$ and columns $F_1(u),\ldots,F_k(u)$ of length greater or equal to $k$ with operator entries depending on spectral parameter $u$ define
	\begin{equation} \label{det_definition}
		\det(F_1(u),\ldots,F_k(u)) \equiv \sum\limits_{\tau\in S_k} \mathrm{sgn}(\tau) F_1(u)^{\tau(1)}\ldots F_k(u)^{\tau(k)},
	\end{equation}
	where $S_k$ is the symmetric group, $F_j(u)^{\tau(j)}$ is the $\tau(j)$-th element in the column $F_j(u)$. For $i\in\{1,\ldots,n\}$ denote the $i$-th column of $L(u)$ by $L_i(u)$. By the definition \eqref{quantum_minor_definition}, the quantum minor $L(u)^{1\ldots m}_{1\ldots m}$ is expressed in terms of $\det$ as
	\begin{equation} \label{L(u)_1..m_1..m_in_terms_of_det}
		\begin{split}
			& L(u)^{1\ldots m}_{1\ldots m} = \det(L_1(u-m+1),L_2(u-m+2),\ldots,L_m(u)) \\
			& = (u-\sigma_1+n-m)\,\det\left(\mathrm{z},\mathcal{L}'_1(u-m+2)-\mathrm{z}\partial_{21},\ldots,\mathcal{L}'_{m-1}(u)-\mathrm{z}\partial_{m1}\right) \\
			& - \sum\limits_{b=1}^{n-1} z_{b+1,1}\,\det\left(\mathcal{L}'_b(u-m+1)-\mathrm{z}\partial_{b+1,1}, \mathcal{L}'_1(u-m+2)-\mathrm{z}\partial_{21}, \ldots, \mathcal{L}'_{m-1}(u)-\mathrm{z}\partial_{m1}\right).
		\end{split}
	\end{equation}
	
	Consider the $\det$ from the first summand after the last equality sign in \eqref{L(u)_1..m_1..m_in_terms_of_det}:
	\begin{multline} \label{L(u)_1..m_1..m_in_terms_of_det_first_summand}
		\det\left(\mathrm{z},\mathcal{L}'_1(u-m+2)-\mathrm{z}\partial_{21},\mathcal{L}'_2(u-m+3)-\mathrm{z}\partial_{31},\ldots,\mathcal{L}'_{m-1}(u)-\mathrm{z}\partial_{m1}\right) \\
		= \det\left(\mathrm{z},\mathcal{L}'_1(u-m+2),\mathcal{L}'_2(u-m+3)-\mathrm{z}\partial_{31},\ldots,\mathcal{L}'_{m-1}(u)-\mathrm{z}\partial_{m1}\right) \\
		- \det\left(\mathrm{z},\mathrm{z}\partial_{21},\mathcal{L}'_2(u-m+3)-\mathrm{z}\partial_{31},\ldots,\mathcal{L}'_{m-1}(u)-\mathrm{z}\partial_{m1}\right).
	\end{multline}
	The $\det$ from the second summand in the RHS of \eqref{L(u)_1..m_1..m_in_terms_of_det_first_summand} is zero, because of the definition \eqref{det_definition} and proportionality of the second column to the first one. Doing similar manipulations with $3$-rd, $\ldots$, $n$-th arguments of the first summand in \eqref{L(u)_1..m_1..m_in_terms_of_det_first_summand} one obtains
	\begin{multline} \label{L(u)_1..m_1..m_in_terms_of_det_first_summand_calculated}
		\det\left(\mathrm{z},\mathcal{L}'_1(u-m+2)-\mathrm{z}\partial_{21},\mathcal{L}'_2(u-m+3)-\mathrm{z}\partial_{31},\ldots,\mathcal{L}'_{m-1}(u)-\mathrm{z}\partial_{m1}\right) \\
		= \det\left(\mathrm{z},\mathcal{L}'_1(u-m+2),\mathcal{L}'_2(u-m+3),\ldots,\mathcal{L}'_{m-1}(u)\right) \\
		= \det\left(\mathcal{L}_1(u-m+2),\mathcal{L}_2(u-m+3),\ldots,\mathcal{L}_{m-1}(u)\right) = \mathcal{L}(u)^{1,\ldots,m-1}_{1,\ldots,m-1}.
	\end{multline}
	The next to the last equality sign in \eqref{L(u)_1..m_1..m_in_terms_of_det_first_summand_calculated} follows from the fact that the first element in $\mathrm{z}$ is $1$ and the first element in $\mathcal{L}'_p(u)$ is $0$, see \eqref{mathrm_z_and_mathcal_L'_p_definition}.
	
	Consider the $\det$ from the $b$-th term in $\sum_{b=1}^{n-1}$ after the the last equality sign in \eqref{L(u)_1..m_1..m_in_terms_of_det}:
	\begin{equation} \label{L(u)_1..m_1..m_in_terms_of_det_b_th_summand}
		\begin{split}
			& \det\left(\mathcal{L}'_b(u-m+1)-\mathrm{z}\partial_{b+1,1}, \mathcal{L}'_1(u-m+2)-\mathrm{z}\partial_{21}, \ldots, \mathcal{L}'_{m-1}(u)-\mathrm{z}\partial_{m1}\right) \\
			& = \det\left(\mathcal{L}'_b(u-m+1)-\mathrm{z}\partial_{b+1,1}, \mathcal{L}'_1(u-m+2), \ldots, \mathcal{L}'_{m-1}(u)\right) \\
			& \begin{aligned}
				+ \sum\limits_{a=1}^{m-1}\det\left[\mathcal{L}'_b(u-m+1)-\mathrm{z}\partial_{b+1,1}, \mathcal{L}'_1(u-m+2)-\mathrm{z}\partial_{21}, \ldots, \mathcal{L}'_{a-1}(u-m+a)-\mathrm{z}\partial_{a1}, \right. \\
				\left.-\mathrm{z}\partial_{a+1,1},\mathcal{L}'_{a+1}(u-m+a+2),\ldots,\mathcal{L}'_{m-1}(u)\right].
			\end{aligned}
		\end{split}
	\end{equation}
	Observe that
	\begin{equation} \label{mathrm_z_commutation_with_partial}
		\mathcal{L}'_k(u-m+k+1)-\mathrm{z}\partial_{k+1,1} = \mathcal{L}'_k(u-m+k+1) - \partial_{k+1,1}\mathrm{z} + e_{k+1} = \mathcal{L}'_k(u-m+k+2) - \partial_{k+1,1}\mathrm{z},
	\end{equation}
	where $e_{k+1}$ is the column with zero entries except the $(k+1)$-th one which is equal to $1$. Decomposing by linearity the first term in the RHS of \eqref{L(u)_1..m_1..m_in_terms_of_det_b_th_summand} and using \eqref{mathrm_z_commutation_with_partial} in every term of $\sum_{a=1}^{m-1}$ one obtains
	\begin{multline} \label{L(u)_1..m_1..m_in_terms_of_det_b_th_summand_iteration_1}
		\det\left(\mathcal{L}'_b(u-m+1)-\mathrm{z}\partial_{b+1,1}, \mathcal{L}'_1(u-m+2)-\mathrm{z}\partial_{21}, \ldots, \mathcal{L}'_{m-1}(u)-\mathrm{z}\partial_{m1}\right) \\
		= \det\left(\mathcal{L}'_b(u-m+1), \mathcal{L}'_1(u-m+2), \ldots, \mathcal{L}'_{m-1}(u)\right) - \det\left(\mathrm{z}\partial_{b+1,1}, \mathcal{L}'_1(u-m+2), \ldots, \mathcal{L}'_{m-1}(u)\right) \\
		\begin{aligned}
			& + \sum\limits_{a=1}^{m-1}\det\left[\mathcal{L}'_b(u-m+2)-\partial_{b+1,1}\mathrm{z}, \mathcal{L}'_1(u-m+3)-\partial_{21}\mathrm{z}, \ldots, \right. \\
			& \left.\mathcal{L}'_{a-1}(u-m+a+1)-\partial_{a1}\mathrm{z},-\mathrm{z}\partial_{a+1,1},\mathcal{L}'_{a+1}(u-m+a+2),\ldots,\mathcal{L}'_{m-1}(u)\right].
		\end{aligned}
	\end{multline}
	Now it is possible to tell that in the $a$-th term of $\sum_{a=1}^{m-1}$ in \eqref{L(u)_1..m_1..m_in_terms_of_det_b_th_summand_iteration_1} ``$-\partial\mathrm{z}$'' vanish from $1$-st, $\ldots$, $a$-th arguments because they are proportional to $\mathrm{z}$ from the $(a+1)$-th argument $-\mathrm{z}\partial_{a+1,1}$. The first summand in the RHS of \eqref{L(u)_1..m_1..m_in_terms_of_det_b_th_summand_iteration_1} disappears because the columns in all arguments of $\det$ have $0$ as the first entry. So, \eqref{L(u)_1..m_1..m_in_terms_of_det_b_th_summand_iteration_1} transforms to
	\begin{equation} \label{L(u)_1..m_1..m_in_terms_of_det_b_th_summand_iteration_2}
		\begin{split}
			& \det\left(\mathcal{L}'_b(u-m+1)-\mathrm{z}\partial_{b+1,1}, \mathcal{L}'_1(u-m+2)-\mathrm{z}\partial_{21}, \ldots, \mathcal{L}'_{m-1}(u)-\mathrm{z}\partial_{m1}\right) \\
			& = -\det\left(\mathrm{z}\partial_{b+1,1}, \mathcal{L}'_1(u-m+2), \ldots, \mathcal{L}'_{m-1}(u)\right) \\
			& \begin{aligned}
				+ \sum\limits_{a=1}^{m-1}\det\left[\mathcal{L}'_b(u-m+2), \mathcal{L}'_1(u-m+3), \ldots, \mathcal{L}'_{a-1}(u-m+a+1), \right. \\
				\left.-\mathrm{z}\partial_{a+1,1},\mathcal{L}'_{a+1}(u-m+a+2),\ldots,\mathcal{L}'_{m-1}(u)\right].
			\end{aligned}
		\end{split}
	\end{equation}
	The first entry in $\mathrm{z}$ is $1$, the first entry in $\mathcal{L}'_p(u)$ is $0$. Consequently, \eqref{L(u)_1..m_1..m_in_terms_of_det_b_th_summand_iteration_2} can be rewritten as
	\begin{equation} \label{L(u)_1..m_1..m_in_terms_of_det_b_th_summand_calculated}
		\begin{split}
			& \det\left(\mathcal{L}'_b(u-m+1)-\mathrm{z}\partial_{b+1,1}, \mathcal{L}'_1(u-m+2)-\mathrm{z}\partial_{21}, \ldots, \mathcal{L}'_{m-1}(u)-\mathrm{z}\partial_{m1}\right) \\
			& = -\det\left(\mathcal{L}_1(u-m+2), \ldots, \mathcal{L}_{m-1}(u)\right)\partial_{b+1,1} \\
			& \begin{aligned}
				+ \sum\limits_{a=1}^{m-1} (-1)^{a+1}\det\left[\mathcal{L}_b(u-m+2), \mathcal{L}'_1(u-m+3), \ldots, \mathcal{L}'_{a-1}(u-m+a+1), \right. \\
				\left. \mathcal{L}'_{a+1}(u-m+a+2),\ldots,\mathcal{L}'_{m-1}(u)\right]\partial_{a+1,1}
			\end{aligned} \\
			& = -\mathcal{L}(u)^{1,\ldots,m-1}_{1,\ldots,m-1}\,\partial_{b+1,1} + \sum\limits_{a=1}^{m-1} (-1)^{a+1}\mathcal{L}(u)^{1,\ldots,m-1}_{b,1,\ldots,a-1,a+1,\ldots,m-1}\partial_{a+1,1} \\
			& = -\mathcal{L}(u)^{1,\ldots,m-1}_{1,\ldots,m-1}\,\partial_{b+1,1} + \sum\limits_{a=1}^{m-1} (-1)^{a+m+1}\mathcal{L}(u)^{1,\ldots,m-1}_{1,\ldots,a-1,a+1,\ldots,m-1,b}\partial_{a+1,1}.
		\end{split}
	\end{equation}
	Substituting \eqref{L(u)_1..m_1..m_in_terms_of_det_first_summand_calculated} and \eqref{L(u)_1..m_1..m_in_terms_of_det_b_th_summand_calculated} into \eqref{L(u)_1..m_1..m_in_terms_of_det} one obtains the desired formula \eqref{L_m_th_corner_minor_in_terms_of_mathcal_L}.

	Consider $i_1,\ldots, i_{m-1}$ such that $2\leq i_1<\ldots<i_{m-1}\leq n$. The expression of the minor $L(u)^{1\ldots m}_{1,i_1,\ldots, i_{m-1}}$ in terms of $\det$ reads
	\begin{equation*}
		\begin{aligned}
			L(u)^{1\ldots m}_{1,i_1,\ldots, i_{m-1}} = \det[(u-\sigma_1+n-m)\mathrm{z}-\sum\limits_{b=1}^{n-1} z_{b+1,1}(\mathcal{L}'_b(u-m+1)-\mathrm{z}\partial_{b+1,1}),\\
			\mathcal{L}'_{i_1-1}(u-m+2)-\mathrm{z}\partial_{i_11},\ldots,\mathcal{L}'_{i_{m-1}-1}(u)-\mathrm{z}\partial_{i_{m-1}1}],
		\end{aligned}
	\end{equation*}
	which is similar to \eqref{L(u)_1..m_1..m_in_terms_of_det}. Calculations analogous to (\ref{L(u)_1..m_1..m_in_terms_of_det}--\ref{L(u)_1..m_1..m_in_terms_of_det_b_th_summand_calculated}) lead to the formula \eqref{L_m_th_non_corner_minor_1_in_terms_of_mathcal_L}.
	
	Consider $i_1,\ldots, i_m$ such that $2\leq i_1<\ldots<i_m\leq n$. The expression of the minor $L(u)^{1\ldots m}_{i_1,\ldots, i_m}$ in terms of $\det$ reads
	\begin{equation*}
		L(u)^{1\ldots m}_{1,i_1,\ldots, i_{m-1}} = \det[\mathcal{L}'_{i_1-1}(u-m+1)-\mathrm{z}\partial_{i_11},\ldots,\mathcal{L}'_{i_m-1}(u)-\mathrm{z}\partial_{i_m1}]
	\end{equation*}
	which is similar to $\det$ from any term in $\sum_{b=1}^{n-1}$ in \eqref{L(u)_1..m_1..m_in_terms_of_det}. Calculations analogous to (\ref{L(u)_1..m_1..m_in_terms_of_det_b_th_summand}--\ref{L(u)_1..m_1..m_in_terms_of_det_b_th_summand_calculated}) lead to the formula \eqref{L_m_th_non_corner_minor_2_in_terms_of_mathcal_L}.

	\section{Derivation of formulas \eqref{B_rab_acts_on_Psi_lambda_bar_lambda}, \eqref{B_r_acts_on_Psi_lambda_bar_lambda} and \eqref{B_r,a1,a2,b1,b2(lambda_ir)_action_on_Psi_lambda}}   \label{appendix_B_action}

	Consider the L-operator for $gl_{n'}(\mathbb{C})$. In order to derive the formulas \eqref{B_rab_acts_on_Psi_lambda_bar_lambda}, \eqref{B_r_acts_on_Psi_lambda_bar_lambda} and \eqref{B_r,a1,a2,b1,b2(lambda_ir)_action_on_Psi_lambda} one needs the commutation relations between the quantum minors \cite[\S~1.15,~Section~8]{Molev_Yangians_and_classical_Lie_algebras}:
	\begin{multline} \label{quantum_minors_commutation_relations_final}
		\left[L(u)^{a_1\ldots a_r}_{b_1\ldots b_r},L(v)^{a_1^\prime\ldots a_m^\prime}_{b_1^\prime\ldots b_m^\prime}\right] \\
		= \sum\limits_{k=1}^{\min(r,m)}\frac{k!}{\prod\limits_{i=1}^k(u-v+m-i)} \sum\limits_{\begin{smallmatrix}
				i_1<\ldots<i_k\leq r \\
				j_1<\ldots<j_k\leq m
		\end{smallmatrix}}\left(L^{a_1^\prime\ldots a_m^\prime}_{b'_1\ldots b_{i_1}\ldots b_{i_k}\ldots b'_m}(v)L^{a_1\ldots a_r}_{b_1\ldots b'_{j_1}\ldots b'_{j_k}\ldots b_r}(u) \right. \\
		\left.- L^{a_1\ldots a'_{j_1}\ldots a'_{j_k}\ldots a_r}_{b_1\ldots b_r}(u)L^{a'_1\ldots a_{i_1}\ldots a_{i_k}\ldots a'_m}_{b_1^\prime\ldots b_m^\prime}(v)\right).
	\end{multline}
	It is worth mentioning that in \cite{Molev_Yangians_and_classical_Lie_algebras} the R-matrix has the form $\widetilde{R}(u)=\mathds{1}-u^{-1}P$ and a bit differs from \eqref{Yang_R_matrix}. Comparing the definition of quantum minors from \S~1.6 of the book with \eqref{quantum_minor_definition} and using the fact that $\widetilde{L}(u)\equiv L(-u)$ obeys \eqref{Yangian_defining_relation_RTT} with $\widetilde{R}$ one can obtain \eqref{quantum_minors_commutation_relations_final} from the corresponding relation in \cite{Molev_Yangians_and_classical_Lie_algebras}.
	
	Setting $(a_1,\ldots,a_r)=(1,\ldots,r)$, $(a'_1,\ldots,a'_m)=(1,\ldots,m)$ in \eqref{quantum_minors_commutation_relations_final} and using the antisymmetry of quantum minors \eqref{antisymmetry_of_quantum_minors} one can derive
	\begin{multline}  \label{type_B'_minors_commutation_relations}
		(u-v+m-\mu+1)_\mu\,T^{1\ldots r}_{b_1\ldots b_r}(u)T^{1\ldots m}_{b'_1\ldots b'_m}(v) = (u-v+m-\mu)_\mu\,T^{1\ldots m}_{b'_1\ldots b'_m}(v)T^{1\ldots r}_{b_1\ldots b_r}(u) \\
		+ \sum\limits_{k=1}^\mu k!(u-v+m-\mu)_{\mu-k}\sum\limits_{\begin{smallmatrix}
				i_1<\ldots<i_k\leq r \\
				j_1\ldots<j_k\leq m
		\end{smallmatrix}} T^{1\ldots m}_{b'_1\ldots b_{i_1}\ldots b_{i_k}\ldots b'_m}(v)T^{1\ldots r}_{b_1\ldots b'_{j_1}\ldots b'_{j_k}\ldots b_r}(u),
	\end{multline}
	where $\mu\equiv\min(r,m)$, the notation $(u-v+m-\mu+1)_\mu$ means the Pochhammer symbol (the rising factorial).
	
	Setting in \eqref{type_B'_minors_commutation_relations} $(b_1,\ldots,b_r)=(1,\ldots,\widehat{a},\ldots,r,b)$, $(b'_1,\ldots,b'_m)=(1,\ldots,m)$, where $1\leq a\leq r<b\leq n$, and using \eqref{antisymmetry_of_quantum_minors} one obtains the commutation relations between the corner minors $A_m(u)$ and the minors $B_{rab}(u)$ from \eqref{B_rab_minor_definition}: for $m$ such that $a\leq m<r$:
	\begin{multline} \label{B_rab_A_m_commutation_a_leq_m_<_r}
		(u-v+1)B_{rab}(u)A_m(v) = (u-v)A_m(v)B_{rab}(u) \\
		+ (-1)^{m-r}B_{mab}(v)A_r(u) + \sum\limits_{c=m+1}^r (-1)^{m-c-1}B_{mac}(v)B_{rcb}(u),
	\end{multline}
	for $m$ such that $r\leq m<b$:
	\begin{multline} \label{B_rab_A_m_commutation_r_leq_m_<_b}
		(u-v+m-r+1)B_{rab}(u)A_m(v) = (u-v+m-r)A_m(v)B_{rab}(u) \\
		+ (-1)^{m-r}B_{mab}(v)A_r(u) + \sum\limits_{c=r+1}^m (-1)^{m-c}B_{mcb}(v)B_{rac}(u),
	\end{multline}
	for $m$ such that $m<a$ or $m\geq b$:
	\begin{equation} \label{B_rab_A_m_commutation_m<a_or_m_geq_b}
		[B_{rab}(u),A_m(v)]=0.
	\end{equation}

	Setting in \eqref{B_rab_A_m_commutation_r_leq_m_<_b} $a=m=r$, $b=r+1$, $u=\lambda_{ri}$, acting by both sides of this equation on $\Psi_{\boldsymbol{\lambda}}$ and using $A_r(\lambda_{ri})\Psi_{\boldsymbol{\lambda}}=0$ (consequence of \eqref{Gelfand_Tsetlin_basis_definition}) one obtains the first relation in \eqref{B_r_acts_on_Psi_lambda_bar_lambda}. Other relations in \eqref{B_r_acts_on_Psi_lambda_bar_lambda} follow from the commutation of $B_r(u)=B_{r,r,r+1}(u)$ and $A_j(v)$ when $j\neq r$ -- the formula \eqref{B_rab_A_m_commutation_m<a_or_m_geq_b}.
	
	Consider the integer numbers $a$, $b$, $r$, $m$ such that $1\leq a\leq r<m<b\leq n$. Consider the element $\Psi_{\boldsymbol{\lambda}}$ of the Gelfand-Tsetlin basis \eqref{Gelfand_Tsetlin_basis_definition}. Acting on $\Psi_{\boldsymbol{\lambda}}$ by \eqref{B_rab_A_m_commutation_r_leq_m_<_b} with $u=\lambda_{rk}$, $v=\lambda_{mi}, \; i\in\{1,\ldots,m\}$ one obtains
	\begin{equation} \label{A_m_B_rab(lambda_kr)_Psi_lambda}
		A_m(\lambda_{mi})B_{rab}(\lambda_{rk})\Psi_{\boldsymbol{\lambda}} = -\frac{1}{\lambda_{rk}-\lambda_{mi}+m-r}\sum\limits_{c=r+1}^m (-1)^{m-c}B_{mcb}(\lambda_{mi})B_{rac}(\lambda_{rk})\Psi_{\boldsymbol{\lambda}}.
	\end{equation}
	From the definition of quantum minors \eqref{quantum_minor_definition} and the L-operator \eqref{L_operators_holomorphic_and_antiholomorphic} it follows that $B_{rab}(u)$ is a polynomial in $u$ of degree $r-1$. Using \eqref{A_m_B_rab(lambda_kr)_Psi_lambda} with $k=1,\ldots,r$ and Lagrange interpolation one finds
	\begin{multline} \label{A_m_B_rab(u)_Psi_lambda}
		A_m(\lambda_{mi})B_{rab}(u)\Psi_{\boldsymbol{\lambda}} \\
		= -\sum\limits_{k=1}^r\left(\prod\limits_{l\neq k}\frac{u-\lambda_{rl}}{\lambda_{rk}-\lambda_{rl}}\right)\frac{1}{\lambda_{rk}-\lambda_{mi}+m-r}\sum\limits_{c=r+1}^m (-1)^{m-c}B_{mcb}(\lambda_{mi})B_{rac}(\lambda_{rk})\Psi_{\boldsymbol{\lambda}}.
	\end{multline}
	Acting on $\Psi_{\boldsymbol{\lambda}}$ by \eqref{B_rab_A_m_commutation_r_leq_m_<_b} with $v=\lambda_{mi}$ and substituting the expression \eqref{A_m_B_rab(u)_Psi_lambda} for $A_m(\lambda_{mi})B_{rab}(u)\Psi_{\boldsymbol{\lambda}}$ one obtains
	\begin{equation} \label{B_mab(lambda_im)Psi_lambda_i.t.o._lower_graduation_minors}
		B_{mab}(\lambda_{mi})\Psi_{\boldsymbol{\lambda}} = \sum\limits_{k=1}^r\sum\limits_{c=r+1}^m \frac{(-1)^{c-r}B_{mcb}(\lambda_{mi})B_{rac}(\lambda_{rk})}{(\lambda_{rk}-\lambda_{mi}+m-r)\prod\limits_{l\neq k}(\lambda_{rk}-\lambda_{rl})}\,\Psi_{\boldsymbol{\lambda}}.
	\end{equation}
	
	Now it is useful to find the expression for $B_{rab}(\lambda_{ri})\Psi_{\boldsymbol{\lambda}}$ similar to \eqref{B_mab(lambda_im)Psi_lambda_i.t.o._lower_graduation_minors}. Acting on $\Psi_{\boldsymbol{\lambda}}$ by \eqref{B_rab_A_m_commutation_a_leq_m_<_r} for $B_{mab}(u)$ and $A_r(v)$ with $u=\lambda_{mk}$ and $v=\lambda_{ri}$ one obtains
	\begin{equation} \label{A_r(lambda_ri)B_mab(lambda_mk)Psi_lambda}
		A_r(\lambda_{ri})B_{mab}(\lambda_{mk})\Psi_{\boldsymbol{\lambda}} = -\frac{1}{\lambda_{mk}-\lambda_{ri}}\sum\limits_{c=r+1}^m (-1)^{r-c-1}B_{rac}(\lambda_{ri})B_{mcb}(\lambda_{mk})\Psi_{\boldsymbol{\lambda}}.
	\end{equation}
	From \eqref{A_r(lambda_ri)B_mab(lambda_mk)Psi_lambda} with $k=1,\ldots,m$ and Lagrange interpolation it follows that
	\begin{multline} \label{A_r(lambda_{ir})B_{mab}(u)Psi_lambda}
		A_r(\lambda_{ri})B_{mab}(u)\Psi_{\boldsymbol{\lambda}} \\
		= -\sum\limits_{k=1}^m\left(\prod\limits_{l\neq k}\frac{u-\lambda_{ml}}{\lambda_{mk}-\lambda_{ml}}\right)\frac{1}{\lambda_{mk}-\lambda_{ri}}\sum\limits_{c=r+1}^m (-1)^{r-c-1}B_{rac}(\lambda_{ri})B_{mcb}(\lambda_{mk})\Psi_{\boldsymbol{\lambda}}.
	\end{multline}
	Acting on $\Psi_{\boldsymbol{\lambda}}$ by \eqref{B_rab_A_m_commutation_a_leq_m_<_r} for $B_{mab}(u)$ and $A_r(v)$ with $v=\lambda_{ri}$ and substituting the expression \eqref{A_r(lambda_{ir})B_{mab}(u)Psi_lambda} for $A_r(\lambda_{ri})B_{mab}(u)\Psi_{\boldsymbol{\lambda}}$ one finds
	\begin{equation} \label{B_rab(lambda_ir)Psi_lambda_i.t.o._lower_graduation_minors}
		B_{rab}(\lambda_{ri})\Psi_{\boldsymbol{\lambda}} = \sum\limits_{k=1}^m\sum\limits_{c=r+1}^m \frac{(-1)^{m-c-1}B_{rac}(\lambda_{ri})B_{mcb}(\lambda_{mk})}{(\lambda_{mk}-\lambda_{ri})\prod\limits_{l\neq k}(\lambda_{mk}-\lambda_{ml})}\,\Psi_{\boldsymbol{\lambda}}.
	\end{equation}
	
	Consider $B_{rab}(\lambda_{ri})\Psi_{\boldsymbol{\lambda}}$, use \eqref{B_rab(lambda_ir)Psi_lambda_i.t.o._lower_graduation_minors} with $m=r+1$:
	\begin{equation} \label{B_{rab}(lambda_{ir})Psi_lambda}
		B_{rab}(\lambda_{ri})\Psi_{\boldsymbol{\lambda}} = -\sum\limits_{s_{r+1}=1}^{r+1} \frac{B_{r,a,r+1}(\lambda_{ri})B_{r+1,r+1,b}(\lambda_{r+1,s_{r+1}})\Psi_{\boldsymbol{\lambda}}}{(\lambda_{r+1,s_{r+1}}-\lambda_{ri})\,p_{r+1,s_{r+1}}(\lambda)},
	\end{equation}
	where
	\begin{equation*}
		p_{\beta\alpha}(\lambda)\equiv\prod\limits_{\begin{smallmatrix}
				\gamma=1 \\ \gamma\neq\alpha
		\end{smallmatrix}}^\beta (\lambda_{\beta\alpha}-\lambda_{\beta\gamma}).
	\end{equation*}
	Using \eqref{B_mab(lambda_im)Psi_lambda_i.t.o._lower_graduation_minors} repeatedly one obtains
	\begin{multline} \label{B_{r,a,r+1}(lambda_{ir})Psi_lambda}
		B_{r,a,r+1}(\lambda_{ri})\Psi_{\boldsymbol{\lambda}} \\
		= (-1)^{r-a}\sum\limits_{s_{r-1}=1}^{r-1}\ldots\sum\limits_{s_a=1}^a \frac{B_{r,r,r+1}(\lambda_{ri})\prod\limits_{t=a}^{\overset{r-1}{\longleftarrow}}B_{t,t,t+1}(\lambda_{ts_t})\;\Psi_{\boldsymbol{\lambda}}}{(\lambda_{r-1,s_{r-1}}-\lambda_{ri}+1)\prod\limits_{q=a}^{r-2}(\lambda_{qs_q}-\lambda_{q+1,s_{q+1}}+1) \prod\limits_{q=a}^{r-1}p_{qs_q}(\lambda)},
	\end{multline}
	where the notation for the ordered product of operators from the numerator was introduced in \eqref{ordered_operator_products_and_1_k<r_definitions}. Since $\{A_j(u)\}_{j=1}^r$ commute with $B_{r+1,r+1,b}(v)$ by \eqref{B_rab_A_m_commutation_m<a_or_m_geq_b}, the action of $B_{r+1,r+1,b}$ on $\Psi_{\boldsymbol{\lambda}}$ does not change the eigenvalues of $A_j(u)$, $j\leq r$. Therefore, if one replaces $\Psi_{\boldsymbol{\lambda}}$ in \eqref{B_{r,a,r+1}(lambda_{ir})Psi_lambda} by $B_{r+1,r+1,b}(\lambda_{r+1,s_{r+1}})\Psi_{\boldsymbol{\lambda}}$, the obtained expression
	\begin{multline} \label{B_r,a,r+1(lambda_ir)B_r+1,r+1,b(lambda_r+1,s_r+1)Psi}
		B_{r,a,r+1}(\lambda_{ri})B_{r+1,r+1,b}(\lambda_{r+1,s_{r+1}})\Psi_{\boldsymbol{\lambda}} \\
		= (-1)^{r-a}\sum\limits_{s_{r-1}=1}^{r-1}\ldots\sum\limits_{s_a=1}^a \frac{B_{r,r,r+1}(\lambda_{ri})\prod\limits_{t=a}^{\overset{r-1}{\longleftarrow}}B_{t,t,t+1}(\lambda_{ts_t})\;B_{r+1,r+1,b}(\lambda_{r+1,s_{r+1}})\Psi_{\boldsymbol{\lambda}}}{(\lambda_{r-1,s_{r-1}}-\lambda_{ri}+1)\prod\limits_{q=a}^{r-2}(\lambda_{qs_q}-\lambda_{q+1,s_{q+1}}+1) \prod\limits_{q=a}^{r-1}p_{qs_q}(\lambda)}.
	\end{multline}
	is still correct. The expression $B_{r+1,r+1,b}(\lambda_{r+1,s_{r+1}})\Psi_{\boldsymbol{\lambda}}$ from \eqref{B_r,a,r+1(lambda_ir)B_r+1,r+1,b(lambda_r+1,s_r+1)Psi} can be rewritten with the help of repeated usage of \eqref{B_rab(lambda_ir)Psi_lambda_i.t.o._lower_graduation_minors}: first with $m=r+2$, then with $m=r+3$, and so on till $m=b-1$. One finds
	\begin{multline} \label{B_{r+1,r+1,b}(lambda_{k_1,r+1})Psi_lambda}
		B_{r+1,r+1,b}(\lambda_{r+1,s_{r+1}})\Psi_{\boldsymbol{\lambda}} \\
		= (-1)^{b-r}\sum\limits_{s_{r+2}=1}^{r+2}\ldots\sum\limits_{s_{b-1}=1}^{b-1}\frac{\prod\limits_{c=r+1}^{\overset{b-1}{\longrightarrow}}B_{c,c,c+1}(\lambda_{cs_c})\;\Psi_{\boldsymbol{\lambda}}}{(\lambda_{r+1,s_{r+1}}-\lambda_{ri})\prod\limits_{d=r+2}^{b-1}(\lambda_{ds_d}-\lambda_{d-1,s_{d-1}}) \prod\limits_{d=r+1}^{b-1}p_{ds_d}(\lambda)}.
	\end{multline}
	
	Substituting \eqref{B_{r+1,r+1,b}(lambda_{k_1,r+1})Psi_lambda} into \eqref{B_r,a,r+1(lambda_ir)B_r+1,r+1,b(lambda_r+1,s_r+1)Psi}, then \eqref{B_r,a,r+1(lambda_ir)B_r+1,r+1,b(lambda_r+1,s_r+1)Psi} into \eqref{B_{rab}(lambda_{ir})Psi_lambda} and remembering the notation \eqref{B_r_definition} one obtains
	\begin{multline} \label{B_{rab}(lambda_{ir})Psi_lambda_explicit_formula}
		B_{rab}(\lambda_{ri})\Psi_{\boldsymbol{\lambda}} = (-1)^{b-a-1}\sum\limits_{s_{r-1}=1}^{r-1}\ldots\sum\limits_{s_a=1}^a\sum\limits_{s_{r+1}=1}^{r+1}\ldots\sum\limits_{s_{b-1}=1}^{b-1}\frac{1}{(\lambda_{r-1,s_{r-1}}-\lambda_{ri}+1)(\lambda_{r+1,s_{r+1}}-\lambda_{ri})} \\
		\times \frac{B_r(\lambda_{ri})\prod\limits_{t=a}^{\overset{r-1}{\longleftarrow}}B_t(\lambda_{ts_t})\;\prod\limits_{c=r+1}^{\overset{b-1}{\longrightarrow}}B_c(\lambda_{cs_c})\;\Psi_{\boldsymbol{\lambda}}}{\prod\limits_{q=a}^{r-2}(\lambda_{qs_q}-\lambda_{q+1,s_{q+1}}+1) \; \prod\limits_{d=r+2}^{b-1}(\lambda_{ds_d}-\lambda_{d-1,s_{d-1}}) \; \prod\limits_{q=a}^{r-1}p_{qs_q}(\lambda) \; \prod\limits_{d=r+1}^{b-1}p_{ds_d}(\lambda)}.
	\end{multline}
	Since $B_{rab}(u)$ is a polynomial in $u$ of degree $r-1$, one obtains the desired formula \eqref{B_rab_acts_on_Psi_lambda_bar_lambda} using \eqref{B_{rab}(lambda_{ir})Psi_lambda_explicit_formula} with $i=1,\ldots,r$ and Lagrange interpolation.
	
	Now derive the formula \eqref{B_r,a1,a2,b1,b2(lambda_ir)_action_on_Psi_lambda} for the action of $L(u)^{1\ldots r}_{1,\ldots,\widehat{a_1},\ldots,\widehat{a_2},\ldots,r,b_1,b_2}$ on $\Psi_{\boldsymbol{\lambda}}$. Consider $a=(a_1,a_2)$, where $1\leq a_1<a_2\leq r$, and $b=(b_1,b_2)$, where $r<b_1<b_2\leq n'$. Denote $B_{rab}(u)\equiv L(u)^{1\ldots r}_{1,\ldots,\widehat{a_1},\ldots,\widehat{a_2},\ldots,r,b_1,b_2}$. Representing $L(u)^{1\ldots r}_{1\ldots r} = (-1)^{a_1+a_2+1}L(u)^{1\ldots r}_{1,\ldots,\widehat{a_1},\ldots,\widehat{a_2},\ldots,r,a_1,a_2}$,
	setting in \eqref{type_B'_minors_commutation_relations}
	\begin{equation*}
		\begin{split}
			& r=m, \\
			& (b_1,\ldots,b_r)=(1,\ldots,\widehat{a_1},\ldots,\widehat{a_2},\ldots,r,b_1,b_2), \\
			& (b_1',\ldots,b_r')=(1,\ldots,\widehat{a_1},\ldots,\widehat{a_2},\ldots,r,a_1,a_2)
		\end{split}
	\end{equation*}
	and using the antisymmetry of quantum minors \eqref{antisymmetry_of_quantum_minors} one can derive the following:
	\begin{multline} \label{commutation_rel_B_rab_A_r_length_a_and_b_=2}
		(u-v+1)(u-v+2)B_{rab}(u)A_r(v) = (u-v)(u-v+1)A_r(v)B_{rab}(u) \\
		+ 2B_{rab}(v)A_r(u) + (u-v)\sum\limits_{\alpha=1}^2[B_{ra_\alpha b_{\widetilde{\alpha}}}(v)B_{ra_{\widetilde{\alpha}}b_\alpha}(u)-B_{ra_\alpha b_\alpha}(v)B_{ra_{\widetilde{\alpha}}b_{\widetilde{\alpha}}}(u)],
	\end{multline}
	where $\widetilde{1}=2$, $\widetilde{2}=1$. Acting on $\Psi_{\boldsymbol{\lambda}}$ by \eqref{commutation_rel_B_rab_A_r_length_a_and_b_=2} with $v=\lambda_{ri}$, $u=\lambda_{rj}$, $j\neq i$ one obtains
	\begin{equation} \label{A_r(lambda_ri)B_{rab}(lambda_rj)Psi_lambda}
		A_r(\lambda_{ri})B_{rab}(\lambda_{rj})\Psi_{\boldsymbol{\lambda}} = -\frac{\sum\limits_{\alpha=1}^2[B_{ra_\alpha b_{\widetilde{\alpha}}}(\lambda_{ri})B_{ra_{\widetilde{\alpha}}b_\alpha}(\lambda_{rj})-B_{ra_\alpha b_\alpha}(\lambda_{ri})B_{ra_{\widetilde{\alpha}}b_{\widetilde{\alpha}}}(\lambda_{rj})]\Psi_{\boldsymbol{\lambda}}}{\lambda_{rj}-\lambda_{ri}+1}.
	\end{equation}
	From the definition it follows that $B_{rab}(u)$ is a polynomial in $u$ of degree $r-2$. Since there are $r-1$ parameters $\lambda_{rj}$ with $j\neq i$, the quantity $A_r(\lambda_{ri})B_{rab}(u)\Psi_{\boldsymbol{\lambda}}$ can be found by the usage of \eqref{A_r(lambda_ri)B_{rab}(lambda_rj)Psi_lambda} and Lagrange interpolation between the points $\lambda_{rj}, \, j\neq i$. One obtains
	\begin{multline} \label{A_r(lambda_{ir})B_{rab}(u)Psi_lambda}
		A_r(\lambda_{ri})B_{rab}(u)\Psi_{\boldsymbol{\lambda}} \\
		= -\sum\limits_{\begin{smallmatrix}
				j=1 \\ j\neq i
		\end{smallmatrix}}^r \left(\prod\limits_{\begin{smallmatrix}
				k=1 \\ k\neq i,j
		\end{smallmatrix}}^r \frac{u-\lambda_{rk}}{\lambda_{rj}-\lambda_{rk}}\right) \frac{\sum\limits_{\alpha=1}^2[B_{ra_\alpha b_{\widetilde{\alpha}}}(\lambda_{ri})B_{ra_{\widetilde{\alpha}}b_\alpha}(\lambda_{rj})-B_{ra_\alpha b_\alpha}(\lambda_{ri})B_{ra_{\widetilde{\alpha}}b_{\widetilde{\alpha}}}(\lambda_{rj})]\Psi_{\boldsymbol{\lambda}}}{\lambda_{rj}-\lambda_{ri}+1}.
	\end{multline}
	Acting on $\Psi_{\boldsymbol{\lambda}}$ by \eqref{commutation_rel_B_rab_A_r_length_a_and_b_=2} with $v=\lambda_{ir}$ and substituting the expression \eqref{A_r(lambda_{ir})B_{rab}(u)Psi_lambda} for $A_r(\lambda_{ir})B_{rab}(u)\Psi_{\boldsymbol{\lambda}}$, one derives after some algebra the desired formula \eqref{B_r,a1,a2,b1,b2(lambda_ir)_action_on_Psi_lambda}.

	\section{Solution of the system (\ref{gl_4_system_K_gamma_hol_1}--\ref{gl_4_system_K_gamma_hol_3}) and of its antiholomorphic counterpart}   \label{appendix_gl4_system_solution}

	Consider
	\begin{multline} \label{K_gamma_prefactor_extraction}
		K_{\boldsymbol{\gamma}}(\boldsymbol{\lambda}) = (-1)^{\sum(\gamma_{2j}-\bar{\gamma}_{2j})}(\gamma_{21}-\gamma_{22})(\bar{\gamma}_{21}-\bar{\gamma}_{22}) \prod\limits_{k,j} \Gamma^\mathbb{C}(\boldsymbol{\lambda}_{3k}-\boldsymbol{\gamma}_{2j}) \\
		\times \Gamma^\mathbb{C}\left(\boldsymbol{\gamma}_{11}+\boldsymbol{\lambda}_{11}-\sum\boldsymbol{\lambda}_{2j}+\boldsymbol{1}\right) \Gamma^\mathbb{C}\left(-\boldsymbol{\gamma}_{11}+\sum\boldsymbol{\gamma}_{2j}+\sum\boldsymbol{\lambda}_{2j}-\sum\boldsymbol{\lambda}_{3j}+\boldsymbol{1}\right) P_{\boldsymbol{\gamma}}(\boldsymbol{\lambda}),
	\end{multline}
	where $P_{\boldsymbol{\gamma}}(\boldsymbol{\lambda})$ depends on $\gamma$, $\bar{\gamma}$, $\lambda$, $\bar{\lambda}$, the function $\Gamma^\mathbb{C}$ is defined in \eqref{Gamma^C}. Then the system (\ref{gl_4_system_K_gamma_hol_1}--\ref{gl_4_system_K_gamma_hol_3}) takes the form
	\begin{gather}
		\begin{split}
			\left(-\gamma_{11}+\sum\gamma_{2j}+\sum\lambda_{2j}-\sum\lambda_{3j}\right)P_{\boldsymbol{\gamma}} = \prod\limits_{k=1}^2 (\gamma_{11}-\lambda_{2k}+1) \; P_{\boldsymbol{\gamma}+e_{11}} \\
			+ \frac{\prod\limits_{k=1}^3(\gamma_{21}-\lambda_{3k})\: P_{\boldsymbol{\gamma}-e_{21}} - \prod\limits_{k=1}^3(\gamma_{22}-\lambda_{3k})\:P_{\boldsymbol{\gamma}-e_{22}}}{\gamma_{21}-\gamma_{22}},  \label{gl_4_P_system_1}
		\end{split} \\
		(\gamma_{21}-\gamma_{11})P_{\boldsymbol{\gamma}} = P_{\boldsymbol{\gamma}-e_{11}} + P_{\boldsymbol{\gamma}+e_{21}},  \label{gl_4_P_system_2} \\
		(\gamma_{22}-\gamma_{11})P_{\boldsymbol{\gamma}} = P_{\boldsymbol{\gamma}-e_{11}} + P_{\boldsymbol{\gamma}+e_{22}},  \label{gl_4_P_system_3}
	\end{gather}
	its analogue with the shifts of antiholomorphic parameters takes the similar form up to the signs of some terms in equations.
	
	Express $P_{\boldsymbol{\gamma}}$ as the multivariable complex Mellin transform \cite[Section~1.2.4]{Neretin_func_complex_field} of some function $P(x)\equiv P(x_{11},x_{21},x_{22})$:
	\begin{equation} \label{ansatz_for_P_gamma}
		P_{\boldsymbol{\gamma}} = \int\limits_{\mathbb{C}} d^2x_{11}\,d^2x_{21}\,d^2x_{22}\,x_{11}^{\boldsymbol{\gamma}_{11}-\boldsymbol{1}}x_{21}^{\boldsymbol{\gamma}_{21}-\boldsymbol{1}}x_{22}^{\boldsymbol{\gamma}_{22}-\boldsymbol{1}}\, P(x_{11},x_{21},x_{22}).
	\end{equation}
	Substituting \eqref{ansatz_for_P_gamma} into (\ref{gl_4_P_system_1}--\ref{gl_4_P_system_3}) and into its antiholomorphic analogue, using the property
	\begin{equation*}
		\mu\int\limits_{\mathbb{C}} d^2x \, x^{\boldsymbol{\mu}-\boldsymbol{1}}\,\mathscr{F}(x) = -\int\limits_{\mathbb{C}} d^2x \, x^{\boldsymbol{\mu}-\boldsymbol{1}}\,x\partial_{x}\mathscr{F}(x), \qquad \boldsymbol{\mu}=(\mu,\bar{\mu})\in\Lambda_\mathbb{C}
	\end{equation*}
	(see \eqref{Lambda_C} for the definition of $\Lambda_\mathbb{C}$) and the similar property for the antiholomorphic sector one obtains the following systems of equations for $P(x)$:
	\begin{gather}
		\begin{split}
			\left(x_{11}\partial_{11}-\sum\limits_{k=1}^2 x_{2k}\partial_{2k}+\sum\limits_{k=1}^2\lambda_{2k}-\sum\limits_{k=1}^3\lambda_{3k}\right)(x_{21}\partial_{21}-x_{22}\partial_{22})P = \prod\limits_{k=1}^3(x_{21}\partial_{21}+\lambda_{3k})\:\frac{P}{x_{21}} \\
			- \prod\limits_{k=1}^3(x_{22}\partial_{22}+\lambda_{3k})\:\frac{P}{x_{22}} + \prod\limits_{k=1}^2 (x_{11}\partial_{11}+\lambda_{2k}-1) \: x_{11}(x_{21}\partial_{21}-x_{22}\partial_{22})P,
		\end{split} \label{gl_4_P_system_1_Mellin} \\
		(x_{21}\partial_{21}-x_{11}\partial_{11})P = -x_{11}^{-1}P - x_{21}P,  \label{gl_4_P_system_2_Mellin} \\
		(x_{22}\partial_{22}-x_{11}\partial_{11})P = -x_{11}^{-1}P - x_{22}P. \label{gl_4_P_system_3_Mellin}
	\end{gather}
	and
	\begin{gather}
		\begin{split}
			\left(\bar{x}_{11}\bar{\partial}_{11}-\sum\limits_{k=1}^2\bar{x}_{2k}\bar{\partial}_{2k}+\sum\limits_{k=1}^2\bar{\lambda}_{2k}-\sum\limits_{k=1}^3\bar{\lambda}_{3k}\right)(\bar{x}_{21}\bar{\partial}_{21}-\bar{x}_{22}\bar{\partial}_{22})P = \prod\limits_{k=1}^3(\bar{x}_{22}\bar{\partial}_{22}+\bar{\lambda}_{3k})\:\frac{P}{\bar{x}_{22}} \\
			-\prod\limits_{k=1}^3(\bar{x}_{21}\bar{\partial}_{21}+\bar{\lambda}_{3k})\:\frac{P}{\bar{x}_{21}} - \prod\limits_{k=1}^2 (\bar{x}_{11}\bar{\partial}_{11}+\bar{\lambda}_{2k}-1) \: \bar{x}_{11}(\bar{x}_{21}\bar{\partial}_{21}-\bar{x}_{22}\bar{\partial}_{22})P,
		\end{split} \label{gl_4_P_system_antiholomorphic_1_Mellin} \\
		(\bar{x}_{21}\bar{\partial}_{21}-\bar{x}_{11}\bar{\partial}_{11})P = \bar{x}_{11}^{-1}P + \bar{x}_{21}P,  \label{gl_4_P_system_antiholomorphic_2_Mellin} \\
		(\bar{x}_{22}\bar{\partial}_{22}-\bar{x}_{11}\bar{\partial}_{11})P = \bar{x}_{11}^{-1}P + \bar{x}_{22}P, \label{gl_4_P_system_antiholomorphic_3_Mellin}
	\end{gather}
	where $\bar{\partial}_{ij}\equiv\frac{\partial}{\partial \bar{x}_{ij}}$. In order to solve \eqref{gl_4_P_system_2_Mellin}, \eqref{gl_4_P_system_3_Mellin}, \eqref{gl_4_P_system_antiholomorphic_2_Mellin} and \eqref{gl_4_P_system_antiholomorphic_3_Mellin} consider the following change of variables:
	\begin{equation*}
		\xi\equiv x_{21}, \quad \eta\equiv x_{22}, \quad \zeta\equiv x_{11}x_{21}x_{22}.
	\end{equation*}
	The homogenity operators appearing in (\ref{gl_4_P_system_1_Mellin}--\ref{gl_4_P_system_antiholomorphic_3_Mellin}) take the form
	\begin{equation*}
		x_{21}\partial_{21} = \xi\partial_\xi + \zeta\partial_\zeta, \qquad x_{22}\partial_{22} = \eta\partial_\eta + \zeta\partial_\zeta, \qquad x_{11}\partial_{11} = \zeta\partial_\zeta
	\end{equation*}
	(similar formulas for the antiholomorphic ones). The pairs of equations (\ref{gl_4_P_system_2_Mellin}, \ref{gl_4_P_system_3_Mellin}) and (\ref{gl_4_P_system_antiholomorphic_2_Mellin}, \ref{gl_4_P_system_antiholomorphic_3_Mellin}) can be rewritten as
	\begin{equation*}
		\begin{cases}
			\partial_\xi P = -\eta\zeta^{-1}P - P \\
			\partial_\eta P = -\xi\zeta^{-1}P - P
		\end{cases}, \qquad 
		\begin{cases}
			\partial_{\bar{\xi}} P = \bar{\eta}\bar{\zeta}^{-1}P + P \\
			\partial_{\bar{\eta}} P = \bar{\xi}\bar{\zeta}^{-1}P + P
		\end{cases},
	\end{equation*}
	correspondingly. Therefore,
	\begin{multline} \label{P_in_terms_of_A(x11x12x22)}
		P(x) = A(\zeta)\exp\left(-\xi\eta\zeta^{-1}-\xi-\eta\right)\exp\left(\bar{\xi}\bar{\eta}\bar{\zeta}^{-1}+\bar{\xi}+\bar{\eta}\right) \\
		= A(x_{11}x_{21}x_{22})\exp\left(-x_{11}^{-1}-x_{21}-x_{22}\right)\exp\left(\bar{x}_{11}^{-1}+\bar{x}_{21}+\bar{x}_{22}\right),
	\end{multline}
	where $A(\zeta)$ is a function of $\zeta$ one needs to find. Substituting \eqref{P_in_terms_of_A(x11x12x22)} into \eqref{gl_4_P_system_1_Mellin} and \eqref{gl_4_P_system_antiholomorphic_1_Mellin} one obtains after some algebra the following system of equations for $A(\zeta)$:
	\begin{gather}
		\left[\zeta\prod\limits_{k=1}^2(\zeta\partial_\zeta+\lambda_{2k})+\prod\limits_{k=1}^3(\zeta\partial_\zeta-1+\lambda_{3k})\right] A(\zeta) = 0,  \label{Meijer_equation_for_A(zeta)} \\
		\left[\bar{\zeta}\prod\limits_{k=1}^2(\bar{\zeta}\partial_{\bar{\zeta}}+\bar{\lambda}_{2k})-\prod\limits_{k=1}^3(\bar{\zeta}\partial_{\bar{\zeta}}-1+\bar{\lambda}_{3k})\right] A(\zeta) = 0.  \label{Meijer_equation_for_A(zeta)_antiholomorphic}
	\end{gather}
	In accordance with formulae (3.10a--3.11b) from \cite{Neretin_func_complex_field}, the function
	\begin{equation} \label{A(zeta)}
		A(\zeta) = c(\boldsymbol{\lambda})\,{}_2G^{\mathbb{C}}_3\left[\begin{array}{c}
			\boldsymbol{1}-\boldsymbol{\lambda}_{31}, \boldsymbol{1}-\boldsymbol{\lambda}_{32}, \boldsymbol{1}-\boldsymbol{\lambda}_{33} \\
			\boldsymbol{\lambda}_{21}, \boldsymbol{\lambda}_{22}
		\end{array}; \zeta\right]
	\end{equation}
	(see (1.15) in \cite{Neretin_func_complex_field}) is a solution of the system (\ref{Meijer_equation_for_A(zeta)}--\ref{Meijer_equation_for_A(zeta)_antiholomorphic}). The quantity $c(\boldsymbol{\lambda})$ is some constant depending on $\lambda, \bar{\lambda}$. Substituting \eqref{A(zeta)} into \eqref{P_in_terms_of_A(x11x12x22)}, \eqref{P_in_terms_of_A(x11x12x22)} into \eqref{ansatz_for_P_gamma}, using the relation (3.6) and (at least, formally) Theorem 1.7 from \cite{Neretin_func_complex_field} one obtains
	\begin{equation} \label{P_gamma}
		P_{\boldsymbol{\gamma}}(\boldsymbol{\lambda}) = c(\boldsymbol{\lambda})\,{}_4G^{\mathbb{C}}_4\left[\begin{array}{c}
			\boldsymbol{\lambda}_{21}, \boldsymbol{\lambda}_{22}, \boldsymbol{\gamma}_{21}, \boldsymbol{\gamma}_{22} \\
			\boldsymbol{1}-\boldsymbol{\lambda}_{31}, \boldsymbol{1}-\boldsymbol{\lambda}_{32}, \boldsymbol{1}-\boldsymbol{\lambda}_{33}, -\boldsymbol{\gamma}_{11}
		\end{array}; 1\right].
	\end{equation}
	Substitution of $P_{\boldsymbol{\gamma}}(\boldsymbol{\lambda})$ from \eqref{P_gamma} into \eqref{K_gamma_prefactor_extraction} gives the formula \eqref{K_gamma_} for the kernel $K_{\boldsymbol{\gamma}}(\boldsymbol{\lambda})$.
	
	Now it is time to verify that ${}_4G^{\mathbb{C}}_4$ from \eqref{P_gamma} indeed solves the system (\ref{gl_4_P_system_1}--\ref{gl_4_P_system_3}) and its antiholomorphic analogue. Equations (\ref{gl_4_P_system_2}--\ref{gl_4_P_system_3}) and their antiholomorphic counterparts follow from the second relation after (3.11b) in \cite{Neretin_func_complex_field} and from its antiholomorphic analogue.
	
	Consider ${}_pG^{\mathbb{C}}_p$ \cite[eq.~(1.15)]{Neretin_func_complex_field}:
	\begin{equation} \label{p_G_p}
		{}_pG^{\mathbb{C}}_p\left[\begin{array}{c}
			\mathbf{a} \\ \mathbf{b}
		\end{array}; z\right] = \frac{1}{2\pi i}\int D\boldsymbol{s} \,
		\mathcal{K}\left[\begin{array}{c}
			\mathbf{a} \\ \mathbf{b}
		\end{array}; \boldsymbol{s}\right] \, z^{-\boldsymbol{s}},
	\end{equation}
	where $\mathbf{a} \equiv (\boldsymbol{a}_1,\ldots,\boldsymbol{a}_p), \; \boldsymbol{a}_j=(a_j,\bar{a}_j)\in\Lambda_\mathbb{C}$ (see \eqref{Lambda_C} for the definition of $\Lambda_\mathbb{C}$), and $\mathbf{b} \equiv (\boldsymbol{b}_1,\ldots,\boldsymbol{b}_p), \; \boldsymbol{b}_j\equiv(b_j,\bar{b}_j)\in\Lambda_\mathbb{C}$. The integrand reads
	\begin{equation} \label{mathcal_K}
		\mathcal{K}\left[\begin{array}{c}
			\mathbf{a} \\ \mathbf{b}
		\end{array}; \boldsymbol{s}\right] \equiv \prod\limits_{\alpha=1}^p\Gamma^\mathbb{C}(\boldsymbol{a}_\alpha+\boldsymbol{s})\prod\limits_{\beta=1}^p\Gamma^\mathbb{C}(\boldsymbol{b}_\beta-\boldsymbol{s}).
	\end{equation}
	The variable $\boldsymbol{s}$ and the integration measure have the form 
	\begin{equation*}
		 s \equiv \frac{k+v}{2}, \; \bar{s} \equiv \frac{-k+v}{2}, \; \boldsymbol{s}\equiv(s,\bar{s}), \qquad \int D\boldsymbol{s} \equiv \sum\limits_{k\in\mathbb{Z}}\int\limits_{\delta+i\mathbb{R}} dv, \; \delta\in\mathbb{R},
	\end{equation*}
	the number $\delta\in\mathbb{R}$ must be such that the corresponding contour separates two groups of poles appearing from two products in the definition of $\mathcal{K}$ in \eqref{mathcal_K} (see \cite[Section~1.4]{Neretin_func_complex_field}). Denote by $G$ the function \eqref{p_G_p}, by $\mathcal{K}$ -- the function \eqref{mathcal_K}, by $G(a_j\pm1)$ -- the function \eqref{p_G_p} with the parameter $a_j$ shifted by $\pm1$, similarly for $\mathcal{K}(a_j\pm1)$ and for other parameters.

	Derive for ${}_pG^{\mathbb{C}}_p$ the analogues of generalized contiguous relations for hypergeometric functions \cite{contiguous_rel_hypergeometric_function}. To do this use the technique similar to that from Section 3 of \cite{contiguous_rel_Meijer_G}. Find the coefficients $E,F,D$, $C_j, \; j=1,\ldots,p$, such that the following equality holds:
	\begin{equation} \label{p_G_p_equality_1}
		(E+Fz)G + D\,G(b_1-1) + z\sum\limits_{j=1}^p C_j\,G(a_j-1) = 0.
	\end{equation}
	Consider $zG$. If one makes in \eqref{p_G_p} the change of variables $k$, $v$ equivalent to the substitution
	\begin{equation} \label{p_G_p_integration_variable_positive_shift}
		k\to k+1, \quad v\to v+1,
	\end{equation}
	the integrand is transformed in the following way:
	\begin{equation*}
		\mathcal{K}\,z^{-\boldsymbol{s}+e} \to \frac{\prod\limits_{\alpha=1}^p(a_\alpha+s)}{\prod\limits_{\beta=1}^p(b_\beta-s-1)} \, \mathcal{K}\,z^{-\boldsymbol{s}},
	\end{equation*}
	where $e\in\Lambda_\mathbb{C}$ is introduced in \eqref{e_and_bar_e_definition}. Consider the term $D\,G(b_1-1)$ in \eqref{p_G_p_equality_1}. The corresponding integrand obeys
	\begin{equation*}
		\mathcal{K}(b_1-1)\, = \frac{\mathcal{K}}{b_1-s-1}.
	\end{equation*}
	Consider the term $zG(a_k-1)$. After the change of variable \eqref{p_G_p_integration_variable_positive_shift} in the corresponding integral expression the integrand is transformed in the following way:
	\begin{equation*}
		\mathcal{K}(a_k-1)\,z^{-\boldsymbol{s}+e} \to \frac{\prod\limits_{\alpha\neq j}(a_\alpha+s)}{\prod\limits_{\beta=1}^p(b_\beta-s-1)}\,\mathcal{K}\,z^{-\boldsymbol{s}}.
	\end{equation*}
	Thus, the sufficient condition for \eqref{p_G_p_equality_1} is
	\begin{equation*}
		E + \frac{\prod\limits_{\alpha=1}^p(a_\alpha+s)}{\prod\limits_{\beta=1}^p(b_\beta-s-1)}\,F + \frac{D}{b_1-s-1} + \sum\limits_{j=1}^p\frac{\prod\limits_{\alpha\neq j}(a_\alpha+s)}{\prod\limits_{\beta=1}^p(b_\beta-s-1)}\,C_k = 0,
	\end{equation*}
	or, equivalently,
	\begin{equation} \label{p_G_p_equality_1_rewritten}
		\prod\limits_{\beta=1}^p(b_\beta-s-1)E + \prod\limits_{\alpha=1}^p(a_\alpha+s)\,F + \prod\limits_{\beta=2}^p(b_\beta-s-1)\,D + \sum\limits_{j=1}^p\prod\limits_{\alpha\neq j}(a_\alpha+s)\,C_k = 0.
	\end{equation}
	The last equality is the condition of zeroing of the polynomial of degree $p$ in $s$. The number of unknowns $E,F,D$, $C_j, \; j=1,\ldots,p$ is $p+3$. Fix $E=1$. Considering the highest degree term in \eqref{p_G_p_equality_1_rewritten} one obtains $F = (-1)^{p+1}$. Substituting the particular values $s=-a_l, \; l=1,\ldots,p$ in \eqref{p_G_p_equality_1_rewritten} it is possible to find that
	\begin{equation*}
		C_j = -\frac{(b_1+a_j-1+D)\prod\limits_{\beta=2}^p(b_\beta+a_j-1)}{\prod\limits_{\alpha\neq j}(a_\alpha-a_j)}.
	\end{equation*}
	Then, provided that $D=0$, from \eqref{p_G_p_equality_1} follows the relation
	\begin{equation} \label{p_G_p_rel_1}
		(1+(-1)^{p+1}z)G = z\sum\limits_{j=1}^p\frac{\prod\limits_{\beta=1}^p(b_\beta+a_j-1)}{\prod\limits_{\alpha\neq j}(a_\alpha-a_j)}\,G(a_j-1).
	\end{equation}
	Substituting $D=1$ and using \eqref{p_G_p_rel_1} one obtains from \eqref{p_G_p_equality_1} the relation
	\begin{equation} \label{p_G_p_rel_2}
		G(b_1-1) = z\sum\limits_{j=1}^p\frac{\prod\limits_{\beta=2}^p(b_\beta+a_j-1)}{\prod\limits_{\alpha\neq j}(a_\alpha-a_j)}\,G(a_j-1).
	\end{equation}
	
	By analogy with \eqref{p_G_p_equality_1}, considering the equality
	\begin{equation*}
		(E+Fz)G + (D+D'z)G(b_1+1) + z\sum\limits_{j=1}^p C_jG(a_j-1) = 0
	\end{equation*}
	with unknown coefficients $E, F, D, D', C_j$ and performing the calculations similar to (\ref{p_G_p_integration_variable_positive_shift}--\ref{p_G_p_rel_2}) it is possible to obtain the identity
	\begin{multline} \label{p_G_p_rel_3}
		\left[1+(-1)^{p+1}(\sum\limits_{\beta=1}^p b_\beta+\sum\limits_{\alpha=1}^p a_\alpha-p+2)z\right]G = - (1+(-1)^{p+1}z)G(b_1+1) \\
		+ z\sum\limits_{j=1}^p\frac{(b_1+a_j+1)\prod\limits_{\beta=1}^p(b_\beta+a_j-1)}{\prod\limits_{\alpha\neq j}(a_\alpha-a_j)}\,G(a_j-1).
	\end{multline}
	Considering an even $p$ and $z=1$, adding \eqref{p_G_p_rel_3} with \eqref{p_G_p_rel_1} multiplied by $-(b_1+1)$ one obtains
	\begin{equation} \label{p_G_p_rel_4}
		\left[\sum\limits_{\beta=1}^p b_\beta+\sum\limits_{\alpha=1}^p a_\alpha-p+1\right]G = -\sum\limits_{j=1}^p\frac{a_j\prod\limits_{\beta=1}^p(b_\beta+a_j-1)}{\prod\limits_{\alpha\neq j}(a_\alpha-a_j)}\,G(a_j-1).
	\end{equation}

	Consider $G={}_4G^{\mathbb{C}}_4$ from \eqref{P_gamma} ($p=4, z=1$). Then \eqref{p_G_p_rel_1}, \eqref{p_G_p_rel_2} and \eqref{p_G_p_rel_4} take the form
	\begin{multline} \label{4G4_rel_1}
		\sum\limits_{l=1}^2\frac{(-\gamma_{11}+\lambda_{2l}-1)\prod\limits_{j=1}^3(\lambda_{2l}-\lambda_{3j})}{(\lambda_{2\widetilde{l}}-\lambda_{2l})\prod\limits_{j=1}^2(\gamma_{2j}-\lambda_{2l})}G(\lambda_{2l}-1) \\
		+ \sum\limits_{r=1}^2\frac{(-\gamma_{11}+\gamma_{2r}-1)\prod\limits_{j=1}^3(\gamma_{2r}-\lambda_{3j})}{(\gamma_{2\widetilde{r}}-\gamma_{2r})\prod\limits_{j=1}^2(\lambda_{2j}-\gamma_{2r})}G(\gamma_{2r}-1) = 0
	\end{multline}
	(where $\widetilde{1}=2, \widetilde{2}=1$),
	\begin{multline} \label{4G4_rel_2}
		G(\gamma_{11}+1) \\
		= \sum\limits_{l=1}^2\frac{\prod\limits_{j=1}^3(\lambda_{2l}-\lambda_{3j})}{(\lambda_{2\widetilde{l}}-\lambda_{2l})\prod\limits_{j=1}^2(\gamma_{2j}-\lambda_{2l})}G(\lambda_{2l}-1) + \sum\limits_{r=1}^2\frac{\prod\limits_{j=1}^3(\gamma_{2r}-\lambda_{3j})}{(\gamma_{2\widetilde{r}}-\gamma_{2r})\prod\limits_{j=1}^2(\lambda_{2j}-\gamma_{2r})}G(\gamma_{2r}-1)
	\end{multline}
	and
	\begin{multline} \label{4G4_rel_3}
		(-\gamma_{11}-\sum\lambda_{3j}+\sum\gamma_{2j}+\sum\lambda_{2j})G = -\sum\limits_{l=1}^2\frac{\lambda_{2l}(-\gamma_{11}+\lambda_{2l}-1)\prod\limits_{j=1}^3(\lambda_{2l}-\lambda_{3j})}{(\lambda_{2\widetilde{l}}-\lambda_{2l})\prod\limits_{j=1}^2(\gamma_{2j}-\lambda_{2l})}G(\lambda_{2l}-1) \\
		- \sum\limits_{r=1}^2\frac{\gamma_{2r}(-\gamma_{11}+\gamma_{2r}-1)\prod\limits_{j=1}^3(\gamma_{2r}-\lambda_{3j})}{(\gamma_{2\widetilde{r}}-\gamma_{2r})\prod\limits_{j=1}^2(\lambda_{2j}-\gamma_{2r})}G(\gamma_{2r}-1),
	\end{multline}
	correspondingly. Subtracting from \eqref{4G4_rel_3} the equality \eqref{4G4_rel_2} multiplied by $\prod_{j=1}^2(\gamma_{11}-\lambda_{2j}+1)$ one obtains
	\begin{equation} \label{4G4_lin_comb_rel_2_rel_3}
		\begin{split}
			& (-\gamma_{11}-\sum\lambda_{3j}+\sum\gamma_{2j}+\sum\lambda_{2j})G - \prod\limits_{j=1}^2(\gamma_{11}-\lambda_{2j}+1)\,G(\gamma_{11}+1) \\
			& = (\gamma_{11}-\sum\lambda_{2j}+1)\sum\limits_{l=1}^2\frac{(\gamma_{11}-\lambda_{2l}+1)\prod\limits_{j=1}^3(\lambda_{3j}-\lambda_{2l})}{(\lambda_{2\widetilde{l}}-\lambda_{2l})\prod\limits_{j=1}^2(\gamma_{2j}-\lambda_{2l})}G(\lambda_{2l}-1) \\
			& + \sum\limits_{r=1}^2\frac{\left[\prod\limits_{j=1}^2(\gamma_{11}-\lambda_{2j}+1)-\gamma_{2r}(\gamma_{11}-\gamma_{2r}+1)\right]\prod\limits_{j=1}^3(\lambda_{3j}-\gamma_{2r})}{(\gamma_{2\widetilde{r}}-\gamma_{2r})\prod\limits_{j=1}^2(\lambda_{2j}-\gamma_{2r})}G(\gamma_{2r}-1).
		\end{split}
	\end{equation}
	With the help of \eqref{4G4_rel_1} one can express the quantity
	\begin{equation*}
		\sum\limits_{l=1}^2\frac{(\gamma_{11}-\lambda_{2l}+1)\prod\limits_{j=1}^3(\lambda_{3j}-\lambda_{2l})}{(\lambda_{2\widetilde{l}}-\lambda_{2l})\prod\limits_{j=1}^2(\gamma_{2j}-\lambda_{2l})}G(\lambda_{2l}-1)
	\end{equation*}
	in terms of a linear combination of $G(\gamma_{21}-1)$ and $G(\gamma_{22}-1)$. Substituting this into \eqref{4G4_lin_comb_rel_2_rel_3} and calculating the coefficients of $G(\gamma_{21}-1)$ and $G(\gamma_{22}-1)$ in the resulting expression one obtains the relation \eqref{gl_4_P_system_1}. The antiholomorphic counterpart of \eqref{gl_4_P_system_1} can be verified in the similar way.

	\section{Derivation of conditions \eqref{gl_4_condition_on_c_shift_2i}, \eqref{gl_4_condition_on_c_shift_3i} on $c(\boldsymbol{\lambda})$ from \eqref{K_gamma_}}   \label{appendix_gl4_coefficient_conditions}

	Similarly to the formulas (\ref{gl_3_A_2_action_on_Psi_lambda}--\ref{gl_3_A_2_action_on_Psi_lambda_1}) in $gl_3(\mathbb{C})$, act on $\Psi_{\boldsymbol{\lambda}}$ from \eqref{gl_4_eigenfunction} by the minors $L(\lambda_{2i})^{12}_{13}$ and $L(\lambda_{3i})^{123}_{124}$ (and by the antiholomorphic counterparts) using the recurrence expressions \eqref{gl_4_minors_in_terms_of_gl_3_minors} for $gl_4(\mathbb{C})$ minors in terms of $gl_3(\mathbb{C})$ minors, the relations (\ref{gl_4_system_K_gamma_hol_1}--\ref{gl_4_system_K_gamma_hol_3}) and their antiholomorphic analogues. One obtains the following system of difference equations for $P_{\boldsymbol{\gamma}}(\boldsymbol{\lambda})$ from \eqref{K_gamma_prefactor_extraction}:
	\begin{gather} 
		(\lambda_{2i}-\gamma_{11})P_{\boldsymbol{\gamma}}(\boldsymbol{\lambda}) - P_{\boldsymbol{\gamma}-e_{11}}(\boldsymbol{\lambda}) = - \frac{\mathcal{M}_i(\lambda)}{\lambda_{2i}-\lambda_{11}}P_{\boldsymbol{\gamma}}(\boldsymbol{\lambda}+e_{2i}), \qquad i=1,2,  \label{gl_4_conditions_for_P_gamma_with_lambda_i2}  \\
		\begin{split}
			\left(-\gamma_{11}+\sum\gamma_{2j}+\sum\lambda_{2j}-\sum\lambda_{3j}\right)P_{\boldsymbol{\gamma}}(\boldsymbol{\lambda}) - (\gamma_{21}-\gamma_{11}-1)\prod\limits_{\begin{smallmatrix}
					k=1 \\ k\neq i
			\end{smallmatrix}}^3(\gamma_{21}-\lambda_{3k})\;\frac{P_{\boldsymbol{\gamma}-e_{21}}(\boldsymbol{\lambda})}{\gamma_{21}-\gamma_{22}} \\
			- (\gamma_{22}-\gamma_{11}-1)\prod\limits_{\begin{smallmatrix}
					k=1 \\ k\neq i
			\end{smallmatrix}}^3(\gamma_{22}-\lambda_{3k})\;\frac{P_{\boldsymbol{\gamma}-e_{22}}(\boldsymbol{\lambda})}{\gamma_{22}-\gamma_{21}} = \frac{\mathcal{M}'_i(\lambda)}{\lambda_{3i}-\sigma_1+1}P_{\boldsymbol{\gamma}}(\boldsymbol{\lambda}+e_{3i}), \qquad i=1,2,3, \label{gl_4_conditions_for_P_gamma_with_lambda_i3}
		\end{split}
	\end{gather}
	and its counterpart with shifts of antiholomorphic parameters, which is similar to (\ref{gl_4_conditions_for_P_gamma_with_lambda_i2}--\ref{gl_4_conditions_for_P_gamma_with_lambda_i3}) up to the signs of some terms in equations. The notations with shifts like $P_{\boldsymbol{\gamma}-e_{11}}(\boldsymbol{\lambda})$ are similar to \eqref{gl_2_short_notations}. The quantities $\mathcal{M}_i(\lambda)$, $\mathcal{M}'_i(\lambda)$ are defined in \eqref{mathcal_M_definition}.
	
	Taking into account the explicit form \eqref{P_gamma} of $P_{\boldsymbol{\gamma}}(\boldsymbol{\lambda})$, the second formula after (3.11b) in \cite{Neretin_func_complex_field} and the similar formula for the antiholomorphic parameters following from (3.11a), (3.11b) in \cite{Neretin_func_complex_field}, one finds that the relations \eqref{gl_4_conditions_for_P_gamma_with_lambda_i2} and their antiholomorphic analogues give the conditions \eqref{gl_4_condition_on_c_shift_2i} on $c(\boldsymbol{\lambda})$.
	
	Now consider the relations \eqref{gl_4_conditions_for_P_gamma_with_lambda_i3}. Using \eqref{p_G_p_rel_2} with $G={}_4G_4^\mathbb{C}$ (i.e. $p=4, z=1, \boldsymbol{b}_1=\boldsymbol{1}-\boldsymbol{\lambda}_{3i}$) one finds (using the notations from Appendix C)
	\begin{multline} \label{4G4_rel_4}
		G(\lambda_{3i}+1) = \sum\limits_{l=1}^2\frac{(-\gamma_{11}+\lambda_{2l}-1)\prod\limits_{j\neq i}(\lambda_{2l}-\lambda_{3j})}{(\lambda_{2\widetilde{l}}-\lambda_{2l})\prod\limits_{j=1}^2(\gamma_{2j}-\lambda_{2l})}\,G(\lambda_{2l}-1) \\
		+ \sum\limits_{r=1}^2\frac{(-\gamma_{11}+\gamma_{2r}-1)\prod\limits_{j\neq i}(\gamma_{2r}-\lambda_{3j})}{(\gamma_{2\widetilde{r}}-\gamma_{2r})\prod\limits_{j=1}^2(\lambda_{2j}-\gamma_{2r})}\,G(\gamma_{2r}-1),
	\end{multline}
	where $\tilde{1}=2, \tilde{2}=1$. Subtracting from \eqref{4G4_rel_3} the equality \eqref{4G4_rel_4} multiplied by $\prod_{j=1}^2(\lambda_{3i}-\lambda_{2j})$ one obtains
	\begin{equation} \label{appendix_D_derivation}
		\begin{split}
			& (-\gamma_{11}-\sum\lambda_{3j}+\sum\gamma_{2j}+\sum\lambda_{2j})G - \prod\limits_{j=1}^2(\lambda_{3i}-\lambda_{2j})\,G(\lambda_{3i}+1) \\
			& = (-\sum\lambda_{2j}+\lambda_{3i})\sum\limits_{l=1}^2\frac{(-\gamma_{11}+\lambda_{2l}-1)\prod\limits_{j=1}^3(\lambda_{2l}-\lambda_{3j})}{(\lambda_{2\widetilde{l}}-\lambda_{2l})\prod\limits_{j=1}^2(\gamma_{2j}-\lambda_{2l})}G(\lambda_{2l}-1) \\
			& + \sum\limits_{r=1}^2\frac{(-\gamma_{11}+\gamma_{2r}-1)\prod\limits_{j\neq i}(\gamma_{2r}-\lambda_{3j})\left[-\gamma_{2r}(\gamma_{2r}-\lambda_{3i})-\prod\limits_{j=1}^2(\lambda_{2j}-\lambda_{3i})\right]}{(\gamma_{2\widetilde{r}}-\gamma_{2r})\prod\limits_{j=1}^2(\lambda_{2j}-\gamma_{2r})}G(\gamma_{2r}-1).
		\end{split}
	\end{equation}
	With the help of \eqref{4G4_rel_1} one can express the quantity
	\begin{equation*}
		\sum\limits_{l=1}^2\frac{(-\gamma_{11}+\lambda_{2l}-1)\prod\limits_{j=1}^3(\lambda_{2l}-\lambda_{3j})}{(\lambda_{2\tilde{l}}-\lambda_{2l})\prod\limits_{j=1}^2(\gamma_{2j}-\lambda_{2l})}G(\lambda_{2l}-1)
	\end{equation*}
	in terms of a linear combination of $G(\gamma_{21}-1)$ and $G(\gamma_{22}-1)$. Substituting this into \eqref{appendix_D_derivation} and calculating the coefficients of $G(\gamma_{21}-1)$ and $G(\gamma_{22}-1)$ in the resulting expression one obtains
	\begin{multline} \label{4G4_rel_5}
		(-\gamma_{11}-\sum\lambda_{3j}+\sum\gamma_{2j}+\sum\lambda_{2j})G - \prod\limits_{j=1}^2(\lambda_{3i}-\lambda_{2j})\,G(\lambda_{3i}+1) \\
		= \sum\limits_{r=1}^2\frac{(\gamma_{2r}-\gamma_{11}-1)\prod\limits_{j\neq i}(\gamma_{2r}-\lambda_{3j})}{\gamma_{2r}-\gamma_{2\tilde{r}}}\,G(\gamma_{2r}-1).
	\end{multline}
	Remembering the formula \eqref{P_gamma} for $P_\gamma(\lambda)$ and comparing \eqref{4G4_rel_5} to \eqref{gl_4_conditions_for_P_gamma_with_lambda_i3} one finds that \eqref{gl_4_conditions_for_P_gamma_with_lambda_i3} holds if $c(\boldsymbol{\lambda})$ obeys the first relation from \eqref{gl_4_condition_on_c_shift_3i}.
	
	In a similar way it is possible to obtain from the antiholomorphic counterpart of \eqref{gl_4_conditions_for_P_gamma_with_lambda_i3} the finite-difference equations for $c(\boldsymbol{\lambda})$ in \eqref{gl_4_condition_on_c_shift_3i} with the shifts of antiholomorphic parameters.

	\section{Orthogonality of the obtained sets of basis elements}   \label{appendix_orthogonality}

	Consider the case of $gl_3(\mathbb{C})$. The expression for the eigenfunctions is \eqref{gl_3_eigenfunction_orthogonal_set}, for the corresponding parameters -- \eqref{gl_3_parameters_orthogonal_set}. Consider
	\begin{equation} \label{gl_3_scalar_product}
		\langle\Psi_{\boldsymbol{\lambda}}|\Psi_{\boldsymbol{\lambda}'}\rangle = \int\limits_{\mathbb{C}} d^2x\,d^2y\,d^2z \, \Psi_{\boldsymbol{\lambda}}^\ast(x,y,z)\,\Psi_{\boldsymbol{\lambda}'}(x,y,z).
	\end{equation}
	In $\lambda',\bar{\lambda}'$ the notations $n_{lj}$ and $\mu_{lj}$ should be replaced by $n_{lj}'$ and $\mu_{lj}'$. Let the integration variable in the formula \eqref{gl_3_eigenfunction_orthogonal_set} for $\Psi_{\boldsymbol{\lambda}'}$ be equal to $\gamma_{11}'$, i.e. $k$ and $v$ from \eqref{gl_3_eigenfunction_orthogonal_set} and \eqref{gamma_11_step_from_gl_2_to_gl_3} are replaced by $k'$ and $v'$. Permute the signs of limit and sum in the expressions for $\Psi_{\boldsymbol{\lambda}}$ and $\Psi_{\boldsymbol{\lambda}'}$. Change the order of integration with respect to $x, y, z$ and with respect to $\boldsymbol{\gamma}_{11}, \boldsymbol{\gamma}_{11}'$ in \eqref{gl_3_scalar_product}. Integrate with respect to $x, y, z$ using the orthogonality of $gl_2(\mathbb{C})$ functions \eqref{gl_2_orthogonality}, the relation \eqref{delta_as_integral_of_power_functions} and the properties of $\Gamma^\mathbb{C}$ (see \eqref{Gamma^C} for the definition of this function): for $\mu=\frac{1}{2}(a+ib), \; \bar{\mu}=\frac{1}{2}(-a+ib), \; a\in\mathbb{Z}, b\in\mathbb{R}$ holds
	\begin{equation} \label{unitarity_of_Gamma^C(mu+0.5)}
		\Gamma^\mathbb{C}(\boldsymbol{\mu}+\frac{\boldsymbol{1}}{\boldsymbol{2}})^\ast\,\Gamma^\mathbb{C}(\boldsymbol{\mu}+\frac{\boldsymbol{1}}{\boldsymbol{2}}) = 1, \quad \Gamma^\mathbb{C}(\boldsymbol{\mu}+\frac{\boldsymbol{1}}{\boldsymbol{2}})\Gamma^\mathbb{C}(\frac{\boldsymbol{1}}{\boldsymbol{2}}-\boldsymbol{\mu}) = (-1)^a, \qquad \boldsymbol{\mu}\equiv(\mu,\bar{\mu}).
	\end{equation}
	Then integrate one of appeared delta-functions by $\boldsymbol{\gamma}_{11}'$. As the result, one obtains
	\begin{multline} \label{gl_3_scalar_product_1}
		\langle\Psi_{\boldsymbol{\lambda}}|\Psi_{\boldsymbol{\lambda}'}\rangle = 16\pi^6(-1)^{\sum n_{2j}}\delta_{n_{11},n_{11}'}\delta(\mu_{11}-\mu_{11}')\delta_{\sum n_{2j},\sum n_{2j}'}\delta\left(\sum\mu_{2j}-\sum\mu_{2j}'\right) \\
		\times \lim\limits_{\varepsilon\to0}\sum\limits_{k\in\mathbb{Z}}\int\limits_{-\infty}^\infty dv \, \prod\limits_{l=1}^2\Gamma^\mathbb{C}(\boldsymbol{\gamma}_{11}-\boldsymbol{\lambda}_{2l}+\frac{\boldsymbol{\varepsilon}}{\boldsymbol{2}}) \prod\limits_{l=1}^2\Gamma^\mathbb{C}(\boldsymbol{\lambda}_{2l}'-\boldsymbol{\gamma}_{11}+\frac{\boldsymbol{\varepsilon}}{\boldsymbol{2}}).
	\end{multline}
	
	The sum and integral in \eqref{gl_3_scalar_product_1} is the case $N=2$ in the formula (4.4) from \cite{Der_Man_Val_SL_2_C_Gustafson_integrals}. Using it and the fact
	\begin{equation} \label{gl_3_relation_on_sum_lambda_2l_from_delta_finction}
		\sum\limits_{l=1}^2\lambda_{2l}'-\sum\limits_{l=1}^2\lambda_{2l}=0, \qquad \sum\limits_{l=1}^2\bar{\lambda}_{2l}'-\sum\limits_{l=1}^2\bar{\lambda}_{2l}=0,
	\end{equation}
	(following from the delta-functions in \eqref{gl_3_scalar_product_1}) one obtains that the LHS of the orthogonality relation \eqref{gl_3_orthogonality} is equal to
	\begin{multline} \label{gl_3_scalar_product_2}
		\|\boldsymbol{\lambda}_{21}-\boldsymbol{\lambda}_{22}\|^2\|\boldsymbol{\lambda}_{21}'-\boldsymbol{\lambda}_{22}'\|^2\langle\Psi_{\boldsymbol{\lambda}}|\Psi_{\boldsymbol{\lambda}'}\rangle = 64\pi^7(-1)^{\sum n_{2j}}\delta_{n_{11},n_{11}'}\delta(\mu_{11}-\mu_{11}') \\
		\times \delta_{\sum n_{2j},\sum n_{2j}'}\delta\left(\sum\mu_{2j}-\sum\mu_{2j}'\right)\|\boldsymbol{\lambda}_{21}-\boldsymbol{\lambda}_{22}\|^2\|\boldsymbol{\lambda}_{21}'-\boldsymbol{\lambda}_{22}'\|^2\lim\limits_{\varepsilon\to0} \frac{\prod\limits_{l,j=1}^2\Gamma^{\mathbb{C}}(\boldsymbol{\lambda}_{2l}'-\boldsymbol{\lambda}_{2j}+\boldsymbol{\varepsilon})}{\Gamma^{\mathbb{C}}\left(2\boldsymbol{\varepsilon}\right)}.
	\end{multline}
	Recall that by definition $\|\boldsymbol{\lambda}_{21}-\boldsymbol{\lambda}_{22}\|^2 = -(\lambda_{21}-\lambda_{22})(\bar{\lambda}_{21}-\bar{\lambda}_{22})$. Transform the expression under the sign of limit in \eqref{gl_3_scalar_product_2} using \eqref{gl_3_relation_on_sum_lambda_2l_from_delta_finction} and the identities
	\begin{multline*}
		\Gamma^\mathbb{C}(\boldsymbol{\mu}) = \frac{\Gamma^\mathbb{C}(\boldsymbol{\mu}+e)}{\mu}, \quad (\lambda_{21}-\lambda_{22})(\lambda_{21}'-\lambda_{22}') = \sum\limits_{\tau\in S_2}\mathrm{sgn}(\tau)\,(\lambda_{21}'-\lambda_{2\tau(1)})(\lambda_{22}'-\lambda_{2\tau(2)}),
	\end{multline*}
	where $\boldsymbol{\mu}=(\mu,\bar{\mu})\in\Lambda_\mathbb{C}$ (see \eqref{Lambda_C}), $e\in\Lambda_\mathbb{C}$ is introduced in \eqref{e_and_bar_e_definition}, $S_2$ is the symmetric group. One obtains
	\begin{multline} \label{gl_3_part_with_lim}
		(\lambda_{21}-\lambda_{22})(\lambda_{21}'-\lambda_{22}')\frac{\prod\limits_{l,j=1}^2\Gamma^{\mathbb{C}}(\boldsymbol{\lambda}_{2l}'-\boldsymbol{\lambda}_{2j}+\boldsymbol{\varepsilon})}{\Gamma^{\mathbb{C}}\left(2\boldsymbol{\varepsilon}\right)} = \frac{\prod\limits_{l,j=1}^2\Gamma^{\mathbb{C}}(\boldsymbol{\lambda}_{2l}'-\boldsymbol{\lambda}_{2j}+\boldsymbol{\varepsilon}+e)}{\Gamma^{\mathbb{C}}\left(2\boldsymbol{\varepsilon}+e\right)} \\
		\times \left(\frac{2\varepsilon}{(\lambda_{21}-\lambda_{22}'+\varepsilon)(\lambda_{22}'-\lambda_{21}+\varepsilon)}\frac{(\lambda_{21}'-\lambda_{21})(\lambda_{22}'-\lambda_{22})}{(\lambda_{21}'-\lambda_{21}+\varepsilon)(\lambda_{22}'-\lambda_{22}+\varepsilon)} \right. \\
		\left. - \frac{2\varepsilon}{(\lambda_{21}'-\lambda_{21}+\varepsilon)(\lambda_{21}-\lambda_{21}'+\varepsilon)}\frac{(\lambda_{21}'-\lambda_{22})(\lambda_{22}'-\lambda_{21})}{(\lambda_{21}'-\lambda_{22}+\varepsilon)(\lambda_{22}'-\lambda_{21}+\varepsilon)}\right).
	\end{multline}
	The expressions
	\begin{multline} \label{gl_3_part_with_lim_expressions_with_finite_limit}
		\frac{(\lambda_{21}'-\lambda_{21})(\lambda_{22}'-\lambda_{22})}{(\lambda_{21}'-\lambda_{21}+\varepsilon)(\lambda_{22}'-\lambda_{22}+\varepsilon)}, \qquad \frac{(\lambda_{21}'-\lambda_{22})(\lambda_{22}'-\lambda_{21})}{(\lambda_{21}'-\lambda_{22}+\varepsilon)(\lambda_{22}'-\lambda_{21}+\varepsilon)}, \\
		\frac{\prod\limits_{l,j=1}^2\Gamma^{\mathbb{C}}(\boldsymbol{\lambda}_{2l}'-\boldsymbol{\lambda}_{2j}+\boldsymbol{\varepsilon}+e)}{\Gamma^{\mathbb{C}}\left(2\boldsymbol{\varepsilon}+e\right)}
	\end{multline}
	from \eqref{gl_3_part_with_lim} have finite limits as $\varepsilon\to 0_+$ for all values of $n_{2j}$, $\mu_{2j}$, $n_{2j}'$, $\mu_{2j}'$. The limits of expressions giving the delta-functions:
	\begin{equation} \label{gl_3_part_with_delta}
		\frac{2\varepsilon}{(\lambda_{21}-\lambda_{22}'+\varepsilon)(\lambda_{22}'-\lambda_{21}+\varepsilon)} \overset{\varepsilon\to 0_+}{\longrightarrow} 4\pi\delta_{n_{21},n_{22}'}\delta(\mu_{21}-\mu_{22}'),
	\end{equation}
	and the same for the similar fraction in the second summand in the RHS of \eqref{gl_3_part_with_lim}. In \eqref{gl_3_part_with_delta} the following formula for delta-function is used:
	\begin{equation} \label{formula_for_delta_function}
		\frac{\varepsilon}{x^2+\varepsilon^2} \overset{\varepsilon\to 0_+}{\longrightarrow} \pi\,\delta(x).
	\end{equation}
	Calculating the limits of expressions from \eqref{gl_3_part_with_lim_expressions_with_finite_limit}, substituting \eqref{gl_3_part_with_delta} into \eqref{gl_3_part_with_lim} and \eqref{gl_3_part_with_lim} into \eqref{gl_3_scalar_product_2} one obtains the desired orthogonality relation \eqref{gl_3_orthogonality} for $gl_3(\mathbb{C})$ eigenfunctions.
	
	Now consider the case of $gl_4(\mathbb{C})$. The expression for the Gelfand-Tsetlin basis is \eqref{gl_4_eigenfunction_orthogonal_set}, for the corresponding parameters -- \eqref{gl_4_parameters_orthogonal_set}. The LHS of the orthogonality relation \eqref{gl_4_orthogonality} reads
	\begin{equation*}
		\langle\rho(\boldsymbol{\lambda})\Psi_{\boldsymbol{\lambda}}|\rho(\boldsymbol{\lambda}')\Psi_{\boldsymbol{\lambda}'}\rangle = \int\limits_{\mathbb{C}}\prod\limits_{1\leq j<l\leq 4}d^2z_{lj} \, \rho(\boldsymbol{\lambda})\Psi_{\boldsymbol{\lambda}}^\ast(z_{21},z_{31},z_{41},\mathbf{z}) \, \rho(\boldsymbol{\lambda}')\Psi_{\boldsymbol{\lambda}'}(z_{21},z_{31},z_{41},\mathbf{z}),
	\end{equation*}
	where the notation $\rho(\boldsymbol{\lambda})$ is introduced in \eqref{rho(lambda)}, and $\mathbf{z}=\{z_{32}, z_{42}, z_{43}\}$. In $\lambda',\bar{\lambda}'$ the notations $m_{lj}$ and $\beta_{lj}$ should be replaced with $m_{lj}'$ and $\beta_{lj}'$. Let the integration variables in the formula \eqref{gl_4_eigenfunction_orthogonal_set} for $\Psi_{\boldsymbol{\lambda}'}$ be $\boldsymbol{\gamma}_{11}'$, $\boldsymbol{\gamma}_{21}'$, $\boldsymbol{\gamma}_{22}'$, i.e. $n_{lj}$ and $\mu_{lj}$ from \eqref{gl_4_eigenfunction_orthogonal_set_notations} are replaced with $n_{lj}'$ and $\mu_{lj}'$. Similarly to the case of $gl_3(\mathbb{C})$, change the order of integration with respect to $\{(z_{21},z_{31},z_{41},\mathbf{z})\}$ and with respect to $\boldsymbol{\gamma}, \boldsymbol{\gamma}'$, integrate with respect to $\{(z_{21},z_{31},z_{41},\mathbf{z})\}$ using \eqref{delta_as_integral_of_power_functions}, \eqref{unitarity_of_Gamma^C(mu+0.5)} and the orthogonality relation \eqref{gl_3_orthogonality} for $gl_3(\mathbb{C})$ functions. Then integrate some of appeared delta-functions by $\boldsymbol{\gamma}'$. As the result one obtains
	\begin{multline} \label{gl_4_scalar_product_1}
		\langle\rho(\boldsymbol{\lambda})\Psi_{\boldsymbol{\lambda}}|\rho(\boldsymbol{\lambda}')\Psi_{\boldsymbol{\lambda}'}\rangle \\
		= (2\pi)^{12} \delta_{m_{11},m_{11}'}\delta\left(\beta_{11}-\beta_{11}'\right) \delta_{\sum m_{2j},\sum m_{2j}'}\delta\left(\sum\beta_{2j}-\sum\beta_{2j}'\right) \delta_{\sum m_{3j},\sum m_{3j}'}\delta\left(\sum\beta_{3j}-\sum\beta_{3j}'\right) \\
		\times (-1)^{\sum m_{2j}+\sum m_{3j}} \rho(\boldsymbol{\lambda})\rho(\boldsymbol{\lambda}')\lim\limits_{\varepsilon_1\to 0_+} \int D\tilde{\boldsymbol{s}} \, D\tilde{\boldsymbol{s}}' \, F_{\varepsilon_1}(\tilde{\boldsymbol{\lambda}},\tilde{\boldsymbol{\lambda}}',\tilde{\boldsymbol{s}},\tilde{\boldsymbol{s}}') \lim\limits_{\varepsilon\to0_+} \int D\tilde{\boldsymbol{\gamma}}_{21} \, D\tilde{\boldsymbol{\gamma}}_{22} \, J_\varepsilon(\tilde{\boldsymbol{\lambda}},\tilde{\boldsymbol{\lambda}}',\tilde{\boldsymbol{\gamma}},\tilde{\boldsymbol{s}},\tilde{\boldsymbol{s}}') \\
		\times \lim\limits_{\varepsilon_2\to 0_+} \int D\tilde{\boldsymbol{\gamma}}_{11} \, \Gamma^\mathbb{C}(\tilde{\boldsymbol{\gamma}}_{11}+\tilde{\boldsymbol{s}}+\frac{\boldsymbol{\varepsilon}_2}{\boldsymbol{2}}) \, \Gamma^\mathbb{C}(-\tilde{\boldsymbol{s}}'+\frac{\boldsymbol{\varepsilon}_2}{\boldsymbol{2}}-\tilde{\boldsymbol{\gamma}}_{11}) \, (-1)^{n_{11}},
	\end{multline}
	where
	\begin{align*}
		& \tilde{s} = \frac{k+iv}{2}, \; \bar{\tilde{s}} = \frac{-k+iv}{2}, \quad \tilde{s}' = \frac{k'+iv'}{2}, \; \bar{\tilde{s}}' = \frac{-k'+iv'}{2}, \quad \tilde{\gamma}_{lj} = \frac{n_{lj}+i\mu_{lj}}{2}, \; \bar{\tilde{\gamma}}_{lj} = \frac{-n_{lj}+i\mu_{lj}}{2}, \\
		& \tilde{\lambda}_{lj} = \frac{m_{lj}+i\beta_{lj}}{2}, \; \bar{\tilde{\lambda}}_{lj} = \frac{-m_{lj}+i\beta_{lj}}{2}, \quad \tilde{\lambda}_{lj}' = \frac{m_{lj}'+i\beta_{lj}'}{2}, \; \bar{\tilde{\lambda}}_{lj}' = \frac{-m_{lj}'+i\beta_{lj}'}{2}
	\end{align*}
	with $k, k' \in \mathbb{Z}$, $v, v' \in \mathbb{R}$. The notations like $\int D\tilde{\boldsymbol{s}}$ are similar to \eqref{gl_4_eigenfunction_orthogonal_set_notations}. The infinitesimal parameters $\varepsilon$, $\varepsilon_1$, $\varepsilon_2$ appeared from the corresponding $i\varepsilon$-prescriptions in \eqref{gl_4_eigenfunction_orthogonal_set} and \eqref{4G4}. The expressions for $F_{\varepsilon_1}(\tilde{\boldsymbol{\lambda}},\tilde{\boldsymbol{\lambda}}',\tilde{\boldsymbol{s}},\tilde{\boldsymbol{s}}')$ and $J_\varepsilon(\tilde{\boldsymbol{\lambda}},\tilde{\boldsymbol{\lambda}}',\tilde{\boldsymbol{\gamma}},\tilde{\boldsymbol{s}},\tilde{\boldsymbol{s}}')$ read
	\begin{multline*}
		F_{\varepsilon_1}(\tilde{\boldsymbol{\lambda}},\tilde{\boldsymbol{\lambda}}',\tilde{\boldsymbol{s}},\tilde{\boldsymbol{s}}') \equiv \prod\limits_{l=1}^2\Gamma^\mathbb{C}(\tilde{\boldsymbol{\lambda}}_{2l}'+\frac{\boldsymbol{\varepsilon}_1}{\boldsymbol{2}}+\tilde{\boldsymbol{s}}') \, \prod\limits_{l=1}^2\Gamma^\mathbb{C}(-\tilde{\boldsymbol{\lambda}}_{2l}+\frac{\boldsymbol{\varepsilon}_1}{\boldsymbol{2}}-\tilde{\boldsymbol{s}}) \\
		\times \prod\limits_{l=1}^3\Gamma^\mathbb{C}(\frac{\boldsymbol{1}}{\boldsymbol{2}}+\tilde{\boldsymbol{\lambda}}_{3l}+\tilde{\boldsymbol{s}}) \prod\limits_{l=1}^3\Gamma^\mathbb{C}(\frac{\boldsymbol{1}}{\boldsymbol{2}}-\tilde{\boldsymbol{\lambda}}_{3l}'-\tilde{\boldsymbol{s}}')
	\end{multline*}
	and
	\begin{multline*}
		J_\varepsilon(\tilde{\boldsymbol{\lambda}},\tilde{\boldsymbol{\lambda}}',\tilde{\boldsymbol{\gamma}},\tilde{\boldsymbol{s}},\tilde{\boldsymbol{s}}') \equiv \|\tilde{\boldsymbol{\gamma}}_{21}-\tilde{\boldsymbol{\gamma}}_{22}\|^2  \, \prod\limits_{l=1}^2\Gamma^\mathbb{C}(\tilde{\boldsymbol{\gamma}}_{2l}+\tilde{\boldsymbol{s}}'+\frac{\boldsymbol{1}}{\boldsymbol{2}}) \prod_{l,j}\Gamma^\mathbb{C}(\tilde{\boldsymbol{\gamma}}_{2j}-\tilde{\boldsymbol{\lambda}}_{3l}+\frac{\boldsymbol{\varepsilon}}{\boldsymbol{2}}) \\
		\times \prod_{l,j}\Gamma^\mathbb{C}(\tilde{\boldsymbol{\lambda}}_{3l}'+\frac{\boldsymbol{\varepsilon}}{\boldsymbol{2}}-\tilde{\boldsymbol{\gamma}}_{2j}) \, \prod\limits_{l=1}^2\Gamma^\mathbb{C}(-\tilde{\boldsymbol{s}}+\frac{\boldsymbol{1}}{\boldsymbol{2}}-\tilde{\boldsymbol{\gamma}}_{2l}).
	\end{multline*}
	Using the explicit expression for the gamma-function of the complex field \cite[Chapter~1,~\S~1,~Section~1.4]{Gelfand_Graev_hypergeom_func_arbitrary_field} it is possible to take the integral with respect to $\tilde{\boldsymbol{\gamma}}_{11}$ in \eqref{gl_4_scalar_product_1}:
	\begin{equation} \label{gl_4_scalar_prod_int_gamma_11}
		\lim\limits_{\varepsilon_2\to 0_+} \int D\tilde{\boldsymbol{\gamma}}_{11} \, \Gamma^\mathbb{C}(\tilde{\boldsymbol{\gamma}}_{11}+\tilde{\boldsymbol{s}}+\frac{\boldsymbol{\varepsilon}_2}{\boldsymbol{2}}) \, \Gamma^\mathbb{C}(-\tilde{\boldsymbol{s}}'+\frac{\boldsymbol{\varepsilon}_2}{\boldsymbol{2}}-\tilde{\boldsymbol{\gamma}}_{11}) \, (-1)^{n_{11}} = (-1)^k \delta^{(2)}(\tilde{\boldsymbol{s}}-\tilde{\boldsymbol{s}}'),
	\end{equation}
	where the notation $\delta^{(2)}(\tilde{\boldsymbol{s}}-\tilde{\boldsymbol{s}}')$ is similar to \eqref{parameters'_delta_function_definition}. Integrate with respect to $\tilde{\boldsymbol{s}}'$ taking into account the delta-function from \eqref{gl_4_scalar_prod_int_gamma_11}, i.e. replacing $\tilde{\boldsymbol{s}}'$ by $\tilde{\boldsymbol{s}}$ everywhere in \eqref{gl_4_scalar_product_1}. Then integrate with respect to $\tilde{\boldsymbol{\gamma}}_{21}$ and $\tilde{\boldsymbol{\gamma}}_{22}$. Consider $J_\varepsilon(\tilde{\boldsymbol{\lambda}},\tilde{\boldsymbol{\lambda}}',\tilde{\boldsymbol{\gamma}},\tilde{\boldsymbol{s}},\tilde{\boldsymbol{s}})$, add the infinitesimal quantity $-\frac{\boldsymbol{3}\boldsymbol{\varepsilon}}{\boldsymbol{2}}$ into the arguments of the functions $\Gamma^\mathbb{C}(\tilde{\boldsymbol{\gamma}}_{2l}+\tilde{\boldsymbol{s}}+\frac{\boldsymbol{1}}{\boldsymbol{2}})$ and $\Gamma^\mathbb{C}(-\tilde{\boldsymbol{s}}+\frac{\boldsymbol{1}}{\boldsymbol{2}}-\tilde{\boldsymbol{\gamma}}_{2l})$. As the consequence, the poles originating from these factors will shift by the infinitely small quantity $\frac{3\varepsilon}{2}$ and will not move to the other side of the contours of integration by $\mu_{2l}$. Note that from the delta-functions in \eqref{gl_4_scalar_product_1} follows
	\begin{equation*}
		\left(\tilde{s}+\frac{1}{2}-\frac{3\varepsilon}{2}\right)+\sum\limits_{l=1}^3\left(-\tilde{\lambda}_{3l}+\frac{\varepsilon}{2}\right) + \sum\limits_{l=1}^3\left(\tilde{\lambda}_{3l}'+\frac{\varepsilon}{2}\right) + \left(-\tilde{s}+\frac{1}{2}-\frac{3\varepsilon}{2}\right) = 1,
	\end{equation*}
	and the similar relation with the antiholomorphic parameters also holds. Under these conditions it is possible to to use the formula (3.5) from \cite{Der_Man_on_complex_gamma_func_integrals} (the case $N=3$) to take the integral with respect to $\tilde{\boldsymbol{\gamma}}_{21}$ and $\tilde{\boldsymbol{\gamma}}_{22}$. One obtains
	\begin{multline} \label{gl_4_scal_prod_int_gamma_21_gamma_22}
		\prod\limits_{1\leq l< j\leq 3}\left[(\tilde{\lambda}_{3l}-\tilde{\lambda}_{3j})(\tilde{\lambda}_{3l}'-\tilde{\lambda}_{3j}')\right]\lim\limits_{\varepsilon\to0_+}\int D\tilde{\boldsymbol{\gamma}}_{21} \, D\tilde{\boldsymbol{\gamma}}_{22} \, J_\varepsilon(\tilde{\boldsymbol{\lambda}},\tilde{\boldsymbol{\lambda}}',\tilde{\boldsymbol{\gamma}},\tilde{\boldsymbol{s}},\tilde{\boldsymbol{s}}) \\
		= 8\pi^2 (-1)^{k+\sum m_{3l}'} \prod\limits_{1\leq l< j\leq 3}\left[(\tilde{\lambda}_{3l}-\tilde{\lambda}_{3j})(\tilde{\lambda}_{3l}'-\tilde{\lambda}_{3j}')\right] \lim\limits_{\varepsilon\to 0_+}\left[ \Gamma^\mathbb{C}(\boldsymbol{1}-3\boldsymbol{\varepsilon}) \prod\limits_{l,j=1}^3\Gamma^\mathbb{C}(\tilde{\boldsymbol{\lambda}}_{3l}'-\tilde{\boldsymbol{\lambda}}_{3j}+\boldsymbol{\varepsilon}) \right. \\
		\left. \times \prod\limits_{l=1}^3\Gamma^\mathbb{C}(-\tilde{\boldsymbol{\lambda}}_{3l}-\tilde{\boldsymbol{s}}+\frac{\boldsymbol{1}}{\boldsymbol{2}}-\boldsymbol{\varepsilon}) \, \prod\limits_{l=1}^3\Gamma^\mathbb{C}(\tilde{\boldsymbol{s}}+\tilde{\boldsymbol{\lambda}}_{3l}'+\frac{\boldsymbol{1}}{\boldsymbol{2}}-\boldsymbol{\varepsilon})\right],
	\end{multline}
	the expression with Vandermondes of $\tilde{\lambda}_{3i},\tilde{\lambda}_{3j}'$ appears from $\rho(\boldsymbol{\lambda})$, $\rho(\boldsymbol{\lambda}')$ (see \eqref{gl_4_scalar_product_1}). From the formula
	\begin{equation*}
		\prod\limits_{1\leq l< j\leq 3}\left[(\tilde{\lambda}_{3l}-\tilde{\lambda}_{3j})(\tilde{\lambda}_{3l}'-\tilde{\lambda}_{3j}')\right] = -\sum\limits_{\tau\in S_3}\mathrm{sgn}(\tau)\prod\limits_{l=1}^3\prod\limits_{\begin{smallmatrix}
				j=1 \\ j\neq \tau(l)
		\end{smallmatrix}}^3 (\tilde{\lambda}_{3l}'-\tilde{\lambda}_{3j})
	\end{equation*}
	(where $S_3$ is the permutation group) it follows that
	\begin{multline} \label{gl_4_scal_prod_int_gamma_21_gamma_22_part_with_delta}
		\prod\limits_{1\leq l< j\leq 3}\left[(\tilde{\lambda}_{3l}-\tilde{\lambda}_{3j})(\tilde{\lambda}_{3l}'-\tilde{\lambda}_{3j}')\right] \Gamma^\mathbb{C}(\boldsymbol{1}-3\boldsymbol{\varepsilon})\prod\limits_{l,j=1}^3\Gamma^\mathbb{C}(\tilde{\boldsymbol{\lambda}}_{3l}'-\tilde{\boldsymbol{\lambda}}_{3j}+\boldsymbol{\varepsilon}) = -\frac{\Gamma(1-3\varepsilon)}{\Gamma(1+3\varepsilon)} \\
		\times \prod\limits_{l,j=1}^3\Gamma^\mathbb{C}(\tilde{\boldsymbol{\lambda}}_{3l}'-\tilde{\boldsymbol{\lambda}}_{3j}+\boldsymbol{\varepsilon}+e)\sum\limits_{\tau\in S_3}\mathrm{sgn}(\tau)\,\frac{3\varepsilon}{\prod\limits_{r=1}^3(\tilde{\lambda}_{3r}'-\tilde{\lambda}_{3\tau(r)}+\varepsilon)}\,\prod\limits_{l=1}^3\prod\limits_{\begin{smallmatrix}
				j=1 \\ j\neq \tau(l)
		\end{smallmatrix}}^3 \frac{\tilde{\lambda}_{3l}'-\tilde{\lambda}_{3j}}{\tilde{\lambda}_{3l}'-\tilde{\lambda}_{3j}+\varepsilon},
	\end{multline}
	where $e\in\Lambda_\mathbb{C}$ is introduced in \eqref{e_and_bar_e_definition}. With the exception of
	\begin{equation} \label{gl_4_scal_prod_int_gamma_21_gamma_22_emerging_delta}
		\frac{3\varepsilon}{\prod\limits_{r=1}^3(\tilde{\lambda}_{3r}'-\tilde{\lambda}_{3\tau(r)}+\varepsilon)},
	\end{equation}
	all expressions in \eqref{gl_4_scal_prod_int_gamma_21_gamma_22_part_with_delta} and in the remaining part of \eqref{gl_4_scal_prod_int_gamma_21_gamma_22} have finite limits as $\varepsilon\to 0_+$ for all values of $m_{lj}$, $\beta_{lj}$, $m_{lj}'$, $\beta_{lj}'$. If for some $l$ and $j$ holds $n_{3l}'\neq n_{3\tau(l)}, \; n_{3j}'\neq n_{3\tau(j)}$, the limit of \eqref{gl_4_scal_prod_int_gamma_21_gamma_22_emerging_delta} as $\varepsilon\to0_+$ is equal to zero. So, the following holds in \eqref{gl_4_scal_prod_int_gamma_21_gamma_22_emerging_delta}: $n_{3j}'=n_{3\tau(j)}, \; j=1,2,3$ (it follows from $\sum n_{3j}'=\sum n_{3j}$ and from $n_{3l}'=n_{3\tau(l)}, \; n_{3j}'=n_{3\tau(j)}$ for any $l,j, \, l\neq j$). Therefore, taking into account $\sum\beta_{3j}=\sum\beta_{3j}'$ (which follows from the corresponding delta-function in \eqref{gl_4_scalar_product_1}) one obtains
	\begin{equation} \label{gl_4_scal_prod_int_gamma_21_gamma_22_emerging_delta_rewritten}
		\begin{split}
			& \frac{3\varepsilon}{\prod\limits_{r=1}^3(\tilde{\lambda}_{3r}'-\tilde{\lambda}_{3\tau(r)}+\varepsilon)} = \frac{48\varepsilon^2}{[(\beta_{3\tau(1)}-\beta_{31}')^2+4\varepsilon^2][(\beta_{3\tau(2)}-\beta_{32}')^2+4\varepsilon^2]} \\
			& + \frac{24i\varepsilon(\beta_{3\tau(1)}-\beta_{31}')^2(\beta_{3\tau(2)}-\beta_{32}')}{[(\beta_{3\tau(1)}-\beta_{31}')^2+4\varepsilon^2][(\beta_{3\tau(2)}-\beta_{32}')^2+4\varepsilon^2][(\beta_{3\tau(1)}+\beta_{3\tau(2)}-\beta_{31}'-\beta_{32}')^2+4\varepsilon^2]} \\
			& + \frac{24i\varepsilon(\beta_{3\tau(1)}-\beta_{31}')(\beta_{3\tau(2)}-\beta_{32}')^2}{[(\beta_{3\tau(1)}-\beta_{31}')^2+4\varepsilon^2][(\beta_{3\tau(2)}-\beta_{32}')^2+4\varepsilon^2][(\beta_{3\tau(1)}+\beta_{3\tau(2)}-\beta_{31}'-\beta_{32}')^2+4\varepsilon^2]} \\
			& - \frac{24\varepsilon^2[(\beta_{3\tau(1)}+\beta_{3\tau(2)}-\beta_{31}'-\beta_{32}')^2-(\beta_{3\tau(1)}-\beta_{31}')^2-(\beta_{3\tau(2)}-\beta_{32}')^2]}{[(\beta_{3\tau(1)}-\beta_{31}')^2+4\varepsilon^2][(\beta_{3\tau(2)}-\beta_{32}')^2+4\varepsilon^2][(\beta_{3\tau(1)}+\beta_{3\tau(2)}-\beta_{31}'-\beta_{32}')^2+4\varepsilon^2]}.
		\end{split}
	\end{equation}
	From the formula \eqref{formula_for_delta_function} it follows that the first term in \eqref{gl_4_scal_prod_int_gamma_21_gamma_22_emerging_delta_rewritten} tends to $12\pi^2\delta(\beta_{3\tau(1)}-\beta_{31}')\delta(\beta_{3\tau(2)}-\beta_{32}')$ as $\varepsilon\to0_+$ and the fourth term -- to $6\pi^2\delta(\beta_{3\tau(1)}-\beta_{31}')\delta(\beta_{3\tau(2)}-\beta_{32}')$. From \eqref{formula_for_delta_function} and
	\begin{equation*}
		\frac{x}{x^2+\varepsilon^2} \overset{\varepsilon\to0_+}{\longrightarrow} \mathcal{P}\frac{1}{x}, \qquad \left(\mathcal{P}\frac{1}{x},f\right) \equiv \mathrm{v.p.}\int\limits_{-\infty}^\infty dx \, \frac{f(x)}{x},
	\end{equation*}
	it follows that the limits of the second and third terms in \eqref{gl_4_scal_prod_int_gamma_21_gamma_22_emerging_delta_rewritten} as $\varepsilon\to0_+$ are equal to
	\begin{equation*}
		\begin{split}
			& 12\pi i\,\delta(\beta_{3\tau(1)}+\beta_{3\tau(2)}-\beta_{31}'-\beta_{32}')\,\mathcal{P}\frac{1}{\beta_{3\tau(1)}-\beta_{31}'}, \\
			& 12\pi i\,\delta(\beta_{3\tau(1)}+\beta_{3\tau(2)}-\beta_{31}'-\beta_{32}')\,\mathcal{P}\frac{1}{\beta_{31}'-\beta_{3\tau(1)}},
		\end{split}
	\end{equation*}
	correspondingly, and cancel out. Therefore,
	\begin{equation} \label{gl_4_scal_prod_int_gamma_21_gamma_22_emerged_delta}
		\frac{3\varepsilon}{\prod\limits_{r=1}^3(\tilde{\lambda}_{3r}'-\tilde{\lambda}_{3\tau(r)}+\varepsilon)} \overset{\varepsilon\to0_+}{\longrightarrow} 18\pi^2\left(\prod\limits_{j=1}^3\delta_{n_{3j}',n_{3\tau(j)}}\right) \delta(\beta_{3\tau(1)}-\beta_{31}')\delta(\beta_{3\tau(2)}-\beta_{32}').
	\end{equation}
	Substitute \eqref{gl_4_scal_prod_int_gamma_21_gamma_22_emerged_delta} into \eqref{gl_4_scal_prod_int_gamma_21_gamma_22_part_with_delta}, then \eqref{gl_4_scal_prod_int_gamma_21_gamma_22_part_with_delta} into \eqref{gl_4_scal_prod_int_gamma_21_gamma_22} and take the limit of the remaining part as $\varepsilon\to0_+$. As the result,
	\begin{multline} \label{gl_4_scal_prod_int_gamma_21_gamma_22_result}
		\delta_{\sum m_{3j},\sum m_{3j}'}\delta\left(\sum\beta_{3j}-\sum\beta_{3j}'\right) \\
		\times \prod\limits_{1\leq l< j\leq 3}\left(\|\tilde{\boldsymbol{\lambda}}_{3l}-\tilde{\boldsymbol{\lambda}}_{3j}\|^2\,\|\tilde{\boldsymbol{\lambda}}_{3l}'-\tilde{\boldsymbol{\lambda}}_{3j}'\|^2\right)\lim\limits_{\varepsilon\to0_+}\int D\tilde{\boldsymbol{\gamma}}_{21} \, D\tilde{\boldsymbol{\gamma}}_{22} \, J_\varepsilon(\tilde{\boldsymbol{\lambda}},\tilde{\boldsymbol{\lambda}}',\tilde{\boldsymbol{\gamma}},\tilde{\boldsymbol{s}},\tilde{\boldsymbol{s}}) \\
		= 18\pi^4 \prod\limits_{1\leq l< j\leq 3}\|\boldsymbol{\lambda}_{3l}-\boldsymbol{\lambda}_{3j}\|^2 \; \sum\limits_{\tau\in S_3}\prod\limits_{j=1}^3\delta^{(2)}(\boldsymbol{\lambda}_{3j}'-\boldsymbol{\lambda}_{3\tau(j)}).
	\end{multline}
	In \eqref{gl_4_scalar_product_1} it remains to take the integral of $F_{\varepsilon_1}(\tilde{\boldsymbol{\lambda}},\tilde{\boldsymbol{\lambda}}',\tilde{\boldsymbol{s}},\tilde{\boldsymbol{s}})$ with respect to $\tilde{s}$. Taking into account the delta-functions from \eqref{gl_4_scal_prod_int_gamma_21_gamma_22_result}, $(-1)^k$ from the RHS of \eqref{gl_4_scalar_prod_int_gamma_11} and the property of $\Gamma^\mathbb{C}$ from \eqref{unitarity_of_Gamma^C(mu+0.5)}, one can transform it into the form
	\begin{multline} \label{gl_4_scal_prod_int_s_result}
		\delta_{\sum m_{2j},\sum m_{2j}'}\delta\left(\sum\beta_{2j}-\sum\beta_{2j}'\right)\|\tilde{\boldsymbol{\lambda}}_{21}-\tilde{\boldsymbol{\lambda}}_{22}\|^2\|\tilde{\boldsymbol{\lambda}}_{21}'-\tilde{\boldsymbol{\lambda}}_{22}'\|^2 \\
		\times (-1)^{\sum m_{3j}}\lim\limits_{\varepsilon_1\to 0_+} \int D\tilde{\boldsymbol{s}} \, \prod\limits_{l=1}^2\Gamma^\mathbb{C}(\tilde{\boldsymbol{\lambda}}_{2l}'+\frac{\boldsymbol{\varepsilon}_1}{\boldsymbol{2}}+\tilde{\boldsymbol{s}}') \, \prod\limits_{l=1}^2\Gamma^\mathbb{C}(-\tilde{\boldsymbol{\lambda}}_{2l}+\frac{\boldsymbol{\varepsilon}_1}{\boldsymbol{2}}-\tilde{\boldsymbol{s}}) \\
		= 2\pi^2(-1)^{\sum m_{3j}+\sum m_{2j}}\|\boldsymbol{\lambda}_{21}-\boldsymbol{\lambda}_{22}\|^2 \sum\limits_{\tau\in S_2}\prod\limits_{l=1}^2\delta^{(2)}(\boldsymbol{\lambda}_{2l}-\boldsymbol{\lambda}_{2\tau(l)}')
	\end{multline}
	similar to the integral in \eqref{gl_3_scalar_product_1}. The delta-functions in the LHS of \eqref{gl_4_scal_prod_int_s_result} came from \eqref{gl_4_scalar_product_1}. The factor $\|\tilde{\boldsymbol{\lambda}}_{21}-\tilde{\boldsymbol{\lambda}}_{22}\|^2\|\tilde{\boldsymbol{\lambda}}_{21}'-\tilde{\boldsymbol{\lambda}}_{22}'\|^2$ appeared from $\rho(\boldsymbol{\lambda})\rho(\boldsymbol{\lambda}')$, see \eqref{gl_4_scalar_product_1}. The integral in \eqref{gl_4_scal_prod_int_s_result} is taken with the help of the same calculations as in $gl_3(\mathbb{C})$.
	
	Thus, combining \eqref{gl_4_scalar_product_1}, \eqref{gl_4_scalar_prod_int_gamma_11}, \eqref{gl_4_scal_prod_int_gamma_21_gamma_22_result} and \eqref{gl_4_scal_prod_int_s_result} one obtains the desired orthogonality relation \eqref{gl_4_orthogonality} for $gl_4(\mathbb{C})$ eigenfunctions.

\end{document}